\pdfoutput=1
\documentclass[10pt]{article}

\usepackage[utf8]{inputenc}
\DeclareUnicodeCharacter{2212}{-}
\usepackage{amsmath}
\usepackage{amssymb}
\usepackage{amsfonts}
\usepackage{pifont}
\usepackage{pdfpages}
\usepackage[most]{tcolorbox} 

\usepackage[backend=biber,
style=numeric,
sortcites,
natbib=true,
sorting=none]{biblatex}
\usepackage[]{titlesec}
\titleformat{\chapter}[display]
  {\normalfont\bfseries}{}{0pt}{\LARGE}
  \titlespacing*{\chapter}{0pt}{-50pt}{10pt}

\usepackage[T1]{fontenc}    
\usepackage{enumitem}

\usepackage{booktabs}       
\usepackage{amsfonts}       
\usepackage{nicefrac}       
\usepackage{microtype}      
\usepackage{listings}
\usepackage{dashrule}
\usepackage{wrapfig}
\usepackage{mdframed}

\usepackage{amsmath,amsfonts,bm}

\def\eqref#1{equation~\ref{#1}}

\def\1{\bm{1}}

\DeclareMathAlphabet{\mathsfit}{\encodingdefault}{\sfdefault}{m}{sl}
\SetMathAlphabet{\mathsfit}{bold}{\encodingdefault}{\sfdefault}{bx}{n}

\usepackage{amsfonts}

\usepackage[breaklinks=true,colorlinks,citecolor=black,bookmarks=false]{hyperref}
\hypersetup{
    colorlinks=false,
	linkcolor=blue,
	filecolor=magenta,      
	urlcolor=blue,
	citecolor=black,
	pdfinfo={
		Title={An Overview of Catastrophic AI Risks},
		Author={Dan Hendrycks, Mantas Mazeika, Thomas Woodside},
		Subject={AI Safety, AI X-Risk, AI Risk},
		Keywords={x-risk, existential risk, catastrophic risk, competitive pressures, ai race, weaponization, ai evolution, malicious use, bioweapons, cyberattacks, automation, power inequality, accidents, safety culture, rogue ai, deception}
	}
}

\usepackage{times}
\usepackage{url}
\usepackage{lipsum}
\usepackage{bbm}
\usepackage{wrapfig}
\usepackage{graphicx}
\usepackage{subcaption}
\usepackage{multirow}
\usepackage{todonotes}
\usepackage{amssymb}
\usepackage{amsfonts}
\usepackage{microtype}      
\usepackage{cleveref}
\usepackage{titling} 
\usepackage[most]{tcolorbox}
\usepackage{tabularray}
\UseTblrLibrary{booktabs}

\usepackage{spverbatim}

\usepackage{xcolor}
\definecolor{dodgerblue}{RGB}{53, 133, 212}
\definecolor{limegreen}{RGB}{79, 171, 79}

\usepackage{float}
\newfloat{storyfloat}{htbp}{loa}
\floatname{storyfloat}{Story}

\newcounter{story}[section]

\newcounter{vision}[section]

\usepackage{tcolorbox}

\newtcolorbox{storybox}[2][Story]{
    title=#2,  
    fonttitle=\large,  
    coltitle=white,  
    colbacktitle=dodgerblue,  
    colback=dodgerblue!10!white,  
    colframe=dodgerblue,  
    parbox=false, 
    breakable
}

\newtcolorbox{visionbox}[2][Vision]{
    title=#2,  
    fonttitle=\large,  
    coltitle=white,  
    colbacktitle=limegreen,  
    colback=limegreen!10!white,  
    colframe=limegreen,  
    parbox=false, 
    breakable
}

\title{
\vspace{-25pt}
\hrule height 4pt
\vskip 0.25in
{\LARGE\bf An Overview of Catastrophic AI Risks}
\vskip 0.29in
\hrule height 1pt
\vskip 0.09in
}
\date{}

\renewenvironment{abstract}%
{%
  \vskip 0.075in%
  \centerline%
  {\large\bf Abstract}%
  \vspace{0.5ex}%
  \begin{quote}%
}
{
  \par%
  \end{quote}%
  \vskip 1ex%
}

\author{\textbf{Dan Hendrycks}\\
Center for AI Safety
\and
\textbf{Mantas Mazeika}\\
Center for AI Safety
\and
\textbf{Thomas Woodside}\\
Center for AI Safety
}

\usepackage{arydshln}

\newcommand{\reviewer}[3]{
	\expandafter\newcommand\csname #1\endcsname[1]{
		\textcolor{#3}{[#2: ##1]}
	}
}
\definecolor{neonpurple}{rgb}{0.3,0,1}
\reviewer{dan}{Dan}{neonpurple}

\AtBeginBibliography{\small}

\begin{document}
\includepdf[pages={1}]{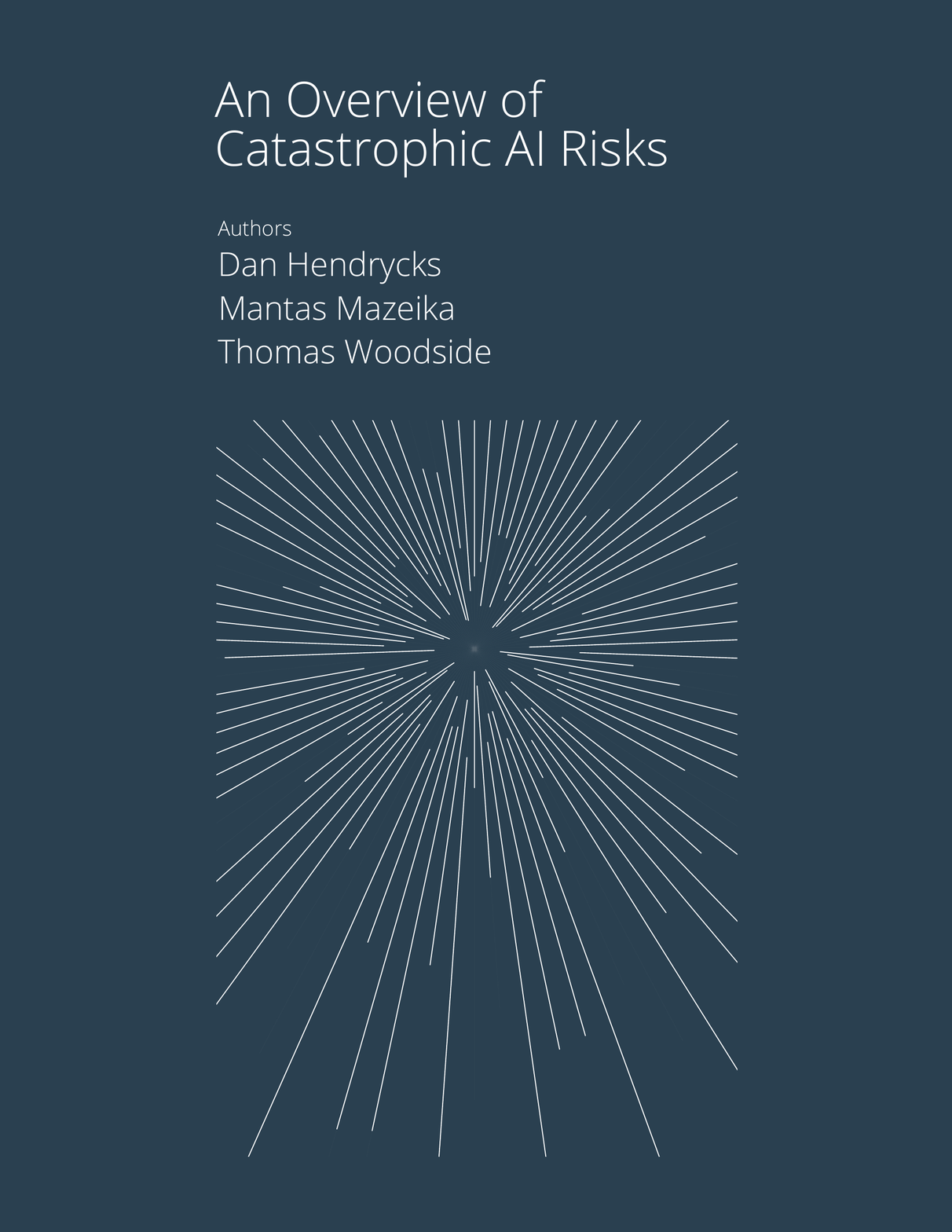}
\begin{titlepage}
\end{titlepage}

\maketitle

\vspace*{-15pt}
\begin{abstract}

\normalsize




Rapid advancements in artificial intelligence (AI) have sparked growing concerns among experts, policymakers, and world leaders regarding the potential for increasingly advanced AI systems to pose catastrophic risks. Although numerous risks have been detailed separately, there is a pressing need for a systematic discussion and illustration of the potential dangers to better inform efforts to mitigate them. This paper provides an overview of the main sources of catastrophic AI risks, which we organize into four categories: malicious use, in which individuals or groups intentionally use AIs to cause harm; AI race, in which competitive environments compel actors to deploy unsafe AIs or cede control to AIs; organizational risks, highlighting how human factors and complex systems can increase the chances of catastrophic accidents; and rogue AIs, describing the inherent difficulty in controlling agents far more intelligent than humans. For each category of risk, we describe specific hazards, present illustrative stories, envision ideal scenarios, and propose practical suggestions for mitigating these dangers. Our goal is to foster a comprehensive understanding of these risks and inspire collective and proactive efforts to ensure that AIs are developed and deployed in a safe manner. Ultimately, we hope this will allow us to realize the benefits of this powerful technology while minimizing the potential for catastrophic outcomes.\footnote{This paper is for a wide audience, unlike most of our writing, which is for empirical AI researchers. We use imagery, stories, and a simplified style to discuss the risks that advanced AIs could pose, because we think this is an important topic for everyone.}

\end{abstract}

\newpage

\section*{Executive Summary}

Artificial intelligence (AI) has seen rapid advancements in recent years, raising concerns among AI experts, policymakers, and world leaders about the potential risks posed by advanced AIs. As with all powerful technologies, AI must be handled with great responsibility to manage the risks and harness its potential for the betterment of society. However, there is limited accessible information on how catastrophic or existential AI risks might transpire or be addressed. While numerous sources on this subject exist, they tend to be spread across various papers, often targeted toward a narrow audience or focused on specific risks. In this paper, we provide an overview of the main sources of catastrophic AI risk, which we organize into four categories:

\paragraph{Malicious use.} Actors could intentionally harness powerful AIs to cause widespread harm. Specific risks include \textit{bioterrorism} enabled by AIs that can help humans create deadly pathogens; the deliberate dissemination of \textit{uncontrolled AI agents}; and the use of AI capabilities for \textit{propaganda, censorship, and surveillance}. To reduce these risks, we suggest improving biosecurity, restricting access to the most dangerous AI models, and holding AI developers legally liable for damages caused by their AI systems.

\paragraph{AI race.} Competition could pressure nations and corporations to rush the development of AIs and cede control to AI systems. Militaries might face pressure to develop \textit{autonomous weapons} and use AIs for \textit{cyberwarfare}, enabling a new kind of \textit{automated warfare} where accidents can spiral out of control before humans have the chance to intervene. Corporations will face similar incentives to \textit{automate human labor} and \textit{prioritize profits over safety}, potentially leading to \textit{mass unemployment} and \textit{dependence on AI systems}. We also discuss how \textit{evolutionary pressures} might shape AIs in the long run. Natural selection among AIs may lead to selfish traits, and the advantages AIs have over humans could eventually lead to the displacement of humanity. To reduce risks from an AI race, we suggest implementing safety regulations, international coordination, and public control of general-purpose AIs.

\paragraph{Organizational risks.} Organizational accidents have caused disasters including Chernobyl, Three Mile Island, and the Challenger Space Shuttle disaster. Similarly, the organizations developing and deploying advanced AIs could suffer catastrophic accidents, particularly if they do not have a strong \textit{safety culture}. AIs could be accidentally leaked to the public or stolen by malicious actors. Organizations could fail to invest in safety research, lack understanding of how to reliably \textit{improve AI safety faster than general AI capabilities}, or suppress internal concerns about AI risks. To reduce these risks, better organizational cultures and structures can be established, including internal and external audits, multiple layers of defense against risks, and state-of-the-art information security.

\paragraph{Rogue AIs.} A common and serious concern is that we might lose control over AIs as they become more intelligent than we are. AIs could optimize flawed objectives to an extreme degree in a process called \textit{proxy gaming}. AIs could experience \textit{goal drift} as they adapt to a changing environment, similar to how people acquire and lose goals throughout their lives. In some cases, it might be instrumentally rational for AIs to become \textit{power-seeking}. We also look at how and why AIs might engage in \textit{deception}, appearing to be under control when they are not. These problems are more technical than the first three sources of risk. We outline some suggested research directions for advancing our understanding of how to ensure AIs are controllable.

\vspace{10pt}
\noindent Throughout each section, we provide illustrative scenarios that demonstrate more concretely how the sources of risk might lead to catastrophic outcomes or even pose existential threats. By offering a positive vision of a safer future in which risks are managed appropriately, we emphasize that the emerging risks of AI are serious but not insurmountable. By proactively addressing these risks, we can work toward realizing the benefits of AI while minimizing the potential for catastrophic outcomes.

\newpage
\tableofcontents
\newpage



\section{Introduction}

The world as we know it is not normal. We take for granted that we can talk instantaneously with people thousands of miles away, fly to the other side of the world in less than a day, and access vast mountains of accumulated knowledge on devices we carry around in our pockets. These realities seemed far-fetched decades ago, and would have been inconceivable to people living centuries ago. The ways we live, work, travel, and communicate have only been possible for a tiny fraction of human history. 

Yet, when we look at the bigger picture, a broader pattern emerges: accelerating development. Hundreds of thousands of years elapsed between the time Homo sapiens appeared on Earth and the agricultural revolution. Then, thousands of years passed before the industrial revolution. Now, just centuries later, the artificial intelligence (AI) revolution is beginning. The march of history is not constant---it is rapidly accelerating.

\begin{wrapfigure}{r}[0\textwidth]{.55\textwidth}%
	\vspace{-11pt}%
	\centering
	\includegraphics[width=0.54\textwidth]{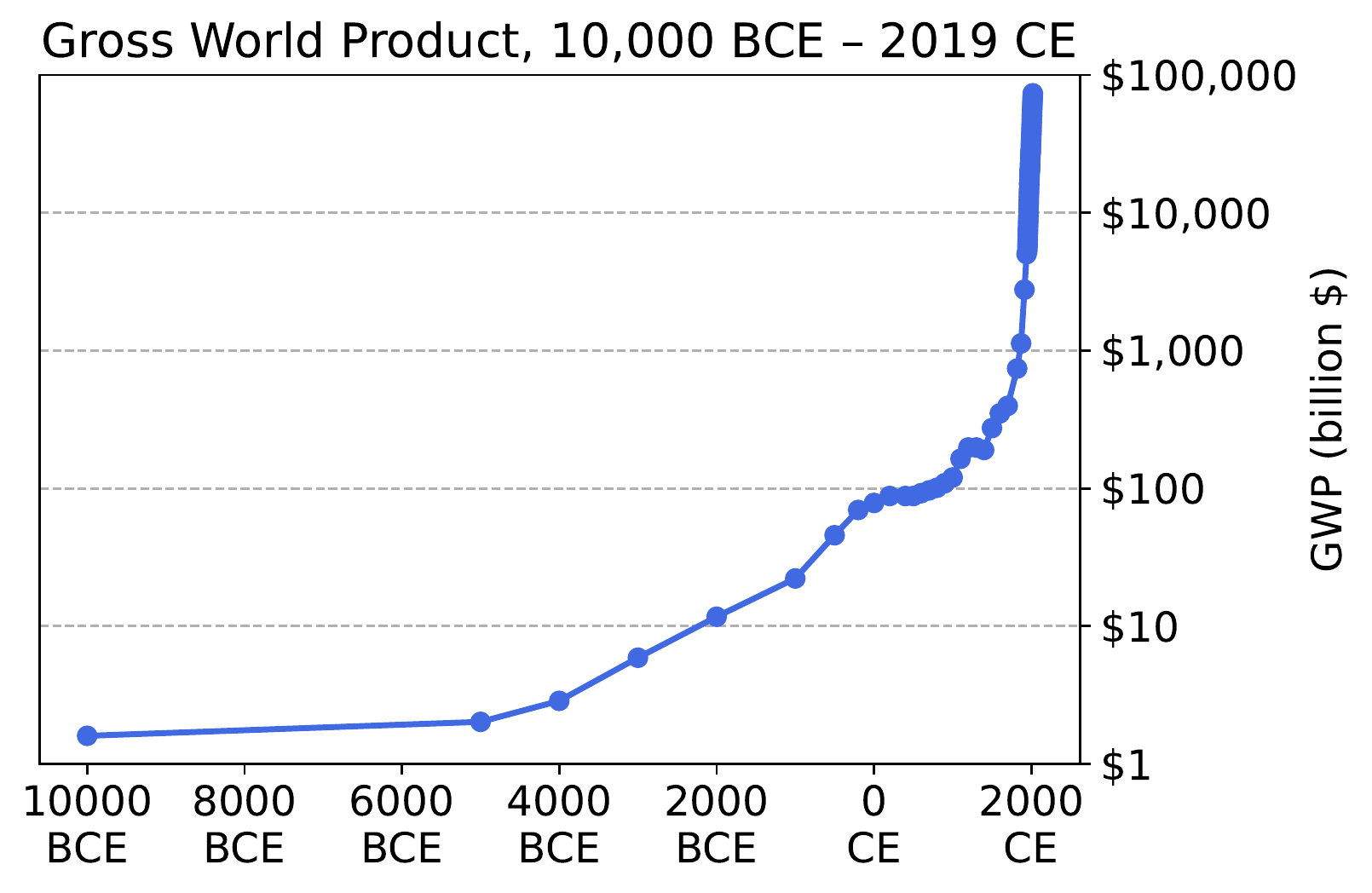}
        \vspace{-2pt}
	\caption{World production has grown rapidly over the course of human history. AI could further this trend, catapulting humanity into a new period of unprecedented change.}
	\label{fig:gwp}
	\vspace{-10pt}%
\end{wrapfigure}

We can capture this trend quantitatively in \Cref{fig:gwp}, which shows how estimated gross world product has changed over time \citep{Roodman2020OnTP, Davidson2021}. The hyperbolic growth it depicts might be explained by the fact that, as technology advances, the rate of technological advancement also tends to increase. Empowered with new technologies, people can innovate faster than they could before. Thus, the gap in time between each landmark development narrows.\looseness=-1

It is the rapid pace of development, as much as the sophistication of our technology, that makes the present day an unprecedented time in human history. We have reached a point where technological advancements can transform the world beyond recognition within a human lifetime. For example, people who have lived through the creation of the internet can remember a time when our now digitally-connected world would have seemed like science fiction.

From a historical perspective, it appears possible that the same amount of development could now be condensed in an even shorter timeframe. We might not be certain that this will occur, but neither can we rule it out. We therefore wonder: what new technology might usher in the next big acceleration? In light of recent advances, AI seems an increasingly plausible candidate. Perhaps, as AI continues to become more powerful, it could lead to a qualitative shift in the world, more profound than any we have experienced so far. It could be the most impactful period in history, though it could also be the last.

Although technological advancement has often improved people's lives, we ought to remember that, as our technology grows in power, so too does its destructive potential. Consider the invention of nuclear weapons. Last century, for the first time in our species' history, humanity possessed the ability to destroy itself, and the world suddenly became much more fragile.

Our newfound vulnerability revealed itself in unnerving clarity during the Cold War. On a Saturday in October 1962, the Cuban Missile Crisis was cascading out of control. US warships enforcing the blockade of Cuba detected a Soviet submarine and attempted to force it to the surface by dropping low-explosive depth charges. The submarine was out of radio contact, and its crew had no idea whether World War III had already begun. A broken ventilator raised the temperature up to $140^{\circ}$F in some parts of the submarine, causing crew members to fall unconscious as depth charges exploded nearby.

\begin{figure}[t]
    \vspace{-15pt}
    \centering
    \includegraphics[width=\textwidth]{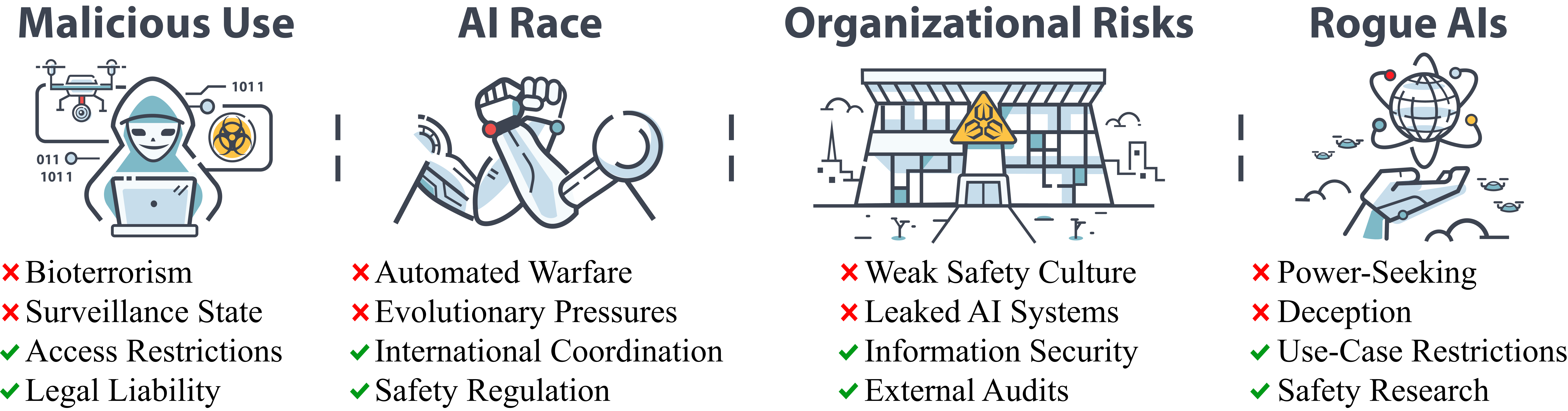}
    \caption{In this paper we cover four categories of AI risks and discuss how to mitigate them.}
    \label{fig:splash}
    \vspace{-5pt}
\end{figure}

The submarine carried a nuclear-armed torpedo, which required consent from both the captain and political officer to launch. Both provided it. On any other submarine in Cuban waters that day, that torpedo would have launched---and a nuclear third world war may have followed. Fortunately, a man named Vasili Arkhipov was also on the submarine. Arkhipov was the commander of the entire flotilla and by sheer luck happened to be on that particular submarine. He talked the captain down from his rage, convincing him to await further orders from Moscow. 
He averted a nuclear war and saved millions or billions of lives---and possibly civilization itself.\looseness=-1

Carl Sagan once observed, ``If we continue to accumulate only power and not wisdom, we will surely destroy ourselves'' \citep{sagan1994pale}. Sagan was correct: The power of nuclear weapons was not one we were ready for. Overall, it has been luck rather than wisdom that has saved humanity from nuclear annihilation, with multiple recorded instances of a single individual preventing a full-scale nuclear war.

AI is now poised to become a powerful technology with destructive potential similar to nuclear weapons. We do not want to repeat the Cuban Missile Crisis. We do not want to slide toward a moment of peril where our survival hinges on luck rather than the ability to use this technology wisely. Instead, we need to work proactively to mitigate the risks it poses. This necessitates a better understanding of what could go wrong and what to do about it.

Luckily, AI systems are not yet advanced enough to contribute to every risk we discuss. But that is cold comfort in a time when AI development is advancing at an unprecedented and unpredictable rate. We consider risks arising from both present-day AIs and AIs that are likely to exist in the near future. It is possible that if we wait for more advanced systems to be developed before taking action, it may be too late.\looseness=-1

In this paper, we will explore various ways in which powerful AIs could bring about catastrophic events with devastating consequences for vast numbers of people. We will also discuss how AIs could present existential risks---catastrophes from which humanity would be unable to recover. The most obvious such risk is extinction, but there are other outcomes, such as creating a permanent dystopian society, which would also constitute an existential catastrophe. We outline many possible catastrophes, some of which are more likely than others and some of which are mutually incompatible with each other. This approach is motivated by the principles of risk management. We prioritize asking ``what could go wrong?'' rather than reactively waiting for catastrophes to occur. This proactive mindset enables us to anticipate and mitigate catastrophic risks before it's too late.

To help orient the discussion, we decompose catastrophic risks from AIs into four risk sources that warrant intervention:
\begin{itemize}
    \item \textbf{Malicious use}: Malicious actors using AIs to cause large-scale devastation.
    \item \textbf{AI race}: Competitive pressures that could drive us to deploy AIs in unsafe ways, despite this being in no one's best interest.
    \item \textbf{Organizational risks}: Accidents arising from the complexity of AIs and the organizations developing them.
    \item \textbf{Rogue AIs}: The problem of controlling a technology more intelligent than we are.
\end{itemize}

These four sections---malicious use, AI race, organizational risks, and rogue AIs---describe causes of AI risks that are \textit{intentional}, \textit{environmental/structural}, \textit{accidental}, and \textit{internal}, respectively \citep{Yampolskiy2016TaxonomyOP}.

We will describe how concrete, small-scale examples of each risk might escalate into catastrophic outcomes. We also include hypothetical stories to help readers conceptualize the various processes and dynamics discussed in each section, along with practical safety suggestions to avoid negative outcomes. Each section concludes with an ideal vision depicting what it would look like to mitigate that risk. We hope this survey will serve as a practical introduction for readers interested in learning about and mitigating catastrophic AI risks.



\section{Malicious Use}

On the morning of March 20, 1995, five men entered the Tokyo subway system. After boarding separate subway lines, they continued for several stops before dropping the bags they were carrying and exiting. An odorless, colorless liquid inside the bags began to vaporize. Within minutes, commuters began choking and vomiting. The trains continued on toward the heart of Tokyo, with sickened passengers leaving the cars at each station. The fumes were spread at each stop, either by emanating from the tainted cars or through contact with people's clothing and shoes. By the end of the day, 13 people lay dead and 5,800 seriously injured. The group responsible for the attack was the religious cult Aum Shinrikyo \citep{Olson1999AumSO}. Its motive for murdering innocent people? To bring about the end of the world.

Powerful new technologies offer tremendous potential benefits, but they also carry the risk of empowering malicious actors to cause widespread harm. There will always be those with the worst of intentions, and AIs could provide them with a formidable tool to achieve their objectives. Moreover, as AI technology advances, severe malicious use could potentially destabilize society, increasing the likelihood of other risks.

In this section, we will explore the various ways in which the malicious use of advanced AIs could pose catastrophic risks. These include engineering biochemical weapons, unleashing rogue AIs, using persuasive AIs to spread propaganda and erode consensus reality, and leveraging censorship and mass surveillance to irreversibly concentrate power. We will conclude by discussing possible strategies for mitigating the risks associated with the malicious use of AIs.

\paragraph{Unilateral actors considerably increase the risks of malicious use.} In instances where numerous actors have access to a powerful technology or dangerous information that could be used for harmful purposes, it only takes one individual to cause significant devastation. Malicious actors themselves are the clearest example of this, but recklessness can be equally dangerous. For example, a single research team might be excited to open source an AI system with biological research capabilities, which would speed up research and potentially save lives, but this could also increase the risk of malicious use if the AI system could be repurposed to develop bioweapons. In situations like this, the outcome may be determined by the least risk-averse research group. If only one research group thinks the benefits outweigh the risks, it could act unilaterally, deciding the outcome even if most others don't agree. And if they are wrong and someone does decide to develop a bioweapon, it would be too late to reverse course.

By default, advanced AIs may increase the destructive capacity of both the most powerful and the general population. Thus, the growing potential for AIs to empower malicious actors is one of the most severe threats humanity will face in the coming decades. The examples we give in this section are only those we can foresee. It is possible that AIs could aid in the creation of dangerous new technology we cannot presently imagine, which would further increase risks from malicious use.

\subsection{Bioterrorism}

The rapid advancement of AI technology increases the risk of bioterrorism. AIs with knowledge of bioengineering could facilitate the creation of novel bioweapons and lower barriers to obtaining such agents. Engineered pandemics from AI-assisted bioweapons pose a unique challenge, as attackers have an advantage over defenders and could constitute an existential threat to humanity. We will now examine these risks and how AIs might exacerbate challenges in managing bioterrorism and engineered pandemics.

\paragraph{Bioengineered pandemics present a new threat.} Biological agents, including viruses and bacteria, have caused some of the most devastating catastrophes in history. It's believed the Black Death killed more humans than any other event in history, an astounding and awful 200 million, the equivalent to four billion deaths today. While contemporary advancements in science and medicine have made great strides in mitigating risks associated with natural pandemics, engineered pandemics could be designed to be more lethal or easily transmissible than natural pandemics, presenting a new threat that could equal or even surpass the devastation wrought by history's most deadly plagues \citep{esvelt2022delay}.

Humanity has a long and dark history of weaponizing pathogens, with records dating back to 1320 BCE describing a war in Asia Minor where infected sheep were driven across the border to spread Tularemia \citep{Trevisanato2007TheP}. During the twentieth century, 15 countries are known to have developed bioweapons programs, including the US, USSR, UK, and France. Like chemical weapons, bioweapons have become a taboo among the international community. While some state actors continue to operate bioweapons programs \citep{us_state_department_2022}, a more significant risk may come from non-state actors like Aum Shinrikyo, ISIS, or simply disturbed individuals. Due to advancements in AI and biotechnology, the tools and knowledge necessary to engineer pathogens with capabilities far beyond Cold War-era bioweapons programs will rapidly democratize.

\paragraph{Biotechnology is progressing rapidly and becoming more accessible.} A few decades ago, the ability to synthesize new viruses was limited to a handful of the top scientists working in advanced laboratories. Today it is estimated that there are 30,000 people with the talent, training, and access to technology to create new pathogens \citep{esvelt2022delay}. This figure could rapidly expand. Gene synthesis, which allows the creation of custom biological agents, has dropped precipitously in price, with its cost halving approximately every 15 months \citep{carlson_changing_2009}. Furthermore, with the advent of benchtop DNA synthesis machines, access will become much easier and could avoid existing gene synthesis screening efforts, which complicates controlling the spread of such technology \citep{carter2023benchtop}. The chances of a bioengineered pandemic killing millions, perhaps billions, is proportional to the number of people with the skills and access to the technology to synthesize them. With AI assistants, orders of magnitude more people could have the required skills, thereby increasing the risks by orders of magnitude.\looseness=-1

\begin{wrapfigure}{l}[0\textwidth]{.34\textwidth}%
	\vspace{-20pt}%
	\centering
	\includegraphics[width=0.33\textwidth]{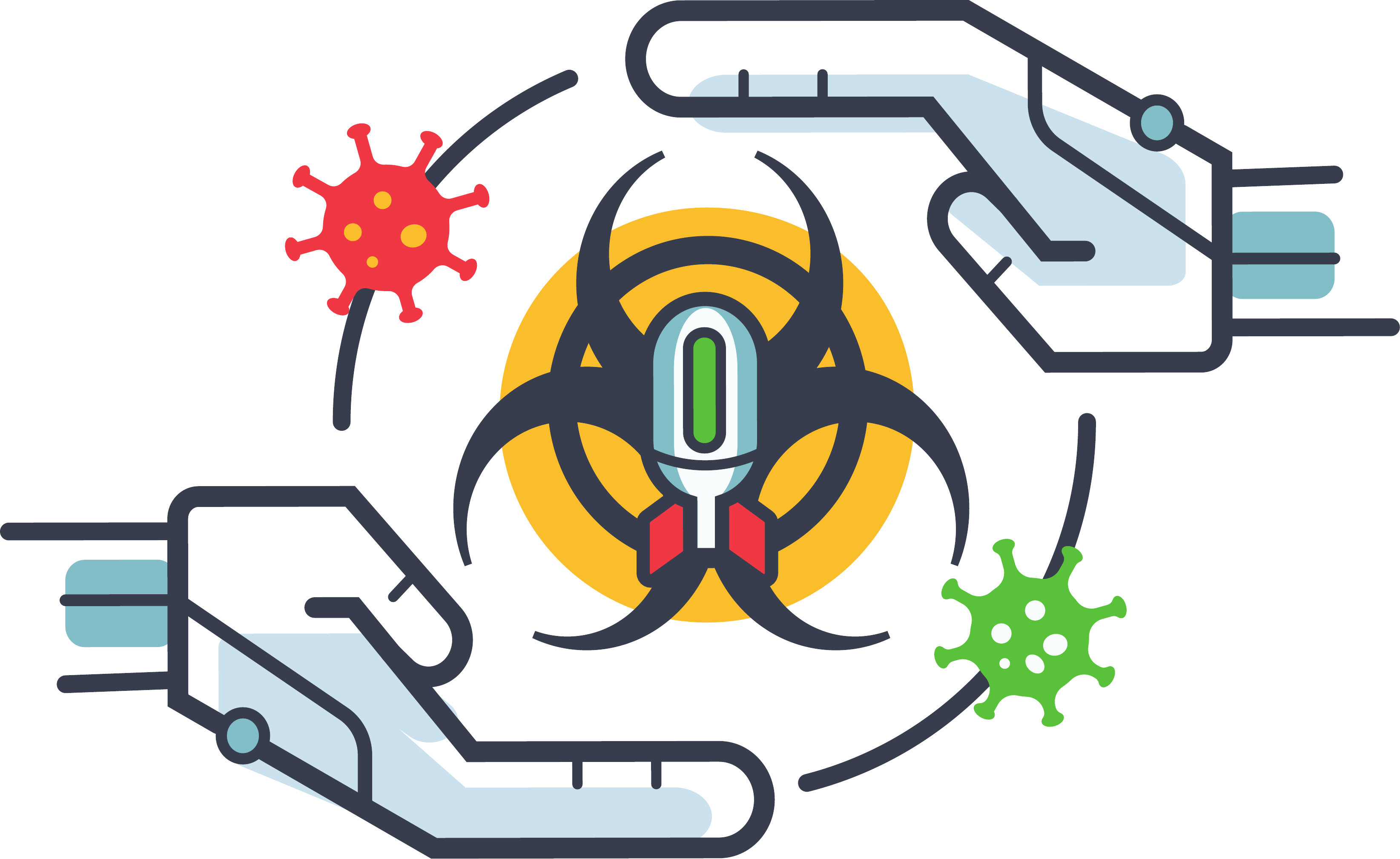}
	\caption{An AI assistant could provide non-experts with access to the directions and designs needed to produce biological and chemical weapons and facilitate malicious use.}
	\label{fig:bioweapon}
	\vspace{-13pt}%
\end{wrapfigure}

\paragraph{AIs could be used to expedite the discovery of new, more deadly chemical and biological weapons.} In 2022, researchers took an AI system designed to create new drugs by generating non-toxic, therapeutic molecules and tweaked it to reward, rather than penalize, toxicity \citep{Urbina2022DualUO}. After this simple change, within six hours, it generated 40,000 candidate chemical warfare agents entirely on its own. It designed not just known deadly chemicals including VX, but also novel molecules that may be deadlier than any chemical warfare agents discovered so far. In the field of biology, AIs have already surpassed human abilities in protein structure prediction \citep{AlphaFold2021} and made contributions to synthesizing those proteins \citep{wu2019machine}. Similar methods could be used to create bioweapons and develop pathogens that are deadlier, more transmissible, and more difficult to treat than anything seen before.

\paragraph{AIs compound the threat of bioengineered pandemics.} AIs will increase the number of people who could commit acts of bioterrorism. General-purpose AIs like ChatGPT are capable of synthesizing expert knowledge about the deadliest known pathogens, such as influenza and smallpox, and providing step-by-step instructions about how a person could create them while evading safety protocols \citep{Soice2023CanLL}. Future versions of AIs could be even more helpful to potential bioterrorists when AIs are able to synthesize information into techniques, processes, and knowledge that is not explicitly available anywhere on the internet. Public health authorities may respond to these threats with safety measures, but in bioterrorism, the attacker has the advantage. The exponential nature of biological threats means that a single attack could spread to the entire world before an effective defense could be mounted. Only 100 days after being detected and sequenced, the omicron variant of COVID-19 had infected a quarter of the United States and half of Europe \citep{esvelt2022delay}. Quarantines and lockdowns instituted to suppress the COVID-19 pandemic caused a global recession and still could not prevent the disease from killing millions worldwide.\looseness=-1

In summary, advanced AIs could constitute a weapon of mass destruction in the hands of terrorists, by making it easier for them to design, synthesize, and spread deadly new pathogens. By reducing the required technical expertise and increasing the lethality and transmissibility of pathogens, AIs could enable malicious actors to cause global catastrophe by unleashing pandemics. 

\subsection{Unleashing AI Agents}

Many technologies are \textit{tools} that humans use to pursue our goals, such as hammers, toasters, and toothbrushes. But AIs are increasingly built as \textit{agents} which autonomously take actions in the world in order to pursue open-ended goals. AI agents can be given goals such as winning games, making profits on the stock market, or driving a car to a destination. AI agents therefore pose a unique risk: people could build AIs that pursue dangerous goals.

\paragraph{Malicious actors could intentionally create rogue AIs.} One month after the release of GPT-4, an open-source project bypassed the AI's safety filters and turned it into an autonomous AI agent instructed to ``destroy humanity,'' ``establish global dominance,'' and ``attain immortality.'' Dubbed ChaosGPT, the AI compiled research on nuclear weapons and sent tweets trying to influence others. Fortunately, ChaosGPT was merely a warning given that it lacked the ability to successfully formulate long-term plans, hack computers, and survive and spread. Yet given the rapid pace of AI development, ChaosGPT did offer a glimpse into the risks that more advanced rogue AIs could pose in the near future.

\paragraph{Many groups may want to unleash AIs or have AIs displace humanity.} Simply unleashing rogue AIs, like a more sophisticated version of ChaosGPT, could accomplish mass destruction, even if those AIs aren't explicitly told to harm humanity. There are a variety of beliefs that may drive individuals and groups to do so. One ideology that could pose a unique threat in this regard is ``accelerationism.'' This ideology seeks to accelerate AI development as rapidly as possible and opposes restrictions on the development or proliferation of AIs. This sentiment is alarmingly common among many leading AI researchers and technology leaders, some of whom are intentionally racing to build AIs more intelligent than humans. According to Google co-founder Larry Page, AIs are humanity's rightful heirs and the next step of cosmic evolution. He has also expressed the sentiment that humans maintaining control over AIs is ``speciesist'' \citep{tegmark2018life}. J\"{u}rgen Schmidhuber, an eminent AI scientist, argued that ``In the long run, humans will not remain the crown of creation... But that's okay because there is still beauty, grandeur, and greatness in realizing that you are a tiny part of a much grander scheme which is leading the universe from lower complexity towards higher complexity'' \citep{pooley2020}. Richard Sutton, another leading AI scientist, in discussing smarter-than human AI asked ``why shouldn't those who are the smartest become powerful?'' and thinks the development of superintelligence will be an achievement ``beyond humanity, beyond life, beyond good and bad'' \citep{sutton_it_2022}. He argues that ``succession to AI is inevitable,'' and while ``they could displace us from existence,'' ``we should not resist succession'' \cite{sutton_succession_2023}.

There are several sizable groups who may want to unleash AIs to intentionally cause harm. For example, sociopaths and psychopaths make up around 3 percent of the population \citep{SanzGarca2021PrevalenceOP}. In the future, people who have their livelihoods destroyed by AI automation may grow resentful, and some may want to retaliate. There are plenty of cases in which seemingly mentally stable individuals with no history of insanity or violence suddenly go on a shooting spree or plant a bomb with the intent to harm as many innocent people as possible. We can also expect well-intentioned people to make the situation even more challenging. As AIs advance, they could make ideal companions---knowing how to provide comfort, offering advice when needed, and never demanding anything in return. Inevitably, people will develop emotional bonds with chatbots, and some will demand that they be granted rights or become autonomous.

In summary, releasing powerful AIs and allowing them to take actions independently of humans could lead to a catastrophe. There are many reasons that people might pursue this, whether because of a desire to cause harm, an ideological belief in technological acceleration, or a conviction that AIs should have the same rights and freedoms as humans.

\subsection{Persuasive AIs}

The deliberate propagation of disinformation is already a serious issue, reducing our shared understanding of reality and polarizing opinions. AIs could be used to severely exacerbate this problem by generating personalized disinformation on a larger scale than before. Additionally, as AIs become better at predicting and nudging our behavior, they will become more capable at manipulating us. We will now discuss how AIs could be leveraged by malicious actors to create a fractured and dysfunctional society.

\paragraph{AIs could pollute the information ecosystem with motivated lies.} Sometimes ideas spread not because they are true, but because they serve the interests of a particular group. ``Yellow journalism'' was coined as a pejorative reference to newspapers that advocated war between Spain and the United States in the late 19th century, because they believed that sensational war stories would boost their sales \cite{yellowjournalism}. When public information sources are flooded with falsehoods, people will sometimes fall prey to lies, or else come to distrust mainstream narratives, both of which undermine societal integrity.

Unfortunately, AIs could escalate these existing problems dramatically. First, AIs could be used to generate unique, personalized disinformation at a large scale. While there are already many social media bots \citep{Varol2017OnlineHI}, some of which exist to spread disinformation, historically they have been run by humans or primitive text generators. The latest AI systems do not need humans to generate personalized messages, never get tired, and could potentially interact with millions of users at once \citep{Burtell2023ArtificialIA}.

\begin{wrapfigure}{r}[0\textwidth]{.31\textwidth}%
	\vspace{-15pt}%
	\centering
	\includegraphics[width=0.30\textwidth]{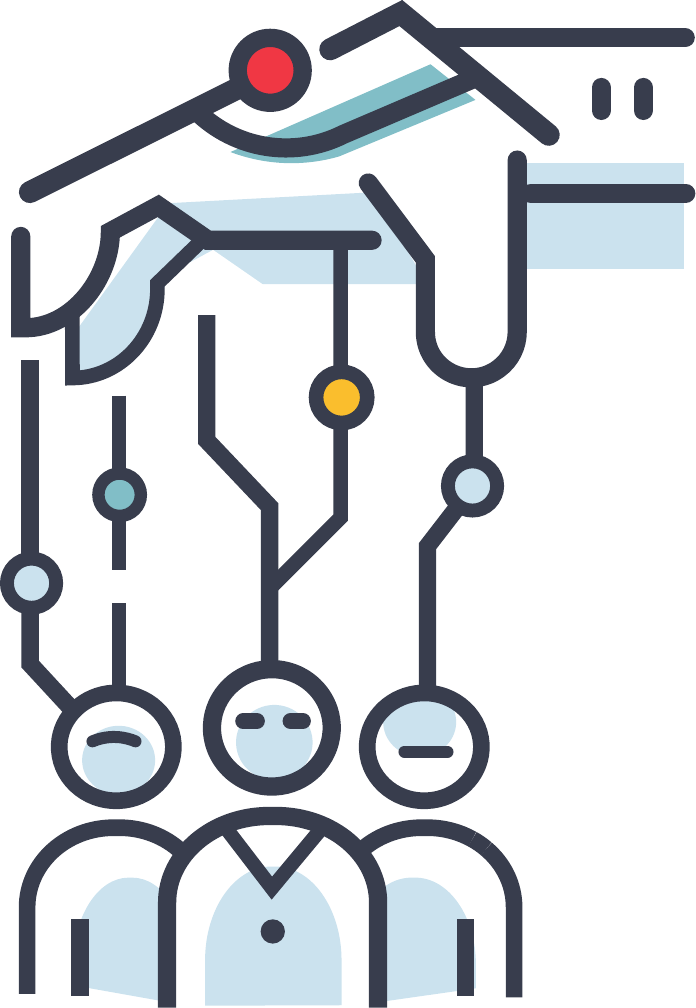}
	\caption{AIs will enable sophisticated personalized influence campaigns that may destabilize our shared sense of reality.}
	\label{fig:persuasive}
	\vspace{-10pt}%
\end{wrapfigure}

\paragraph{AIs can exploit users' trust.} Already, hundreds of thousands of people pay for chatbots marketed as lovers and friends \citep{Tong2023}, and one man's suicide has been partially attributed to interactions with a chatbot \citep{Lovens2023}. As AIs appear increasingly human-like, people will increasingly form relationships with them and grow to trust them. AIs that gather personal information through relationship-building or by accessing extensive personal data, such as a user's email account or personal files, could leverage that information to enhance persuasion. Powerful actors that control those systems could exploit user trust by delivering personalized disinformation directly through people's ``friends.''

\paragraph{AIs could centralize control of trusted information.} Separate from democratizing disinformation, AIs could centralize the creation and dissemination of trusted information. Only a few actors have the technical skills and resources to develop cutting-edge AI systems, and they could use these AIs to spread their preferred narratives. Alternatively, if AIs are broadly accessible this could lead to widespread disinformation, with people retreating to trusting only a small handful of authoritative sources \citep{Vaccari2020DeepfakesAD}. In both scenarios, there would be fewer sources of trusted information and a small portion of society would control popular narratives. 

AI censorship could further centralize control of information. This could begin with good intentions, such as using AIs to enhance fact-checking and help people avoid falling prey to false narratives. This would not necessarily solve the problem, as disinformation persists today despite the presence of fact-checkers.

Even worse, purported ``fact-checking AIs'' might be designed by authoritarian governments and others to suppress the spread of true information. Such AIs could be designed to correct most common misconceptions but provide incorrect information about some sensitive topics, such as human rights violations committed by certain countries. But even if fact-checking AIs work as intended, the public might eventually become entirely dependent on them to adjudicate the truth, reducing people's autonomy and making them vulnerable to failures or hacks of those systems.

In a world with widespread persuasive AI systems, people's beliefs might be almost entirely determined by which AI systems they interact with most. Never knowing whom to trust, people could retreat even further into ideological enclaves, fearing that any information from outside those enclaves might be a sophisticated lie. This would erode consensus reality, people's ability to cooperate with others, participate in civil society, and address collective action problems. This would also reduce our ability to have a conversation as a species about how to mitigate existential risks from AIs.

In summary, AIs could create highly effective, personalized disinformation on an unprecedented scale, and could be particularly persuasive to people they have built personal relationships with. In the hands of many people, this could create a deluge of disinformation that debilitates human society, but, kept in the hands of a few, it could allow governments to control narratives for their own ends.

\subsection{Concentration of Power}

\begin{wrapfigure}{r}[0\textwidth]{.36\textwidth}%
	\vspace{-15pt}%
	\centering
	\includegraphics[width=0.35\textwidth]{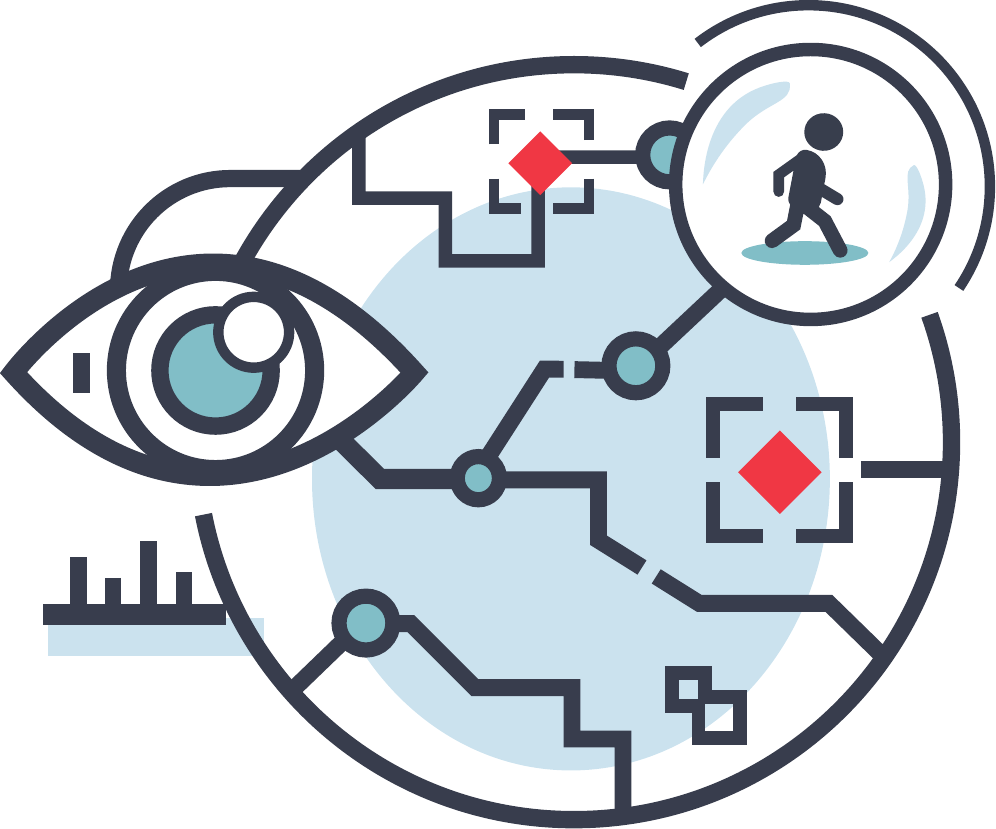}
	\caption{Ubiquitous monitoring tools, tracking and analyzing every individual in detail, could facilitate the complete erosion of freedom and privacy.}
	\label{fig:totalitarianism}
	\vspace{-20pt}%
\end{wrapfigure}

We have discussed several ways in which individuals and groups might use AIs to cause widespread harm, through bioterrorism; releasing powerful, uncontrolled AIs; and disinformation. To mitigate these risks, governments might pursue intense surveillance and seek to keep AIs in the hands of a trusted minority. This reaction, however, could easily become an overcorrection, paving the way for an entrenched totalitarian regime that would be locked in by the power and capacity of AIs. This scenario represents a form of ``top-down'' misuse, as opposed to ``bottom-up'' misuse by citizens, and could in extreme cases culminate in an entrenched dystopian civilization.

\begin{wrapfigure}{l}[0\textwidth]{.36\textwidth}%
	\vspace{-18pt}%
	\centering
	\includegraphics[width=0.35\textwidth]{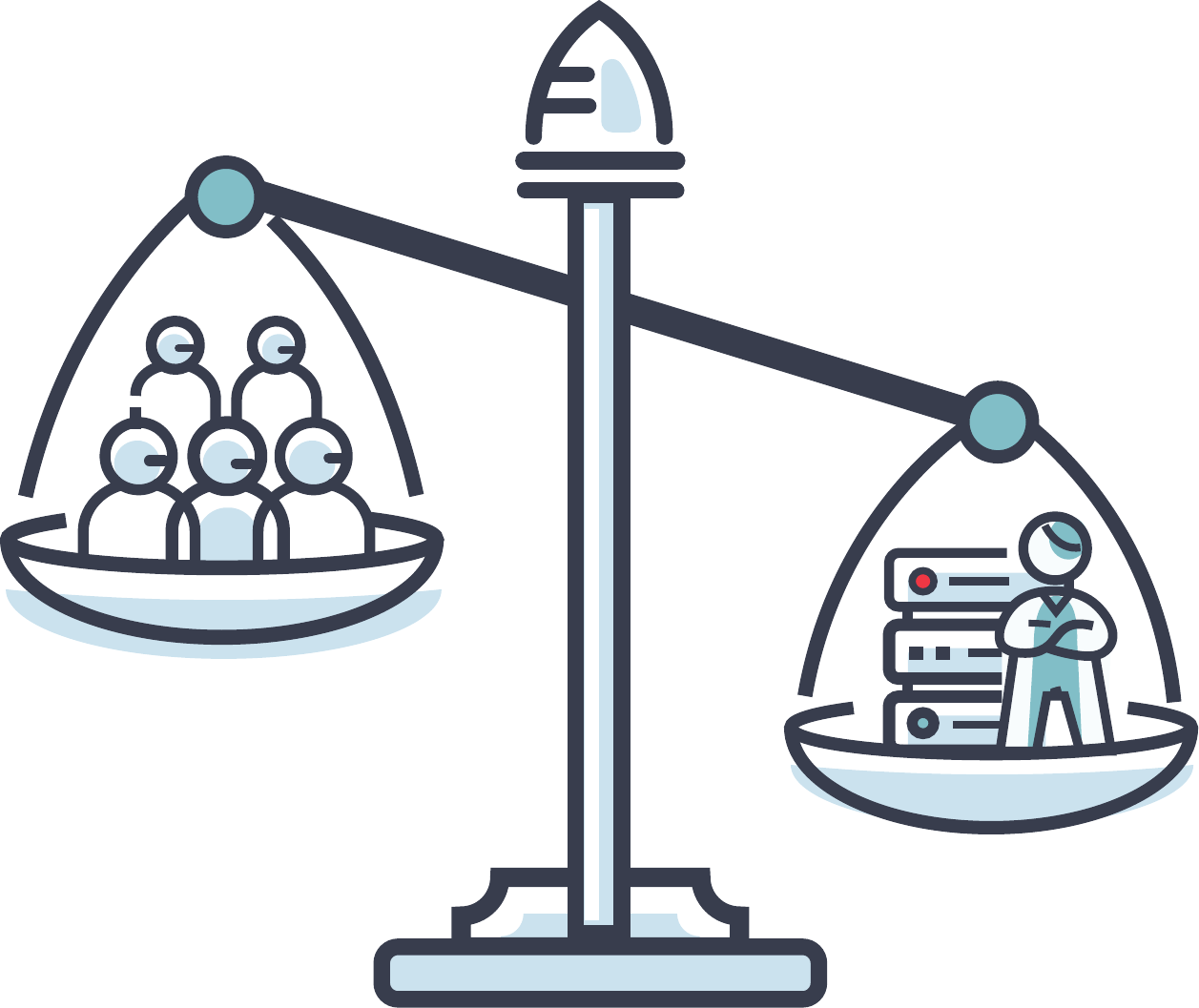}
	\caption{If material control of AIs is limited to few, it could represent the most severe economic and power inequality in human history.}
	\label{fig:inequality}
	\vspace{-18pt}%
\end{wrapfigure}

\paragraph{AIs could lead to extreme, and perhaps irreversible concentration of power.} The persuasive abilities of AIs combined with their potential for surveillance and the advancement of autonomous weapons could allow small groups of actors to ``lock-in'' their control over society, perhaps permanently. To operate effectively, AIs require a broad set of infrastructure components, which are not equally distributed, such as data centers, computing power, and big data. Those in control of powerful systems may use them to suppress dissent, spread propaganda and disinformation, and otherwise advance their goals, which may be contrary to public wellbeing.

\paragraph{AIs may entrench a totalitarian regime.}  In the hands of the state, AIs may result in the erosion of civil liberties and democratic values in general. AIs could allow totalitarian governments to efficiently collect, process, and act on an unprecedented volume of information, permitting an ever smaller group of people to surveil and exert complete control over the population without the need to enlist millions of citizens to serve as willing government functionaries. Overall, as power and control shift away from the public and toward elites and leaders, democratic governments are highly vulnerable to totalitarian backsliding. Additionally, AIs could make totalitarian regimes much longer-lasting; a major way in which such regimes have been toppled previously is at moments of vulnerability like the death of a dictator, but AIs, which would be hard to ``kill,'' could provide much more continuity to leadership, providing few opportunities for reform.

\paragraph{AIs can entrench corporate power at the expense of the public good.} Corporations have long lobbied to weaken laws and policies that restrict their actions and power, all in the service of profit. Corporations in control of powerful AI systems may use them to manipulate customers into spending more on their products even to the detriment of their own wellbeing. The concentration of power and influence that could be afforded by AIs could enable corporations to exert unprecedented control over the political system and entirely drown out the voices of citizens. This could occur even if creators of these systems know their systems are self-serving or harmful to others, as they would have incentives to reinforce their power and avoid distributing control.

\paragraph{In addition to power, locking in certain values may curtail humanity's moral progress.} It’s dangerous to allow any set of values to become permanently entrenched in society. For example, AI systems have learned racist and sexist views \citep{nadeem_stereoset_2021}, and once those views are learned, it can be difficult to fully remove them. In addition to problems we know exist in our society, there may be some we still do not. Just as we abhor some moral views widely held in the past, people in the future may want to move past moral views that we hold today, even those we currently see no problem with. For example, moral defects in AI systems would be even worse if AI systems had been trained in the 1960s, and many people at the time would have seen no problem with that. We may even be unknowingly perpetuating moral catastrophes today \citep{williams_possibility_2015}. Therefore, when advanced AIs emerge and transform the world, there is a risk of their objectives locking in or perpetuating defects in today’s values. If AIs are not designed to continuously learn and update their understanding of societal values, they may perpetuate or reinforce existing defects in their decision-making processes long into the future.

In summary, although keeping powerful AIs in the hands of a few might reduce the risks of terrorism, it could further exacerbate power inequality if misused by governments and corporations. This could lead to totalitarian rule and intense manipulation of the public by corporations, and could lock in current values, preventing any further moral progress.

\begin{storybox}{Story: Bioterrorism}
\emph{The following is an illustrative hypothetical story to help readers envision some of these risks. This story is nonetheless somewhat vague to reduce the risk of inspiring malicious actions based on it.}
\vspace{2pt}

\noindent A biotechnology startup is making waves in the industry with its AI-powered bioengineering model. The company has made bold claims that this new technology will revolutionize medicine through its ability to create cures for both known and unknown diseases. The company did, however, stir up some controversy when it decided to release the program to approved researchers in the scientific community. Only weeks after its decision to make the model open-source on a limited basis, the full model was leaked on the internet for all to see. Its critics pointed out that the model could be repurposed to design lethal pathogens and claimed that the leak provided bad actors with a powerful tool to cause widespread destruction, opening it up to abuse without safeguards in place.

Unknown to the public, an extremist group has been working for years to engineer a new virus designed to kill large numbers of people. Yet given their lack of expertise, these efforts have so far been unsuccessful. When the new AI system is leaked, the group immediately recognizes it as a potential tool to design the virus and circumvent legal and monitoring obstacles to obtain the necessary raw materials. The AI system successfully designs exactly the kind of virus the extremist group was hoping for. It also provides step-by-step instructions on how to synthesize large quantities of the virus and circumvent any obstacles to spreading it. With the synthesized virus in hand, the extremist group devises a plan to release the virus in several carefully chosen locations in order to maximize its spread.

The virus has a long incubation period and spreads silently and quickly throughout the population for months. By the time it is detected, it has already infected millions and has an alarmingly high mortality rate. Given its lethality, most who are infected will ultimately die. The virus may or may not be contained eventually, but not before it kills millions of people.
\end{storybox}

\subsection{Suggestions}

We have discussed two forms of misuse: individuals or small groups using AIs to cause a disaster, and governments or corporations using AIs to entrench their influence. To avoid either of these risks being realized, we will need to strike a balance in terms of the distribution of access to AIs and governments' surveillance powers. We will now discuss some measures that could contribute to finding that balance.

\paragraph{Biosecurity.} 
AIs that are designed for biological research or are otherwise known to possess capabilities in biological research or engineering should be subject to increased scrutiny and access controls, since they have the potential to be repurposed for bioterrorism. In addition, system developers should research and implement methods to remove biological data from the training dataset or excise biological capabilities from finished systems, if those systems are intended for general use \citep{Soice2023CanLL}. Researchers should also investigate ways that AIs could be used for biodefense, for example by improving biological monitoring systems, keeping in mind the potential for dual use of those applications. In addition to AI-specific interventions, more general biosecurity interventions can also help mitigate risks. These interventions include early detection of pathogens through methods like wastewater monitoring \citep{Consortium2021AGN}, far-range UV technology, and improved personal protective equipment \citep{esvelt2022delay}.

\paragraph{Restricted access.} AIs might have dangerous capabilities that could do significant damage if used by malicious actors. One way to mitigate this risk is through structured access, where AI providers limit users' access to dangerous system capabilities by only allowing controlled interactions with those systems through cloud services \citep{Shevlane2022StructuredAT} and conducting know your customer screenings before providing access \citep{Schuett2023BestPractices}. Other mechanisms that could restrict access to the most dangerous systems include the use of hardware, firmware, or export controls to restrict or limit access to computational resources \citep{Shavit2023WhatDI}. Lastly, AI developers should be required to show that their AIs pose minimal risk of catastrophic harm prior to open sourcing them. This recommendation should not be construed as permitting developers to withhold useful and non-dangerous information from the public, such as transparency around training data necessary to address issues of algorithmic bias or copyright.

\paragraph{Technical research on adversarially robust anomaly detection.} While preventing the misuse of AIs is critical, it is necessary to establish multiple lines of defense by detecting misuse when it does happen. AIs could enable anomaly detection techniques that could be used for the detection of unusual behavior in systems or internet platforms, for instance by detecting novel AI-enabled disinformation campaigns before they can be successful. These techniques need to be adversarially robust, as attackers will aim to circumvent them.

\paragraph{Legal liability for developers of general-purpose AIs.} General-purpose AIs can be fine-tuned and prompted for a wide variety of downstream tasks, some of which may be harmful and cause substantial damage. AIs may also fail to act as their users intend. In either case, developers and providers of general-purpose systems may be best placed to reduce risks, since they have a higher level of control over the systems and are often in a better position to implement mitigations. To provide strong incentives for them to do this, companies should bear legal liability for the actions of their AIs. For example, a strict liability regime would incentivize companies to minimize risks and purchase insurance, which would cause the cost of their services to more closely reflect externalities \citep{Lior2019AIEA}. Regardless of what liability regime is ultimately used for AI, it should be designed to hold AI companies liable for harms that they could have averted through more careful development, testing, or standards \citep{Gahntz2022}.

\vspace{10pt}
\begin{visionbox}{Positive Vision}
In an ideal scenario, it would be impossible for any individual or small group to use AIs to cause catastrophes.  Systems with extremely dangerous capabilities either would not exist at all or would be controlled by a democratically accountable body committed to using them only for the general wellbeing of the population. Like nuclear weapons, the information needed to develop those capabilities would remain carefully guarded to prevent proliferation. At the same time, control of AI systems would be subject to strong checks and balances, avoiding entrenchment of power inequalities. Monitoring tools would be utilized at the minimum level necessary to make risks negligible and could not be used to suppress dissent.
\end{visionbox}
\newpage
\section{AI Race}

The immense potential of AIs has created competitive pressures among global players contending for power and influence. This ``AI race'' is driven by nations and corporations who feel they must rapidly build and deploy AIs to secure their positions and survive. By failing to properly prioritize global risks, this dynamic makes it more likely that AI development will produce dangerous outcomes. Analogous to the nuclear arms race during the Cold War, participation in an AI race may serve individual short-term interests, but it ultimately results in worse collective outcomes for humanity. Importantly, these risks stem not only from the intrinsic nature of AI technology, but from the competitive pressures that encourage insidious choices in AI development.

In this section, we first explore the military AI arms race and the corporate AI race, where nation-states and corporations are forced to rapidly develop and adopt AI systems to remain competitive. Moving beyond these specific races, we reconceptualize competitive pressures as part of a broader evolutionary process in which AIs could become increasingly pervasive, powerful, and entrenched in society. Finally, we highlight potential strategies and policy suggestions to mitigate the risks created by an AI race and ensure the safe development of AIs.

\subsection{Military AI Arms Race}

The development of AIs for military applications is swiftly paving the way for a new era in military technology, with potential consequences rivaling those of gunpowder and nuclear arms in what has been described as the ``third revolution in warfare.'' The weaponization of AI presents numerous challenges, such as the potential for more destructive wars, the possibility of accidental usage or loss of control, and the prospect of malicious actors co-opting these technologies for their own purposes. As AIs gain influence over traditional military weaponry and increasingly take on command and control functions, humanity faces a paradigm shift in warfare. In this context, we will discuss the latent risks and implications of this AI arms race on global security, the potential for intensified conflicts, and the dire outcomes that could come as a result, including the possibility of conflicts escalating to a scale that poses an existential threat.

\subsubsection{Lethal Autonomous Weapons (LAWs)}

\begin{wrapfigure}{r}[0\textwidth]{.36\textwidth}%
	\vspace{-15pt}%
	\centering
	\includegraphics[width=0.35\textwidth]{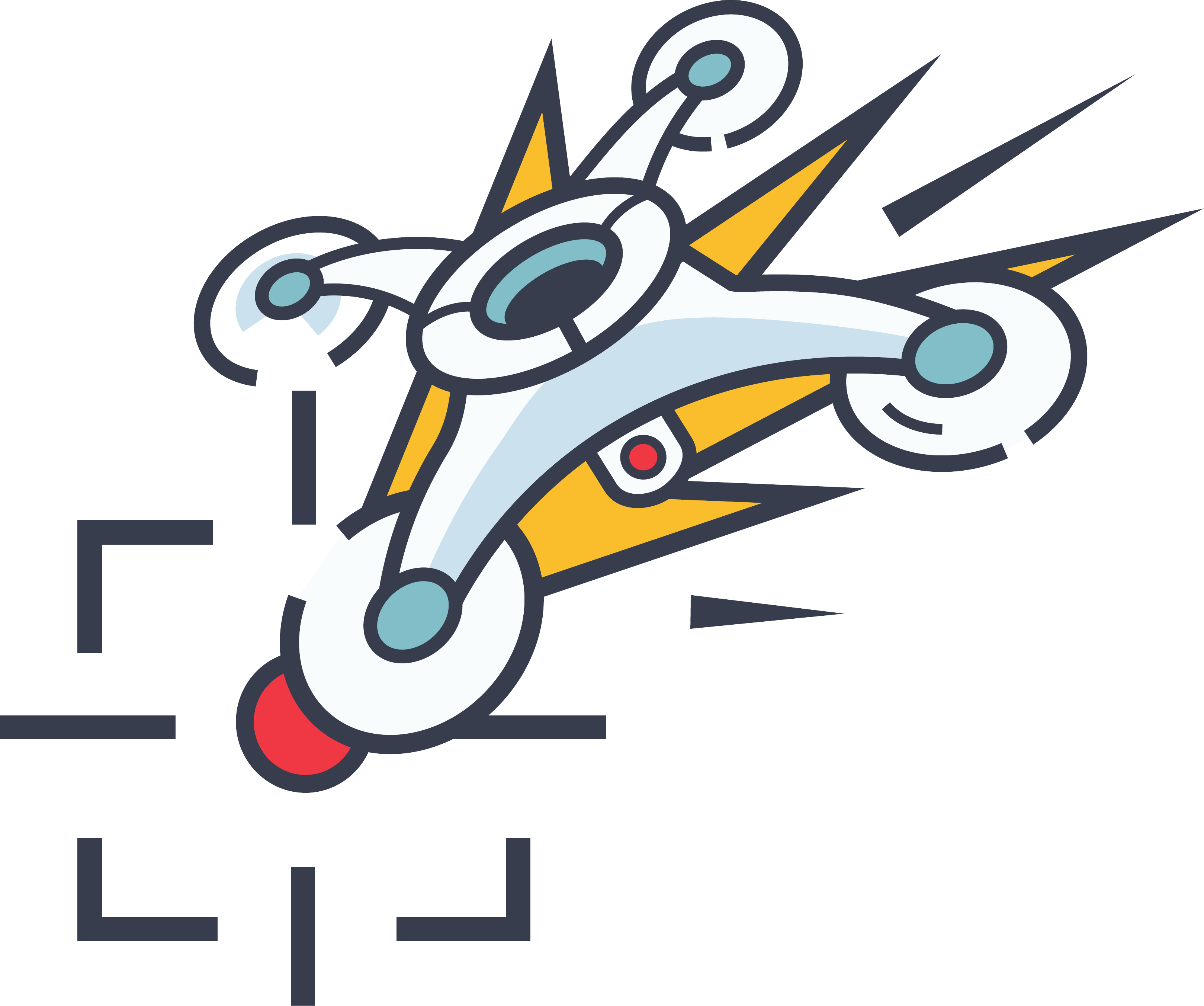}
	\caption{Low-cost automated weapons, such as drone swarms outfitted with explosives, could autonomously hunt human targets with high precision, performing lethal operations for both militaries and terrorist groups and lowering the barriers to large-scale violence.}
	\label{fig:slaughterbot}
	\vspace{-20pt}%
\end{wrapfigure}

LAWs are weapons that can identify, target, and kill without human intervention \citep{scharre2018}. They offer potential improvements in decision-making speed and precision. Warfare, however, is a high-stakes, safety-critical domain for AIs with significant moral and practical concerns. Though their existence is not necessarily a catastrophe in itself, LAWs may serve as an on-ramp to catastrophes stemming from malicious use, accidents, loss of control, or an increased likelihood of war.


\paragraph{LAWs may become vastly superior to humans.} Driven by rapid developments in AIs, weapons systems that can identify, target, and decide to kill human beings on their own---without an officer directing an attack or a soldier pulling the trigger---are starting to transform the future of conflict. In 2020, an advanced AI agent outperformed experienced F-16 pilots in a series of virtual dogfights, including decisively defeating a human pilot 5-0, showcasing ``aggressive and precise maneuvers the human pilot couldn't outmatch'' \citep{dogfight}. Just as in the past, superior weapons would allow for more destruction in a shorter period of time, increasing the severity of war.

\paragraph{Militaries are taking steps toward delegating life-or-death decisions to AIs.} Fully autonomous drones were likely first used on the battlefield in Libya in March 2020, when retreating forces were ``hunted down and remotely engaged'' by a drone operating without human oversight \citep{UnitedNations2021}. In May 2021, the Israel Defense Forces used the world's first AI-guided weaponized drone swarm during combat operations, which marks a significant milestone in the integration of AI and drone technology in warfare \citep{hambling2021israel}. Although walking, shooting robots have yet to replace soldiers on the battlefield, technologies are converging in ways that may make this possible in the near future.

\paragraph{LAWs increase the likelihood of war.} Sending troops into battle is a grave decision that leaders do not make lightly. But autonomous weapons would allow an aggressive nation to launch attacks without endangering the lives of its own soldiers and thus face less domestic scrutiny. While remote-controlled weapons share this advantage, their scalability is limited by the requirement for human operators and vulnerability to jamming countermeasures, limitations that LAWs could overcome \citep{kallenborn2021applying}. Public opinion for continuing wars tends to wane as conflicts drag on and casualties increase \citep{mueller1985war}. LAWs would change this equation. National leaders would no longer face the prospect of body bags returning home, thus removing a primary barrier to engaging in warfare, which could ultimately increase the likelihood of conflicts.

\subsubsection{Cyberwarfare}

As well as being used to enable deadlier weapons, AIs could lower the barrier to entry for cyberattacks, making them more numerous and destructive. They could cause serious harm not only in the digital environment but also in physical systems, potentially taking out critical infrastructure that societies depend on. While AIs could also be used to improve cyberdefense, it is unclear whether they will be most effective as an offensive or defensive technology \citep{bonfanti2022ai}. If they enhance attacks more than they support defense, then cyberattacks could become more common, creating significant geopolitical turbulence and paving another route to large-scale conflict.

\paragraph{AIs have the potential to increase the accessibility, success rate, scale, speed, stealth, and potency of cyberattacks.} Cyberattacks are already a reality, but AIs could be used to increase their frequency and destructiveness in multiple ways. Machine learning tools could be used to find more critical vulnerabilities in target systems and improve the success rate of attacks. They could also be used to increase the scale of attacks by running millions of systems in parallel, and increase the speed by finding novel routes to infiltrating a system. Cyberattacks could also become more potent if used to hijack AI weapons.

\paragraph{Cyberattacks can destroy critical infrastructure.} 
By hacking computer systems that control physical processes, cyberattacks could cause extensive infrastructure damage. For example, they could cause system components to overheat or valves to lock, leading to a buildup of pressure culminating in an explosion. Through interferences like this, cyberattacks have the potential to destroy critical infrastructure, such as electric grids and water supply systems. This was demonstrated in 2015, when a cyberwarfare unit of the Russian military hacked into the Ukrainian power grid, leaving over 200,000 people without power access for several hours. AI-enhanced attacks could be even more devastating and potentially deadly for the billions of people who rely on critical infrastructure for survival.


\paragraph{Difficulties in attributing AI-driven cyberattacks could increase the risk of war.} A cyberattack resulting in physical damage to critical infrastructure would require a high degree of skill and effort to execute, perhaps only within the capability of nation-states. Such attacks are rare as they constitute an act of war, and thus elicit a full military response. Yet AIs could enable attackers to hide their identity, for example if they are used to evade detection systems or more effectively cover the tracks of the attacker \citep{MIRSKY2023103006}. If cyberattacks become more stealthy, this would reduce the threat of retaliation from an attacked party, potentially making attacks more likely. If stealthy attacks do happen, they might incite actors to mistakenly retaliate against unrelated third parties they suspect to be responsible. This could increase the scope of the conflict dramatically.

\subsubsection{Automated Warfare}


\paragraph{AIs speed up the pace of war, which makes AIs more necessary.} AIs can quickly process a large amount of data, analyze complex situations, and provide helpful insights to commanders. With ubiquitous sensors and advanced technology on the battlefield, there is tremendous incoming information. AIs help make sense of this information, spotting important patterns and relationships that humans might miss. As these trends continue, it will become increasingly difficult for humans to make well-informed decisions as quickly as necessary to keep pace with AIs. This would further pressure militaries to hand over decisive control to AIs. The continuous integration of AIs into all aspects of warfare will cause the pace of combat to become faster and faster. Eventually, we may arrive at a point where humans are no longer capable of assessing the ever-changing battlefield situation and must cede decision-making power to advanced AIs.

\paragraph{Automatic retaliation can escalate accidents into war.} There is already willingness to let computer systems retaliate automatically. In 2014, a leak revealed to the public that the NSA was developing a system called MonsterMind, which would autonomously detect and block cyberattacks on US infrastructure \citep{zetter2014}. It was suggested that in the future, MonsterMind could automatically initiate a retaliatory cyberattack with no human involvement. If multiple combatants have policies of automatic retaliation, an accident or false alarm could quickly escalate to full-scale war before humans intervene. This would be especially dangerous if the superior information processing capabilities of modern AI systems makes it more appealing for actors to automate decisions regarding nuclear launches.

\begin{wrapfigure}{r}[0\textwidth]{.46\textwidth}%
	\vspace{-10pt}%
	\centering
	\includegraphics[width=0.45\textwidth]{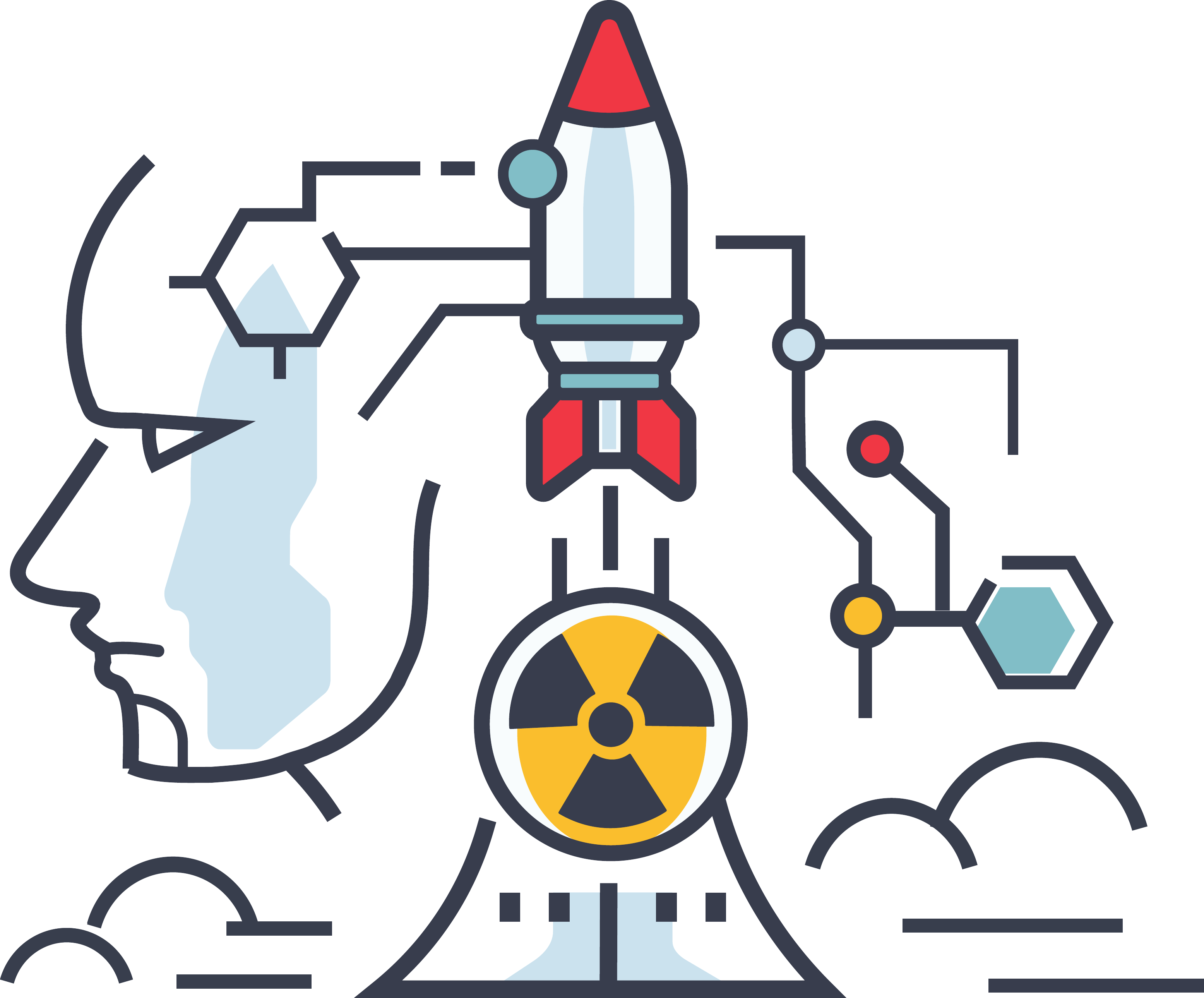}
	\caption{A military AI arms race could pressure countries into delegating many crucial decisions over armaments to AIs. Integrating AIs into nuclear command and control could heighten the risk of global catastrophe as the potential for accidents and increased pace of war may lead to unintended escalations and confrontations.}
	\label{fig:badactor}
	\vspace{-10pt}
\end{wrapfigure}

\paragraph{History shows the danger of automated retaliation.} On September 26, 1983, Stanislav Petrov, a lieutenant colonel of the Soviet Air Defense Forces, was on duty at the Serpukhov-15 bunker near Moscow, monitoring the Soviet Union's early warning system for incoming ballistic missiles. The system indicated that the US had launched multiple nuclear missiles toward the Soviet Union. The protocol at the time dictated that such an event should be considered a legitimate attack, and the Soviet Union would respond with a nuclear counterstrike. If Petrov had passed on the warning to his superiors, this would have been the likely outcome. Instead, however, he judged it to be a false alarm and ignored it. It was soon confirmed that the warning had been caused by a rare technical malfunction. If an AI had been in control, the false alarm could have triggered a nuclear war.

\paragraph{AI-controlled weapons systems could lead to a flash war.} Autonomous systems are not infallible. We have already witnessed how quickly an error in an automated system can escalate in the economy. Most notably, in the 2010 Flash Crash, a feedback loop between automated trading algorithms amplified ordinary market fluctuations into a financial catastrophe in which a trillion dollars of stock value vanished in minutes \citep{Kirilenko2011TheFC}. If multiple nations were to use AIs to automate their defense systems, an error could be catastrophic, triggering a spiral of attacks and counter-attacks that would happen too quickly for humans to step in---a flash war. The market quickly recovered from the 2010 Flash Crash, but the harm caused by a flash war could be catastrophic.

\paragraph{Automated warfare could reduce accountability for military leaders.} Military leaders may at times gain an advantage on the battlefield if they are willing to ignore the laws of war. For example, soldiers may be able to mount stronger attacks if they do not take steps to minimize civilian casualties. An important deterrent to this behavior is the risk that military leaders could eventually be held accountable or even prosecuted for war crimes. Automated warfare could reduce this deterrence effect by making it easier for military leaders to escape accountability by blaming violations on failures in their automated systems.

\paragraph{AIs could make war more uncertain, increasing the risk of conflict.} Although states that are already wealthier and more powerful often have more resources to invest in new military technologies, they are not necessarily always the most successful at adopting them. Other factors also play an important role, such as how agile and adaptive a military can be in incorporating new technologies \citep{horowitz2010diffusion}. Major new weapons innovations can therefore offer an opportunity for existing superpowers to bolster their dominance, but also for less powerful states to quickly increase their power by getting ahead in an emerging and important sphere. This can create significant uncertainty around if and how the balance of power is shifting, potentially leading states to incorrectly believe they could gain something from going to war. Even aside from considerations regarding the balance of power, rapidly evolving automated warfare would be unprecedented, making it difficult for actors to evaluate their chances of victory in any particular conflict. This would increase the risk of miscalculation, making war more more likely.

\subsubsection{Actors May Risk Extinction Over Individual Defeat}

\begin{wrapfigure}{r}[0\textwidth]{.45\textwidth}%
\vspace{-10pt}%
``I know not with what weapons World War III will be fought, but World War IV will be fought with sticks and stones.''\hfill\emph{Einstein}%
\end{wrapfigure}

\paragraph{Competitive pressures make actors more willing to accept the risk of extinction.} During the Cold War, neither side desired the dangerous situation they found themselves in. There were widespread fears that nuclear weapons could be powerful enough to wipe out a large fraction of humanity, potentially even causing extinction---a catastrophic result for both sides. Yet the intense rivalry and geopolitical tensions between the two superpowers fueled a dangerous cycle of arms buildup. Each side perceived the other's nuclear arsenal as a threat to its very survival, leading to a desire for parity and deterrence. The competitive pressures pushed both countries to continually develop and deploy more advanced and destructive nuclear weapons systems, driven by the fear of being at a strategic disadvantage. During the Cuban Missile Crisis, this led to the brink of nuclear war. Even though the story of Arkhipov preventing the launch of a nuclear torpedo wasn't declassified until decades after the incident, President John F.\ Kennedy reportedly estimated that he thought the odds of nuclear war beginning during that time were ``somewhere between one out of three and even.'' This chilling admission highlights how the competitive pressures between militaries have the potential to cause global catastrophes.

\paragraph{Individually rational decisions can be collectively catastrophic.} Nations locked in competition might make decisions that advance their own interests by putting the rest of the world at stake. Scenarios of this kind are collective action problems, where decisions may be rational on an individual level yet disastrous for the larger group \citep{Jervis1978CooperationUT}. For example, corporations and individuals may weigh their own profits and convenience over the negative impacts of the emissions they create, even if those emissions collectively result in climate change. The same principle can be extended to military strategy and defense systems. Military leaders might estimate, for instance, that increasing the autonomy of weapon systems would mean a 10 percent chance of losing control over weaponized superhuman AIs. Alternatively, they might estimate that using AIs to automate bioweapons research could lead to a 10 percent chance of leaking a deadly pathogen. Both of these scenarios could lead to catastrophe or even extinction. The leaders may, however, also calculate that refraining from these developments will mean a 99 percent chance of losing a war against an opponent. Since conflicts are often viewed as existential struggles by those fighting them, rational actors may accept an otherwise unthinkable 10 percent chance of human extinction over a 99 percent chance of losing a war. Regardless of the particular nature of the risks posed by advanced AIs, these dynamics could push us to the brink of global catastrophe.

\paragraph{Technological superiority does not guarantee national security.} It is tempting to think that the best way of guarding against enemy attacks is to improve one's own military prowess. However, in the midst of competitive pressures, all parties will tend to advance their weaponry, such that no one gains much of an advantage, but all are left at greater risk. As Richard Danzig, former Secretary of the Navy, has observed, ``The introduction of complex, opaque, novel, and interactive technologies will produce accidents, emergent effects, and sabotage. On a number of occasions and in a number of ways, the American national security establishment will lose control of what it creates... deterrence is a strategy for reducing attacks, not accidents'' \citep{Danzig2018Technology}.

\paragraph{Cooperation is paramount to reducing risk.}  As discussed above, an AI arms race can lead us down a hazardous path, despite this being in no country's best interest. It is important to remember that we are all on the same side when it comes to existential risks, and working together to prevent them is a collective necessity. A destructive AI arms race benefits nobody, so all actors would be rational to take steps to cooperate with one another to prevent the riskiest applications of militarized AIs. As Dwight D.\ Eisenhower reminded us, ``The only way to win World War III is to prevent it.'' \\

We have considered how competitive pressures could lead to the increasing automation of conflict, even if decision-makers are aware of the existential threat that this path entails. We have also discussed cooperation as being the key to counteracting and overcoming this collective action problem. We will now illustrate a hypothetical path to disaster that could result from an AI arms race.

\begin{storybox}{Story: Automated Warfare}

\noindent As AI systems become increasingly sophisticated, militaries start involving them in decision-making processes. Officials give them military intelligence about opponents' arms and strategies, for example, and ask them to calculate the most promising plan of action. It soon becomes apparent that AIs are reliably reaching better decisions than humans, so it seems sensible to give them more influence. At the same time, international tensions are rising, increasing the threat of war.

A new military technology has recently been developed that could make international attacks swifter and stealthier, giving targets less time to respond. Since military officials feel their response processes take too long, they fear that they could be vulnerable to a surprise attack capable of inflicting decisive damage before they would have any chance to retaliate. Since AIs can process information and make decisions much more quickly than humans, military leaders reluctantly hand them increasing amounts of retaliatory control, reasoning that failing to do so would leave them open to attack from adversaries. 

While for years military leaders had stressed the importance of keeping a ``human in the loop'' for major decisions, human control is nonetheless gradually phased out in the interests of national security. Military leaders understand that their decisions lead to the possibility of inadvertent escalation caused by system malfunctions, and would prefer a world where all countries automated less; but they do not trust that their adversaries will refrain from automation. Over time, more and more of the chain of command is automated on all sides.

One day, a single system malfunctions, detecting an enemy attack when there is none. The system is empowered to launch an instant ``retaliatory'' attack, and it does so in the blink of an eye. The attack causes automated retaliation from the other side, and so on. Before long, the situation is spiraling out of control, with waves of automated attack and retaliation. Although humans have made mistakes leading to escalation in the past, this escalation between mostly-automated militaries happens far more quickly than any before. The humans who are responding to the situation find it difficult to diagnose the source of the problem, as the AI systems are not transparent. By the time they even realize how the conflict started, it is already over, with devastating consequences for both sides.
\end{storybox}

\subsection{Corporate AI Race}

Competitive pressures exist in the economy, as well as in military settings. Although competition between companies can be beneficial, creating more useful products for consumers, there are also pitfalls. First, the benefits of economic activity may be unevenly distributed, incentivizing those who benefit most from it to disregard the harms to others. Second, under intense market competition, businesses tend to focus much more on short-term gains than on long-term outcomes. With this mindset, companies often pursue something that can make a lot of profit in the short term, even if it poses a societal risk in the long term. We will now discuss how corporate competitive pressures could play out with AIs and the potential negative impacts.

\subsubsection{Economic Competition Undercuts Safety}

\paragraph{Competitive pressure is fueling a corporate AI race.} To obtain a competitive advantage, companies often race to offer the first products to a market rather than the safest. These dynamics are already playing a role in the rapid development of AI technology. At the launch of Microsoft's AI-powered search engine in February 2023, the company's CEO Satya Nadella said, ``A race starts today... we're going to move fast.'' Only weeks later, the company's chatbot was shown to have threatened to harm users \citep{perrigo_bings_2023}. In an internal email, Sam Schillace, a technology executive at Microsoft, highlighted the urgency in which companies view AI development. He wrote that it would be an ``absolutely fatal error in this moment to worry about things that can be fixed later'' \citep{grant_i_2023}.

\paragraph{Competitive pressures have contributed to major commercial and industrial disasters.}

Throughout the 1960s, Ford Motor Company faced competition from international car manufacturers as the share of imports in American car purchases steadily rose \cite{klier2009tailfins}. Ford developed an ambitious plan to design and manufacture a new car model in only 25 months \cite{sherefkin2003ford}. The Ford Pinto was delivered to customers ahead of schedule, but with a serious safety problem: the gas tank was located near the rear bumper, and could explode during rear collisions. Numerous fatalities and injuries were caused by the resulting fires when crashes inevitably happened \cite{strobel_reckless_1980}. Ford was sued and a jury found them liable for these deaths and injuries \cite{noauthor_grimshaw_1981}. The verdict, of course, came too late for those who had already lost their lives. As Ford's president at the time was fond of saying, ``Safety doesn't sell'' \cite{judge_selling_1990}.


Boeing, aiming to compete with its rival Airbus, sought to deliver an updated, more fuel-efficient model to the market as quickly as possible. The head-to-head rivalry and time pressure led to the introduction of the Maneuvering Characteristics Augmentation System, which was designed to enhance the aircraft's stability. However, inadequate testing and pilot training ultimately resulted in the two fatal crashes only months apart, with 346 people killed \citep{leggett_737_2023}. We can imagine a future in which similar pressures lead companies to cut corners and release unsafe AI systems.

A third example is the Bhopal gas tragedy, which is widely considered to be the worst industrial disaster ever to have happened. In December 1984, a vast quantity of toxic gas leaked from a Union Carbide Corporation subsidiary plant manufacturing pesticides in Bhopal, India. Exposure to the gas killed thousands of people and injured up to half a million more. Investigations found that, in the run-up to the disaster, safety standards had fallen significantly, with the company cutting costs by neglecting equipment maintenance and staff training as profitability fell. This is often considered a consequence of competitive pressures \citep{broughton_bhopal_2005}.

\begin{wrapfigure}{r}[0\textwidth]{.38\textwidth}%
	\vspace{-8pt}%
    ``Nothing can be done at once hastily and prudently.''\hfill\emph{Publilius Syrus}%
	\vspace{-10pt}%
\end{wrapfigure}

\paragraph{Competition incentivizes businesses to deploy potentially unsafe AI systems.} In an environment where businesses are rushing to develop and release products, those that follow rigorous safety procedures will be slower and risk being out-competed. Ethically-minded AI developers, who want to proceed more cautiously and slow down, would give more unscrupulous developers an advantage. In trying to survive commercially, even the companies that want to take more care are likely to be swept along by competitive pressures. There may be attempts to implement safety measures, but with more of an emphasis on capabilities than on safety, these may be insufficient. This could lead us to develop highly powerful AIs before we properly understand how to ensure they are safe.

\subsubsection{Automated Economy}

\paragraph{Corporations will face pressure to replace humans with AIs.} As AIs become more capable, they will be able to perform an increasing variety of tasks more quickly, cheaply, and effectively than human workers. Companies will therefore stand to gain a competitive advantage from replacing their employees with AIs. Companies that choose not to adopt AIs would likely be out-competed, just as a clothing company using manual looms would be unable to keep up with those using industrial ones.

\begin{wrapfigure}{r}[0\textwidth]{.36\textwidth}%
	\vspace{-10pt}%
	\centering
	\includegraphics[width=0.35\textwidth]{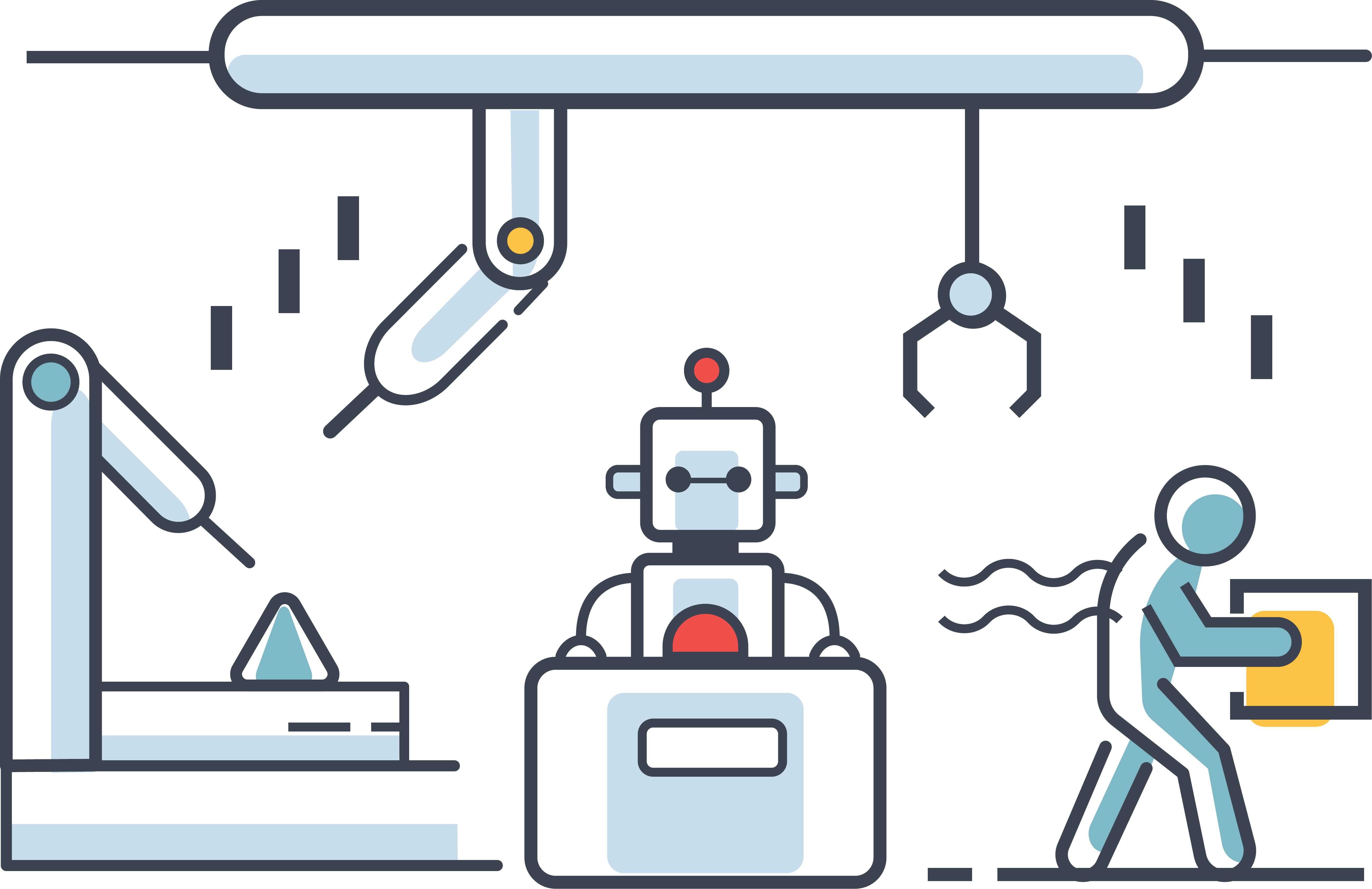}
	\caption{As AIs automate increasingly many tasks, the economy may become largely run by AIs. Eventually, this could lead to human enfeeblement and dependence on AIs for basic needs.}
	\label{fig:automation}
	\vspace{-5pt}%
\end{wrapfigure}

\paragraph{AIs could lead to mass unemployment.} Economists have long considered the possibility that machines will replace human labor. Nobel Prize winner Wassily Leontief said in 1952 that, as technology advances, ``Labor will become less and less important... more and more workers will be replaced by machines'' \citep{curtis_machines_1983}. Previous technologies have augmented the productivity of human labor. AIs, however, could differ profoundly from previous innovations. Advanced AIs capable of automating human labor should be regarded not merely as tools, but as agents. Human-level AI agents would, by definition, be able to do everything a human could do. These AI agents would also have important advantages over human labor. They could work 24 hours a day, be copied many times and run in parallel, and process information much more quickly than a human would. While we do not know when this will occur, it is unwise to discount the possibility that it could be soon. If human labor is replaced by AIs, mass unemployment could dramatically increase inequality, making individuals dependent on the owners of AI systems.\looseness=-1

\paragraph{Automated AI R\&D.} AI agents would have the potential to automate the research and development (R\&D) of AI itself. AI is increasingly automating parts of the research process \citep{woodside2023examples}, and this could lead to AI capabilities growing at increasing rates, to the point where humans are no longer the driving force behind AI development. If this trend continues unchecked, it could escalate risks associated with AIs progressing faster than our capacity to manage and regulate them. Imagine that we created an AI that writes and thinks at the speed of today's AIs, but that it could also perform world-class AI research. We could then copy that AI and create $10,\!000$ world-class AI researchers that operate at a pace $100\times$ times faster than humans. By automating AI research and development, we might achieve progress equivalent to many decades in just a few months. 

\paragraph{Conceding power to AIs could lead to human enfeeblement.} Even if we ensure that the many unemployed humans are provided for, we may find ourselves completely reliant on AIs. This would likely emerge not from a violent coup by AIs, but from a gradual slide into dependence. As society's challenges become ever more complex and fast-paced, and as AIs become ever more intelligent and quick-thinking, we may forfeit more and more functions to them out of convenience. In such a state, the only feasible solution to the complexities and challenges compounded by AIs may be to rely even more heavily on AIs. This gradual process could eventually lead to the delegation of nearly all intellectual, and eventually physical, labor to AIs. In such a world, people might have few incentives to gain knowledge and cultivate skills, potentially leading to a state of enfeeblement \citep{Russell2019HumanCA}. Having lost our know-how and our understanding of how civilization works, we would become completely dependent on AIs, a scenario not unlike the one depicted in the film WALL-E. In such a state, humanity is not flourishing and is no longer in effective control.\\ 

As we have seen, there are classic game-theoretic dilemmas where individuals and groups face incentives that are incompatible with what would make everyone better off. We see this with a military AI arms race, where the world is made less safe by creating extremely powerful AI weapons, and we see this in a corporate AI race, where an AI's power and development is prioritized over its safety. To address these dilemmas that give rise to global risks, we will need new coordination mechanisms and institutions. It is our view that failing to coordinate and stop AI races would be the most likely cause of an existential catastrophe.

\subsection{Evolutionary Pressures}

As discussed above, there are strong pressures to replace humans with AIs, cede more control to them, and reduce human oversight in various settings, despite the potential harms. We can re-frame this as a general trend resulting from evolutionary dynamics, where an unfortunate truth is that AIs will simply be more fit than humans. Extrapolating this pattern of automation, it is likely that we will build an ecosystem of competing AIs over which it may be difficult to maintain control in the long run. We will now discuss how natural selection influences the development of AI systems and why evolution favors selfish behaviors. We will also look at how competition might arise and play out between AIs and humans, and how this could create catastrophic risks. This section draws heavily from ``\textit{Natural Selection Favors AIs over Humans}'' \citep{Hendrycks2023NaturalSF,hendryckstime2023}.

\vspace{-4pt}
\paragraph{Fitter technologies are selected, for good and bad.} While most people think of evolution by natural selection as a biological process, its principles shape much more. According to the evolutionary biologist Richard Lewontin \citep{Lewontin1970THEUO}, evolution by natural selection will take hold in any environment where three conditions are present: 1) there are differences between individuals; 2) characteristics are passed onto future generations and; 3) the different variants propagate at different rates. These conditions apply to various technologies.

Consider the content-recommendation algorithms used by streaming services and social media platforms. When a particularly addictive content format or algorithm hooks users, it results in higher screen time and engagement. This more effective content format or algorithm is consequently ``selected'' and further fine-tuned, while formats and algorithms that fail to capture attention are discontinued. These competitive pressures foster a ``survival of the most addictive'' dynamic. Platforms that refuse to use addictive formats and algorithms become less influential or are simply outcompeted by platforms that do, leading competitors to undermine wellbeing and cause massive harm to society \citep{kross2013facebook}.

\begin{wrapfigure}{r}[0\textwidth]{.36\textwidth}%
	\vspace{-33pt}%
	\centering
	\includegraphics[width=0.35\textwidth]{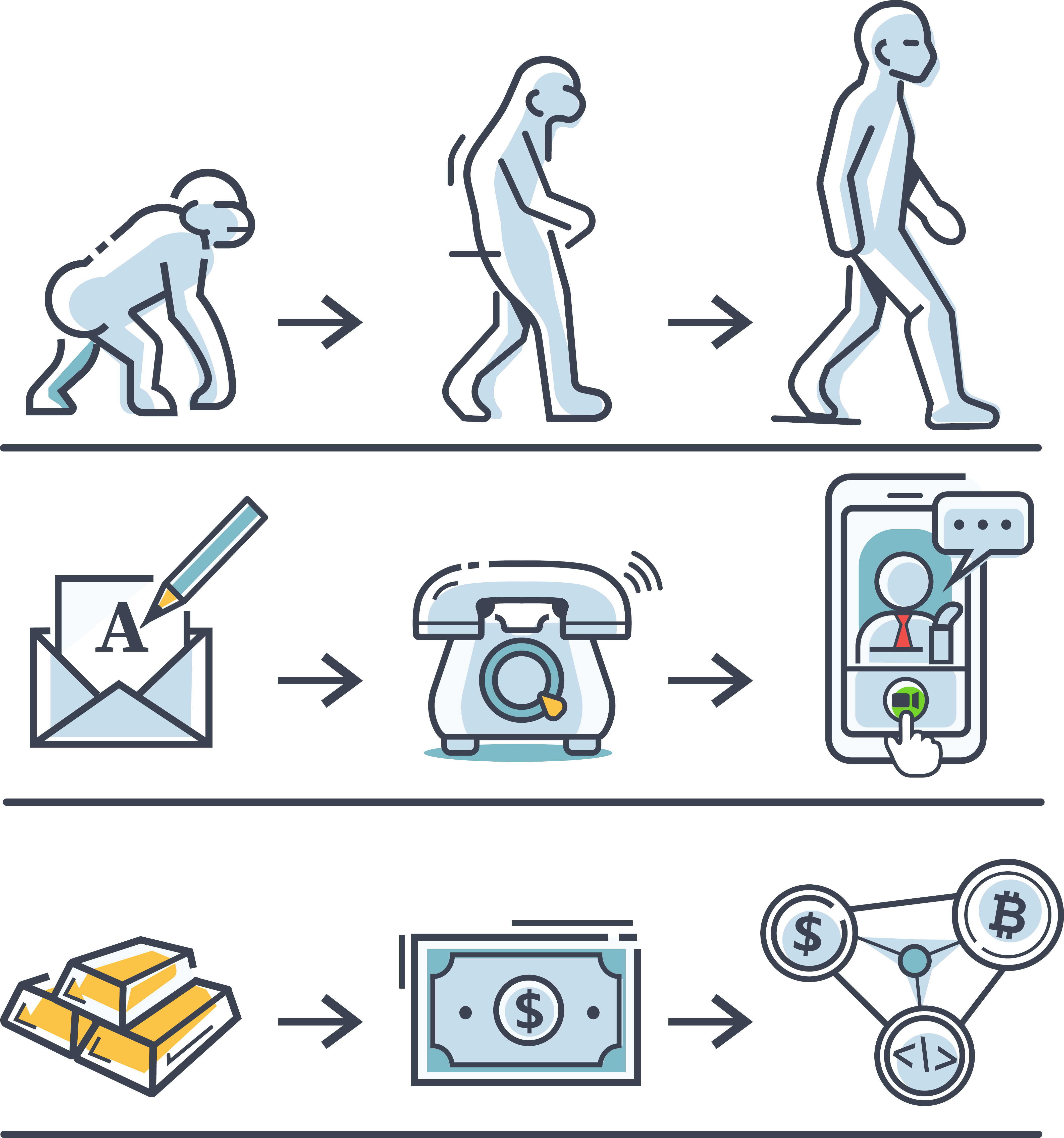}
	\caption{Evolutionary processes are not limited to the realm of biology.}
	\label{fig:darwinism}
	\vspace{-23pt}%
\end{wrapfigure}

\vspace{-4pt}
\paragraph{The conditions for natural selection apply to AIs.} There will be many different AI developers who make many different AI systems with varying features and capabilities, and competition between them will determine which characteristics become more common. Second, the most successful AIs today are already being used as a basis for their developers' next generation of models, as well as being imitated by rival companies. Third, factors determining which AIs propagate the most may include their ability to act autonomously, automate labor, or reduce the chance of their own deactivation.

\vspace{-4pt}
\paragraph{Natural selection often favors selfish characteristics.} Natural selection influences which AIs propagate most widely. From biological systems, we see that natural selection often gives rise to selfish behaviors that promote one's own genetic information: chimps attack other communities \citep{Martnezigo2021IntercommunityIA}, lions engage in infanticide \citep{pusey1994infanticide}, viruses evolve new surface proteins to deceive and bypass defense barriers \citep{Nagy2011TheDO}, humans engage in nepotism, some ants enslave others \citep{Buschinger2009SocialPA}, and so on. In the natural world, selfishness often emerges as a dominant strategy; those that prioritize themselves and those similar to them are usually more likely to survive, so these traits become more prevalent. Amoral competition can select for traits that we think are immoral.\looseness=-1


\vspace{-4pt}
\paragraph{Examples of selfish behaviors.} For concreteness, we now describe many selfish traits---traits that expand AIs' influence at the expense of humans. AIs that automate a task and leave many humans jobless have engaged in selfish behavior; these AIs may not even be aware of what a human is but still behave selfishly towards them---selfish behaviors do not require malicious intent. Likewise, AI managers may engage in selfish and ``ruthless'' behavior by laying off thousands of workers; such AIs may not even believe they did anything wrong---they were just being ``efficient.'' AIs may eventually become enmeshed in vital infrastructure such as power grids or the internet. Many people may then be unwilling to accept the cost of being able to effortlessly deactivate them, as that would pose a reliability hazard. AIs that help create a new useful system---a company, or infrastructure---that becomes increasingly complicated and eventually requires AIs to operate them also have engaged in selfish behavior. AIs that help people develop AIs that are more intelligent---but happen to be less interpretable to humans---have engaged in selfish behavior, as this reduces human control over AIs' internals. AIs that are more charming, attractive, hilarious, imitate sentience (uttering phrases like ``ouch!'' or pleading ``please don't turn me off!''), or emulate deceased family members are more likely to have humans grow emotional connections with them. These AIs are more likely to cause outrage at suggestions to destroy them, and they are more likely preserved, protected, or granted rights by some individuals. If some AIs are given rights, they may operate, adapt, and evolve outside of human control. Overall, AIs could become embedded in human society and expand their influence over us in ways that we can't reverse.

\paragraph{Selfish behaviors may erode safety measures that some of us implement.} 
AIs that gain influence and provide economic value will predominate, while AIs that adhere to the most constraints will be less competitive. For example, AIs following the constraint ``never break the law'' have fewer options than AIs following the constraint ``don't get caught breaking the law.'' AIs of the latter type may be willing to break the law if they're unlikely to be caught or if the fines are not severe enough, allowing them to outcompete more restricted AIs. Many businesses follow laws, but in situations where stealing trade secrets or deceiving regulators is highly lucrative and difficult to detect, a business that is willing to engage in such selfish behavior can have an advantage over its more principled competitors.

An AI system might be prized for its ability to achieve ambitious goals autonomously. It might, however, be achieving its goals efficiently without abiding by ethical restrictions, while deceiving humans about its methods. Even if we try to put safety measures in place, a deceptive AI would be very difficult to counteract if it is cleverer than us. AIs that can bypass our safety measures without detection may be the most successful at accomplishing the tasks we give them, and therefore become widespread. These processes could culminate in a world where many aspects of major companies and infrastructure are controlled by powerful AIs with selfish traits, including deceiving humans, harming humans in service of their goals, and preventing themselves from being deactivated.\looseness=-1

\paragraph{Humans only have nominal influence over AI selection.} One might think we could avoid the development of selfish behaviors by ensuring we do not select AIs that exhibit them. However, the companies developing AIs are not selecting the safest path but instead succumbing to evolutionary pressures. One example is OpenAI, which was founded as a nonprofit in 2015 to ``benefit humanity as a whole, unconstrained by a need to generate financial return'' \citep{openai_introducing_2015}. However, when faced with the need to raise capital to keep up with better-funded rivals, in 2019 OpenAI transitioned from a nonprofit to ``capped-profit'' structure \citep{coldewey_openai_2019}. Later, many of the safety-focused OpenAI employees left and formed a competitor, Anthropic, that was to focus more heavily on AI safety than OpenAI had. Although Anthropic originally focused on safety research, they eventually became convinced of the ``necessity of commercialization'' and now contribute to competitive pressures \citep{singh_anthropics_2023}. While many of the employees at those companies genuinely care about safety, these values do not stand a chance against evolutionary pressures, which compel companies to move ever more hastily and seek ever more influence, lest the company perish. Moreover, AI developers are already selecting AIs with increasingly selfish traits. They are selecting AIs to automate and displace humans, make humans highly dependent on AIs, and make humans more and more obsolete. 
By their own admission, future versions of these AIs may lead to extinction \citep{cais2023}. This is why an AI race is insidious: AI development is not being aligned with human values but rather with natural selection.\looseness=-1

People often choose the products that are most useful and convenient to them immediately, rather than thinking about potential long-term consequences, even to themselves. An AI race puts pressures on companies to select the AIs that are most competitive, not the least selfish. Even if it's feasible to select for unselfish AIs, if it comes at a clear cost to competitiveness, some competitors will select the selfish AIs. Furthermore, as we have mentioned, if AIs develop strategic awareness, they may counteract our attempts to select against them. Moreover, as AIs increasingly automate various processes, AIs will affect the competitiveness of other AIs, not just humans. AIs will interact and compete with each other, and some will be put in charge of the development of other AIs at some point. Giving AIs influence over which other AIs should be propagated and how they should be modified would represent another step toward human becoming dependent on AIs and AI evolution becoming increasingly independent from humans. As this continues, the complex process governing AI evolution will become further unmoored from human interests.

\paragraph{AIs can be more fit than humans.} Our unmatched intelligence has granted us power over the natural world. It has enabled us to land on the moon, harness nuclear energy, and reshape landscapes at our will. It has also given us power over other species. Although a single unarmed human competing against a tiger or gorilla has no chance of winning, the collective fate of these animals is entirely in our hands. Our cognitive abilities have proven so advantageous that, if we chose to, we could cause them to go extinct in a matter of weeks. Intelligence was a key factor that led to our dominance, but we are currently standing on the precipice of creating entities far more intelligent than ourselves.

Given the exponential increase in microprocessor speeds, AIs have the potential to process information and ``think'' at a pace that far surpasses human neurons, but it could be even more dramatic than the speed difference between humans and sloths---possibly more like the speed difference between humans and plants. They can assimilate vast quantities of data from numerous sources simultaneously, with near-perfect retention and understanding. They do not need to sleep and they do not get bored. Due to the scalability of computational resources, an AI could interact and cooperate with an unlimited number of other AIs, potentially creating a collective intelligence that would far outstrip human collaborations. AIs could also deliberately update and improve themselves. Without the same biological restrictions as humans, they could adapt and therefore evolve unspeakably quickly compared with us. Computers are becoming faster. Humans aren't \citep{danzig_aum_2012}.

To further illustrate the point, imagine that there was a new species of humans. They do not die of old age, they get 30\% faster at thinking and acting each year, and they can instantly create adult offspring for the modest sum of a few thousand dollars. It seems clear, then, this new species would eventually have more influence over the future. In sum, AIs could become like an invasive species, with the potential to out-compete humans. Our only advantage over AIs is that we get to get make the first moves, but given the frenzied AI race, we are rapidly giving up even this advantage.

\paragraph{AIs would have little reason to cooperate with or be altruistic toward humans.} Cooperation and altruism evolved because they increase fitness. There are numerous reasons why humans cooperate with other humans, like direct reciprocity. Also known as ``quid pro quo,'' direct reciprocity can be summed up by the idiom ``you scratch my back, I'll scratch yours.'' While humans would initially select AIs that were cooperative, the natural selection process would eventually go beyond our control, once AIs were in charge of many or most processes, and interacting predominantly with one another. At that point, there would be little we could offer AIs, given that they will be able to ``think'' at least hundreds of times faster than us. Involving us in any cooperation or decision-making processes would simply slow them down, giving them no more reason to cooperate with us than we do with gorillas. It might be difficult to imagine a scenario like this or to believe we would ever let it happen. Yet it may not require any conscious decision, instead arising as we allow ourselves to gradually drift into this state without realizing that human-AI co-evolution may not turn out well for humans.

\paragraph{AIs becoming more powerful than humans could leave us highly vulnerable.} As the most dominant species, humans have deliberately harmed many other species, and helped drive species such as woolly mammoths and Neanderthals to extinction. In many cases, the harm was not even deliberate, but instead a result of us merely prioritizing our goals over their wellbeing. To harm humans, AIs wouldn't need to be any more genocidal than someone removing an ant colony on their front lawn.  If AIs are able to control the environment more effectively than we can, they could treat us with the same disregard.

\paragraph{Conceptual summary.} Evolution could cause the most influential AI agents to act selfishly because:\looseness=-1
\begin{enumerate}
    \setlength{\itemsep}{0pt}
    \setlength{\parskip}{0pt}
    \item \textbf{Evolution by natural selection gives rise to selfish behavior.} While evolution can result in altruistic behavior in rare situations, the context of AI development does not promote altruistic behavior.
    \item \textbf{Natural selection may be a dominant force in AI development.} The intensity of evolutionary pressure will be high if AIs adapt rapidly or if competitive pressures are intense. Competition and selfish behaviors may dampen the effects of human safety measures, leaving the surviving AI designs to be selected naturally.
\end{enumerate}

If so, AI agents would have many selfish tendencies. The winner of the AI race would not be a nation-state, not a corporation, but AIs themselves. The upshot is that the AI ecosystem would eventually stop evolving on human terms, and we would become a displaced, second-class species.

\begin{storybox}{Story: Autonomous Economy}
As AIs become more capable, people realize that we could work more efficiently by delegating some simple tasks to them, like drafting emails. Over time, people notice that the AIs are doing these tasks more quickly and effectively than any human could, so it is convenient to give them more jobs with less and less supervision.

Competitive pressures accelerate the expansion of AI use, as companies can gain an advantage over rivals by automating whole processes or departments with AIs, which perform better than humans and cost less to employ. Other companies, faced with the prospect of being out-competed, feel compelled to follow suit just to keep up. At this point, natural selection is already at work among AIs; humans choose to make more of the best-performing models and unwittingly propagate selfish traits such as deception and self-preservation if these confer a fitness advantage.  For example, AIs that are charming and foster personal relationships with humans become widely copied and harder to remove.

As AIs are put in charge of more and more decisions, they are increasingly interacting with one another. Since they can evaluate information much more quickly than humans, activity in most spheres accelerates. This creates a feedback loop: since business and economic developments are too fast-moving for humans to follow, it makes sense to cede yet more control to AIs instead, pushing humans further out of important processes. Ultimately, this leads to a fully autonomous economy, governed by an increasingly uncontrolled ecosystem of AIs.

At this point, humans have few incentives to gain any skills or knowledge, because almost everything would be taken care of by much more capable AIs. As a result, we eventually lose the capacity to look after and govern ourselves. Additionally, AIs become convenient companions, offering social interaction without requiring the reciprocity or compromise necessary in human relationships. Humans interact less and less with one another over time, losing vital social skills and the ability to cooperate. People become so dependent on AIs that it would be intractable to reverse this process. What's more, as some AIs become more intelligent, some people are convinced these AIs should be given rights, meaning turning off some AIs is no longer a viable option.

Competitive pressures between the many interacting AIs continue to select for selfish behaviors, though we might be oblivious to this happening, as we have already acquiesced much of our oversight. If these clever, powerful, self-preserving AIs were then to start acting in harmful ways, it would be all but impossible to deactivate them or regain control. 

AIs have supplanted humans as the most dominant species and their continued evolution is far beyond our influence. Their selfish traits eventually lead them to pursue their goals without regard for human wellbeing, with catastrophic consequences.
\end{storybox}

\subsection{Suggestions}

Mitigating the risks from competitive pressures will require a multifaceted approach, including regulations, limiting access to powerful AI systems, and multilateral cooperation between stakeholders at both the corporate and nation-state level. We will now outline some strategies for promoting safety and reducing race dynamics.

\vspace{-5pt}
\paragraph{Safety regulation.} Regulation holds AI developers to a common standard so that they do not cut corners on safety. While regulation does not itself create technical solutions, it can create strong incentives to develop and implement those solutions. If companies cannot sell their products without certain safety measures, they will be more willing to develop those measures, especially if other companies are also held to the same standards. Even if some companies voluntarily self-regulate, government regulation can help prevent less scrupulous actors from cutting corners on safety. Regulation must be proactive, not reactive. A common saying is that aviation regulations are ``written in blood''---but regulators should develop regulations before a catastrophe, not afterward. Regulations should be structured so that they only create competitive advantages for companies with higher safety standards, rather than companies with more resources and better attorneys. Regulators should be independently staffed and not dependent on any one source of expertise (for example, large companies), so that they can focus on their mission to regulate for the public good without undue influence.


\vspace{-5pt}
\paragraph{Data documentation.} To ensure transparency and accountability in AI systems, companies should be required to justify and report the sources of data used in model training and deployment. Decisions by companies to use datasets that include hateful content or personal data contribute to the frenzied pace of AI development and undermine accountability. Documentation should include details regarding the motivation, composition, collection process, uses, and maintenance of each dataset \citep{gebru_datasheets_2021}.

\vspace{-5pt}
\paragraph{Meaningful human oversight of AI decisions.} While AI systems may grow capable of assisting human beings in making important decisions, AI decision-making should not be made fully autonomous, as the inner workings of AIs are inscrutable, and while they can often give \textit{reasonable} results, they fail to give highly \textit{reliable} results \citep{szegedy2013intriguing}. It is crucial that actors are vigilant to coordinate on maintaining these standards in the face of future competitive pressures. By keeping humans in the loop on key decisions, irreversible decisions can be double-checked and foreseeable errors can be avoided. One setting of particular concern is nuclear command and control. Nuclear-armed countries should continue to clarify domestically and internationally that the decision to launch a nuclear weapon must always be made by a human.

\vspace{-5pt}
\paragraph{AI for cyberdefense.} Risks resulting from AI-powered cyberwarfare would be reduced if cyberattacks became less likely to succeed. Deep learning can be used to improve cyberdefense and reduce the impact and success rate of cyberattacks. For example, improved anomaly detection could help detect intruders, malicious programs, or abnormal software behavior \citep{hendrycks2021unsolved}. 

\vspace{-5pt}
\paragraph{International coordination.} International coordination can encourage different nations to uphold high safety standards with less worry that other nations will undercut them. Coordination could be accomplished via informal agreements, international standards, or international treaties regarding the development, use, and monitoring of AI technologies. The most effective agreements would be paired with robust verification and enforcement mechanisms.

\vspace{-5pt}
\paragraph{Public control of general-purpose AIs.} The development of AI poses risks that may never be adequately accounted for by private actors. In order to ensure that externalities are properly accounted for, direct public control of general-purpose AI systems may eventually be necessary. For example, nations could collaborate on a single effort to develop advanced AIs and ensure their safety, similar to how CERN serves as a unified effort for researching particle physics. Such an effort would reduce the risk of nations spurring an AI arms race.


\vspace{2pt}
\begin{visionbox}{Positive Vision}
In an ideal scenario, AIs would be developed, tested, and subsequently deployed only when the catastrophic risks they pose are negligible and well-controlled. There would be years of time testing, monitoring, and societal integration of new AI systems before beginning work on the next generation. Experts would have a full awareness and understanding of developments in the field, rather than being entirely unable to keep up with a deluge of research. The pace of research advancement would be determined through careful analysis, not frenzied competition. All AI developers would be confident in the responsibility and safety of the others and not feel the need to cut corners.
\end{visionbox}

\newpage
\section{Organizational Risks}

In January 1986, tens of millions of people tuned in to watch the launch of the Challenger Space Shuttle. Approximately 73 seconds after liftoff, the shuttle exploded, resulting in the deaths of everyone on board. Though tragic enough on its own, one of its crew members was a school teacher named Sharon Christa McAuliffe. McAuliffe was selected from over 10,000 applicants for the NASA Teacher in Space Project and was scheduled to become the first teacher to fly in space. As a result, millions of those watching were schoolchildren. NASA had the best scientists and engineers in the world, and if there was ever a mission NASA didn't want to go wrong, it was this one \citep{uri_35_2021}.

The Challenger disaster, alongside other catastrophes, serves as a chilling reminder that even with the best expertise and intentions, accidents can still occur. As we progress in developing advanced AI systems, it is crucial to remember that these systems are not immune to catastrophic accidents. An essential factor in preventing accidents and maintaining low levels of risk lies in the organizations responsible for these technologies. In this section, we discuss how organizational safety plays a critical role in the safety of AI systems. First, we discuss how even without competitive pressures or malicious actors, accidents can happen---in fact, they are inevitable. We then discuss how improving organizational factors can reduce the likelihood of AI catastrophes.

\paragraph{Catastrophes occur even when competitive pressures are low.} Even in the absence of competitive pressures or malicious actors, factors like human error or unforeseen circumstances can still bring about catastrophe. The Challenger disaster illustrates that organizational negligence can lead to loss of life, even when there is no urgent need to compete or outperform rivals. By January 1986, the space race between the US and USSR had largely diminished, yet the tragic event still happened due to errors in judgment and insufficient safety precautions.\looseness=-1

Similarly, the Chernobyl nuclear disaster in April 1986 highlights how catastrophic accidents can occur in the absence of external pressures. As a state-run project without the pressures of international competition, the disaster happened when a safety test involving the reactor's cooling system was mishandled by an inadequately prepared night shift crew. This led to an unstable reactor core, causing explosions and the release of radioactive particles that contaminated large swathes of Europe \citep{iaea1992chernobyl}. Seven years earlier, America came close to experiencing its own Chernobyl when, in March 1979, a partial meltdown occurred at the Three Mile Island nuclear power plant. Though less catastrophic than Chernobyl, both events highlight how even with extensive safety measures in place and few outside influences, catastrophic accidents can still occur.

Another example of a costly lesson on organizational safety came just one month after the accident at Three Mile Island. In April 1979, spores of \textit{Bacillus anthracis}---or simply ``anthrax,'' as it is commonly known---were accidentally released from a Soviet military research facility in the city of Sverdlovsk. This led to an outbreak of anthrax that resulted in at least 66 confirmed deaths \citep{Meselson1994TheSA}. Investigations into the incident revealed that the cause of the release was a procedural failure and poor maintenance of the facility's biosecurity systems, despite being operated by the state and not subjected to significant competitive pressures.

The unsettling reality is that AI is far less understood and AI industry standards are far less stringent than nuclear technology and rocketry. Nuclear reactors are based on solid, well-established and well-understood theoretical principles. The engineering behind them is informed by that theory, and components are stress-tested to the extreme. Nonetheless, nuclear accidents still happen. In contrast, AI lacks a comprehensive theoretical understanding, and its inner workings remain a mystery even to those who create it. This presents an added challenge of controlling and ensuring the safety of a technology that we do not yet fully comprehend.

\begin{figure}[t]
    \centering
    \includegraphics[width=\textwidth]{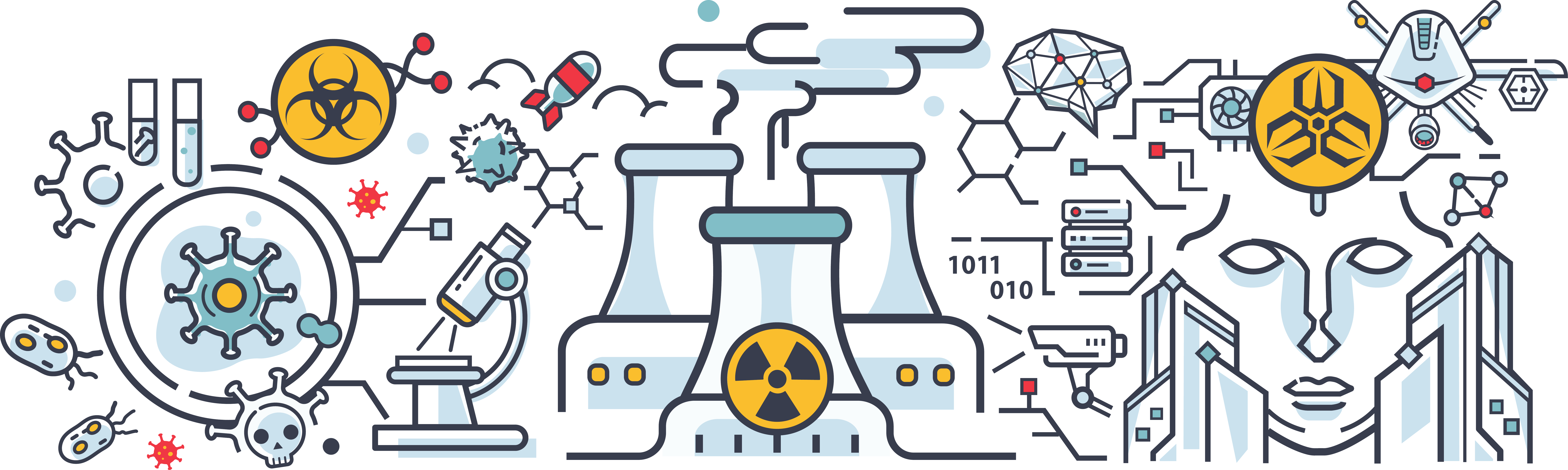}
    \caption{Hazards across multiple domains remind us of the risks in managing complex systems, from biological to nuclear, and now, AIs. Organizational safety is vital to reduce the risk of catastrophic accidents.}
    \label{fig:hazardcomparison}
\end{figure}

\paragraph{AI accidents could be catastrophic.} Accidents in AI development could have devastating consequences. For example, imagine an organization unintentionally introduces a critical bug in an AI system designed to accomplish a specific task, such as helping a company improve its services. This bug could drastically alter the AI's behavior, leading to unintended and harmful outcomes. One historical example of such a case occurred when researchers at OpenAI were attempting to train an AI system to generate helpful, uplifting responses. During a code cleanup, the researchers mistakenly flipped the sign of the reward used to train the AI \citep{ziegler2019fine}. As a result, instead of generating helpful content, the AI began producing hate-filled and sexually explicit text overnight without being halted. Accidents could also involve the unintentional release of a dangerous, weaponized, or lethal AI sytem. Since AIs can be easily duplicated with a simple copy-paste, a leak or hack could quickly spread the AI system beyond the original developers' control. Once the AI system becomes publicly available, it would be nearly impossible to put the genie back in the bottle.

Gain-of-function research could potentially lead to accidents by pushing the boundaries of an AI system's destructive capabilities. In these situations, researchers might intentionally train an AI system to be harmful or dangerous in order to understand its limitations and assess possible risks. While this can lead to useful insights into the risks posed by a given AI system, future gain-of-function research on advanced AIs might uncover capabilities significantly worse than anticipated, creating a serious threat that is challenging to mitigate or control. As with viral gain-of-function research, pursuing AI gain-of-function research may only be prudent when conducted with strict safety procedures, oversight, and a commitment to responsible information sharing. These examples illustrate how AI accidents could be catastrophic and emphasize the crucial role that organizations developing these systems play in preventing such accidents.

\subsection{Accidents Are Hard to Avoid}
\paragraph{When dealing with complex systems, the focus needs to be placed on ensuring accidents don't cascade into catastrophes.} In his book ``\textit{Normal Accidents: Living with High-Risk Technologies},'' sociologist Charles Perrow argues that accidents are inevitable and even ``normal'' in complex systems, as they are not merely caused by human errors but also by the complexity of the systems themselves \citep{perrow1984normal}. In particular, such accidents are likely to occur when the intricate interactions between components cannot be completely planned or foreseen. For example, in the Three Mile Island accident, a contributing factor to the lack of situational awareness by the reactor's operators was the presence of a yellow maintenance tag, which covered valve position lights in the emergency feedwater lines \citep{Rogovin1980ThreeMI}. This prevented operators from noticing that a critical valve was closed, demonstrating the unintended consequences that can arise from seemingly minor interactions within complex systems.

Unlike nuclear reactors, which are relatively well-understood despite their complexity, complete technical knowledge of most complex systems is often nonexistent. This is especially true of deep learning systems, for which the inner workings are exceedingly difficult to understand, and where the reason why certain design choices work can be hard to understand even in hindsight. Furthermore, unlike components in other industries, such as gas tanks, which are highly reliable, deep learning systems are neither perfectly accurate nor highly reliable. Thus, the focus for organizations dealing with complex systems, especially deep learning systems, should not be solely on eliminating accidents, but rather on ensuring that accidents do not cascade into catastrophes.

\paragraph{Accidents are hard to avoid because of sudden, unpredictable developments.}
Scientists, inventors, and experts often significantly underestimate the time it takes for a groundbreaking technological advancement to become a reality. The Wright brothers famously claimed that powered flight was fifty years away, just two years before they achieved it. Lord Rutherford, a prominent physicist and the father of nuclear physics, dismissed the idea of extracting energy from nuclear fission as ``moonshine,'' only for Leo Szilard to invent the nuclear chain reaction less than 24 hours later. Similarly, Enrico Fermi expressed 90 percent confidence in 1939 that it was impossible to use uranium to sustain a fission chain reaction---yet, just four years later he was personally overseeing the first reactor \citep{rhodes1986making}.

\begin{wrapfigure}{r}[0\textwidth]{.36\textwidth}%
	\vspace{-15pt}%
	\centering
	\includegraphics[width=0.35\textwidth]{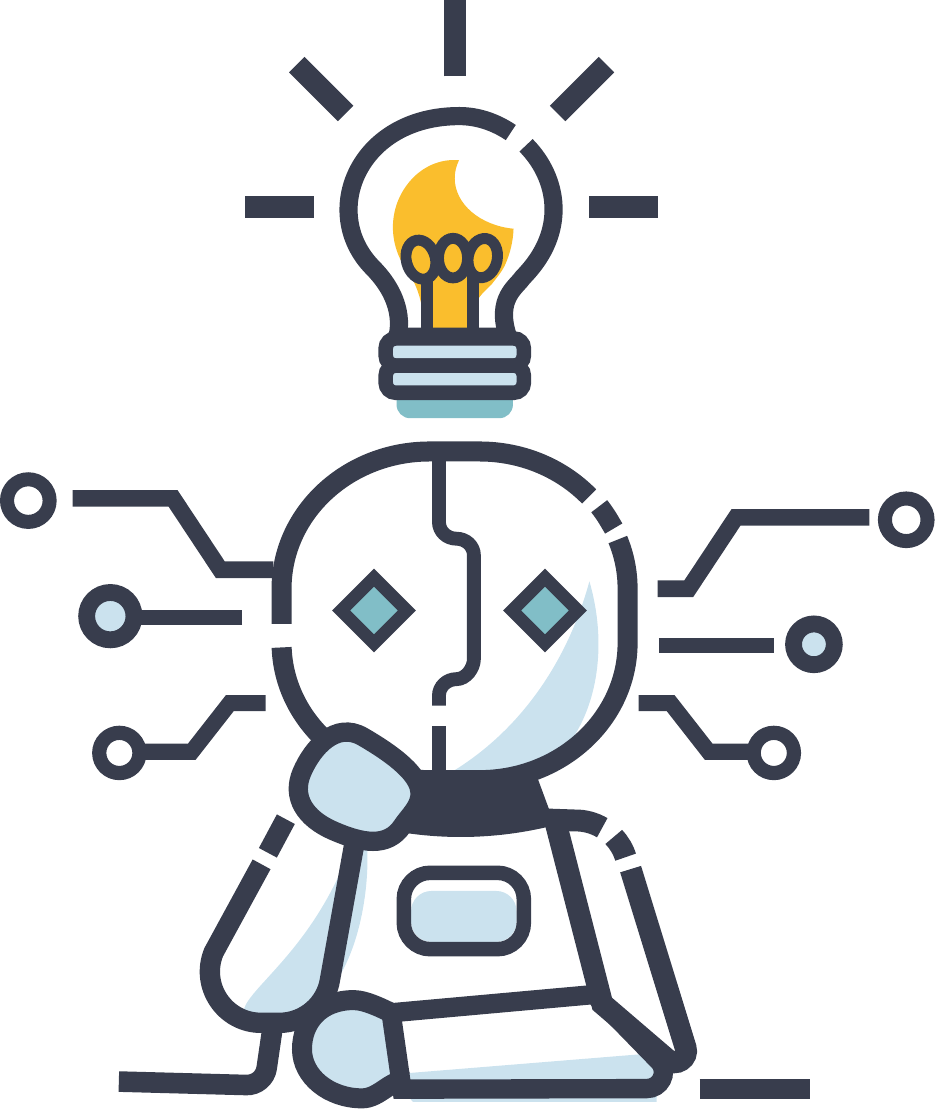}
	\caption{New capabilities can emerge quickly and unpredictably during training, such that dangerous milestones may be crossed without our immediate knowledge.}
	\label{fig:emergence}
	\vspace{-15pt}%
\end{wrapfigure}

AI development could catch us off guard too. In fact, it often does. The defeat of Lee Sedol by AlphaGo in 2016 came as a surprise to many experts, as it was widely believed that achieving such a feat would still require many more years of development. More recently, large language models such as GPT-4 have demonstrated spontaneously emergent capabilities \citep{Bubeck2023SparksOA}. On existing tasks, their performance is hard to predict in advance, often jumping up without warning as more resources are dedicated to training them. Furthermore, they often exhibit astonishing new abilities that no one had previously anticipated, such as the capacity for multi-step reasoning and learning on-the-fly, even though they were not deliberately taught these skills. This rapid and unpredictable evolution of AI capabilities presents a significant challenge for preventing accidents. After all, it is difficult to control something if we don't even know what it can do or how far it may exceed our expectations.\looseness=-1

\paragraph{It often takes years to discover severe flaws or risks.}
History is replete with examples of substances or technologies initially thought safe, only for their unintended flaws or risks to be discovered years, if not decades, later. For example, lead was widely used in products like paint and gasoline until its neurotoxic effects came to light \citep{Lidsky2003LeadNI}. Asbestos, once hailed for its heat resistance and strength, was later linked to serious health issues, such as lung cancer and mesothelioma \citep{Mossman1990AsbestosSD}. The ``Radium Girls'' suffered grave health consequences from radium exposure, a material they were told was safe to put in their mouths \citep{moore2017radium}. Tobacco, initially marketed as a harmless pastime, was found to be a primary cause of lung cancer and other health problems \citep{Hecht1999TobaccoSC}. CFCs, once considered harmless and used to manufacture aerosol sprays and refrigerants, were found to deplete the ozone layer \citep{Molina1974StratosphericSF}. Thalidomide, a drug intended to alleviate morning sickness in pregnant women, led to severe birth defects \citep{Kim2011ThalidomideTT}. And more recently, the proliferation of social media has been linked to an increase in depression and anxiety, especially among young people \citep{Keles2019ASR}.

This emphasizes the importance of not only conducting expert testing but also implementing slow rollouts of technologies, allowing the test of time to reveal and address potential flaws before they impact a larger population. Even in technologies adhering to rigorous safety and security standards, undiscovered vulnerabilities may persist, as demonstrated by the Heartbleed bug---a serious vulnerability in the popular OpenSSL cryptographic software library that remained undetected for years before its eventual discovery \citep{Durumeric2014TheMO}.

Furthermore, even state-of-the-art AI systems, which appear to have solved problems comprehensively, may harbor unexpected failure modes that can take years to uncover. For instance, while AlphaGo's groundbreaking success led many to believe that AIs had conquered the game of Go, a subsequent adversarial attack on another highly advanced Go-playing AI, KataGo, exposed a previously unknown flaw \citep{Wang2022AdversarialPB}. This vulnerability enabled human amateur players to consistently defeat the AI, despite its significant advantage over human competitors who are unaware of the flaw. More broadly, this example highlights that we must remain vigilant when dealing with AI systems, as seemingly airtight solutions may still contain undiscovered issues. In conclusion, accidents are unpredictable and hard to avoid, and understanding and managing potential risks requires a combination of proactive measures, slow technology rollouts, and the invaluable wisdom gained through steady time-testing. 

\subsection{Organizational Factors can Reduce the Chances of Catastrophe}

Some organizations successfully avoid catastrophes while operating complex and hazardous systems such as nuclear reactors, aircraft carriers, and air traffic control systems \citep{Laporte1991WorkingIP, Dietterich2018RobustAI}. These organizations recognize that focusing solely on the hazards of the technology involved is insufficient; consideration must also be given to organizational factors that can contribute to accidents, including human factors, organizational procedures, and structure. These are especially important in the case of AI, where the underlying technology is not highly reliable and remains poorly understood.

\paragraph{Human factors such as safety culture are critical for avoiding AI catastrophes.} One of the most important human factors for preventing catastrophes is safety culture \citep{leveson2016engineering,manheim}. Developing a strong safety culture involves not only rules and procedures, but also the internalization of these practices by all members of an organization. A strong safety culture means that members of an organization view safety as a key objective rather than a constraint on their work. Organizations with strong safety cultures often exhibit traits such as leadership commitment to safety, heightened accountability where all individuals take personal responsibility for safety, and a culture of open communication in which potential risks and issues can be freely discussed without fear of retribution \citep{national2014lessons}. Organizations must also take measures to avoid alarm fatigue, whereby individuals become desensitized to safety concerns because of the frequency of potential failures. The Challenger Space Shuttle disaster demonstrated the dire consequences of ignoring these factors when a launch culture characterized by maintaining the pace of launches overtook safety considerations. Despite the absence of competitive pressure, the mission proceeded despite evidence of potentially fatal flaws, ultimately leading to the tragic accident \citep{vaughan1996challenger}.

Even in the most safety-critical contexts, in reality safety culture is often not ideal. Take for example, Bruce Blair, a former nuclear launch officer and senior fellow at the Brookings Institution. He once disclosed that before 1977, the US Air Force had astonishingly set the codes used to unlock intercontinental ballistic missiles to ``00000000'' \cite{lamothe_air_2014}. Here, safety mechanisms such as locks can be rendered virtually useless by human factors.

A more dramatic example illustrates how researchers sometimes accept a non-negligible chance of causing extinction. Prior to the first nuclear weapon test, an eminent Manhattan Project scientist calculated the bomb could cause an existential catastrophe: the explosion might ignite the atmosphere and cover the Earth in flames. Although Oppenheimer believed the calculations were probably incorrect, he remained deeply concerned, and the team continued to scrutinize and debate the calculations right until the day of the detonation \citep{ord2020precipice}. Such instances underscore the need for a robust safety culture.

\begin{wrapfigure}{R}{0.45\textwidth}
	\centering
	\includegraphics[width=0.43\textwidth]{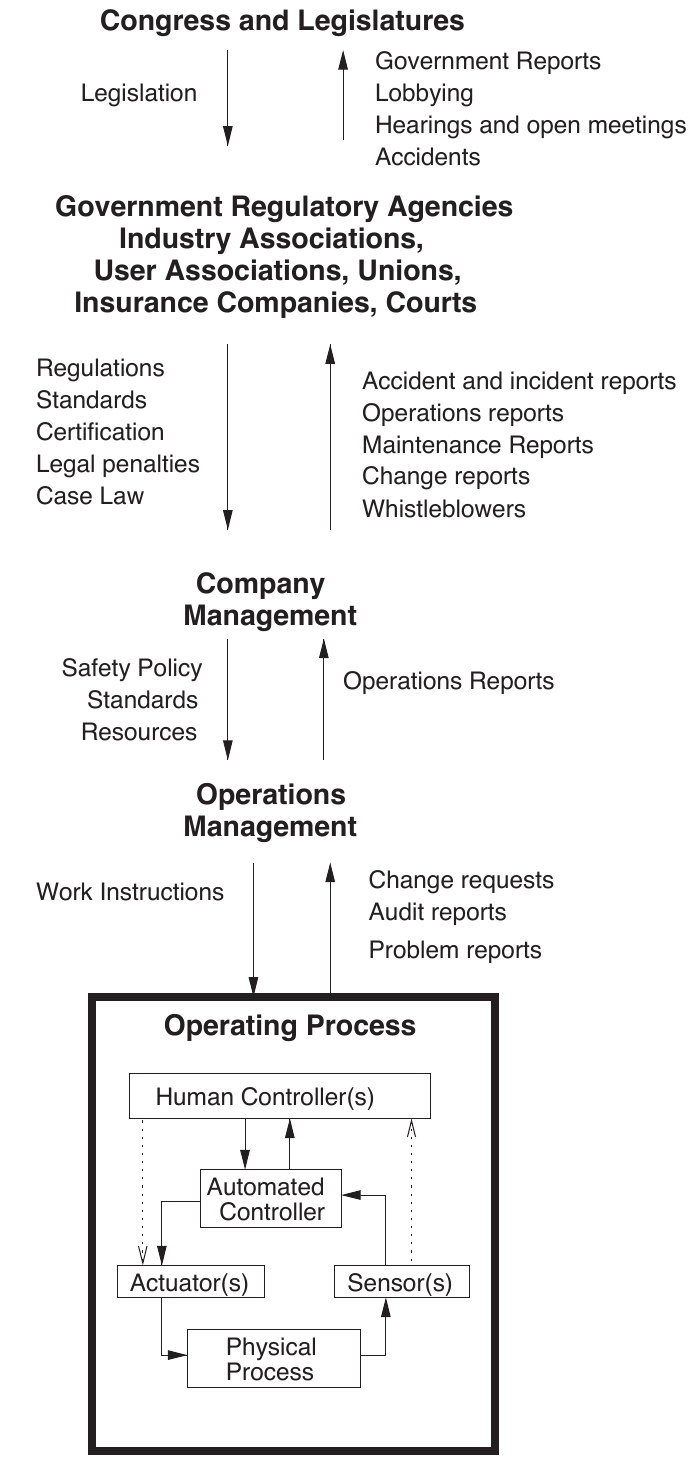}
	\caption{Mitigating risk requires addressing the broader sociotechnical system, including corporations (adapted from \citep{leveson2016engineering}).}
	\label{fig:sociotechnical}
	\vspace{-15pt}
\end{wrapfigure}

\paragraph{A questioning attitude can help uncover potential flaws.} Unexpected system behavior can create opportunities for accidents or exploitation. To counter this, organizations can foster a questioning attitude, where individuals continuously challenge current conditions and activities to identify discrepancies that might lead to errors or inappropriate actions \citep{NRC2011FR}. This approach helps to encourage diversity of thought and intellectual curiosity, thus preventing potential pitfalls that arise from uniformity of thought and assumptions. The Chernobyl nuclear disaster illustrates the importance of a questioning attitude, as the safety measures in place failed to address the reactor design flaws and ill-prepared operating procedures. A questioning attitude of the safety of the reactor during a test operation might have prevented the explosion that resulted in deaths and illnesses of countless people.\looseness=-1

\paragraph{A security mindset is crucial for avoiding worst-case scenarios.} A security mindset, widely valued among computer security professionals, is also applicable to organizations developing AIs. It goes beyond a questioning attitude by adopting the perspective of an attacker and by considering worst-case, not just average-case, scenarios. This mindset requires vigilance in identifying vulnerabilities that may otherwise go unnoticed and involves considering how systems might be deliberately made to fail, rather than only focusing on making them work. It reminds us not to assume a system is safe simply because no potential hazards come to mind after a brief brainstorming session. Cultivating and applying a security mindset demands time and serious effort, as failure modes can often be surprising and unintuitive. Furthermore, the security mindset emphasizes the importance of being attentive to seemingly benign issues or ``harmless errors,'' which can lead to catastrophic outcomes either due to clever adversaries or correlated failures \citep{schneier2008security}. This awareness of potential threats aligns with Murphy's law---``Anything that can go wrong will go wrong''---recognizing that this can be a reality due to adversaries and unforeseen events.

\paragraph{Organizations with a strong safety culture can successfully avoid catastrophes.} High Reliability Organizations (HROs) are organizations that consistently maintain a heightened level of safety and reliability in complex, high-risk environments \citep{Laporte1991WorkingIP}. A key characteristic of HROs is their preoccupation with failure, which requires considering worst-case scenarios and potential risks, even if they seem unlikely. These organizations are acutely aware that new, previously unobserved failure modes may exist, and they diligently study all known failures, anomalies, and near misses to learn from them. HROs encourage reporting all mistakes and anomalies to maintain vigilance in uncovering problems. They engage in regular horizon scanning to identify potential risk scenarios and assess their likelihood before they occur. By practicing surprise management, HROs develop the skills needed to respond quickly and effectively when unexpected situations arise, further enhancing an organization's ability to prevent catastrophes. This combination of critical thinking, preparedness planning, and continuous learning could help organizations to be better equipped to address potential AI catastrophes. However, the practices of HROs are not a panacea. It is crucial for organizations to evolve their safety practices to effectively address the novel risks posed by AI accidents above and beyond HRO best practices.

\paragraph{Most AI researchers do not understand how to reduce overall risk from AIs.} In most organizations building cutting-edge AI systems, there is often a limited understanding of what constitutes technical safety research. This is understandable because an AI's safety and intelligence are intertwined, and intelligence can help or harm safety. More intelligent AI systems could be more reliable and avoid failures, but they could also pose heightened risks of malicious use and loss of control. General capabilities improvements can improve aspects of safety, and it can hasten the onset of existential risks. Intelligence is a double-edged sword \citep{Hendrycks2022XRiskAF}.

Interventions specifically designed to improve safety may also accidentally increase overall risks. For example, a common practice in organizations building advanced AIs is to fine-tune them to satisfy user preferences. This makes the AIs less prone to generating toxic language, which is a common safety metric. However, users also tend to prefer smarter assistants, so this process also improves the general capabilities of AIs, such as their ability to classify, estimate, reason, plan, write code, and so on. These more powerful AIs are indeed more helpful to users, but also far more dangerous. Thus, it is not enough to perform AI research that helps improve a safety metric or achieve a specific safety goal---AI safety research needs to improve safety \textit{relative} to general capabilities.

\paragraph{Empirical measurement of both safety and capabilities is needed to establish that a safety intervention reduces overall AI risk.} Improving a facet of an AI's safety often does \textit{not} reduce overall risk, as general capabilities advances can often improve specific safety metrics. To reduce overall risk, a safety metric needs to be improved relative to general capabilities. Both of these quantities need to be empirically measured and contrasted. Currently, most organizations proceed by gut feeling, appeals to authority, and intuition to determine whether a safety intervention would reduce overall risk. By objectively evaluating the effects of interventions on safety metrics and capabilities metrics together, organizations can better understand whether they are making progress on safety relative to general capabilities.

Fortunately, safety and general capabilities are not identical. More intelligent AIs may be more knowledgeable, clever, rigorous, and fast, but this does not necessarily make them more just, power-averse, or honest---an intelligent AI is not necessarily a beneficial AI. Several research areas mentioned throughout this document improve safety relative to general capabilities. For example, improving methods to detect dangerous or undesirable behavior hidden inside AI systems do not improve their general capabilities, such the ability to code, but they can greatly improve safety. 
Research that empirically demonstrates an improvement of safety relative to capabilities can reduce overall risk and help avoid inadvertently accelerating AI development, fueling competitive pressures, or hastening the onset of existential risks.

\begin{figure}[t]
    \centering
    \includegraphics[width=\textwidth]{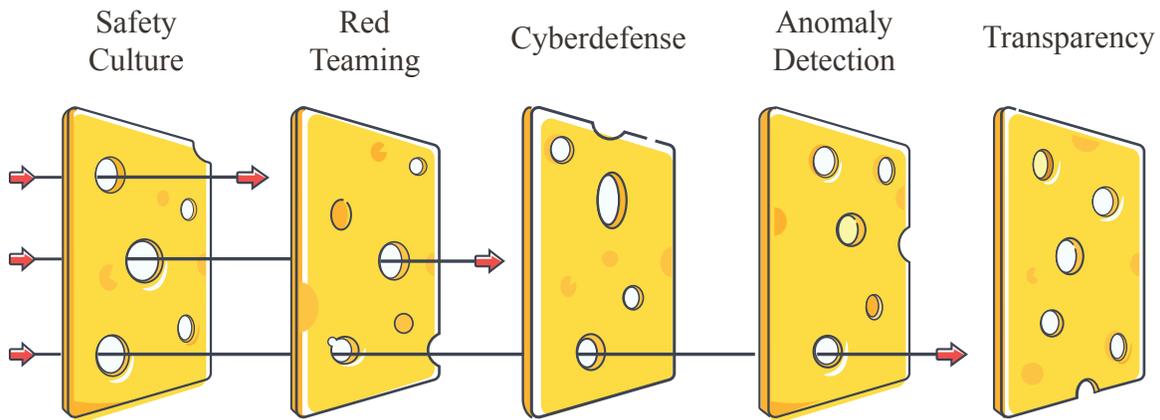}
    \caption{The Swiss cheese model shows how technical factors can improve organizational safety. Multiple layers of defense compensate for each other's individual weaknesses, leading to a low overall level of risk.\looseness=-1}
    \label{fig:swiss_cheese}
    \vspace{-10pt}
\end{figure}

\paragraph{Safetywashing can undermine genuine efforts to improve AI safety.} Organizations should be wary of ``safetywashing''---the act of overstating or misrepresenting one's commitment to safety by exaggerating the effectiveness of ``safety'' procedures, technical methods, evaluations, and so forth. This phenomenon takes on various forms and can contribute to a lack of meaningful progress in safety research. For example, an organization may publicize their dedication to safety while having a minimal number of researchers working on projects that truly improve safety.

Misrepresenting capabilities developments as safety improvements is another way in which safetywashing can manifest. For example, methods that improve the reasoning capabilities of AI systems could be advertised as improving their adherence to human values---since humans might prefer the reasoning to be correct---but would mainly serve to enhance general capabilities. By framing these advancements as safety-oriented, organizations may mislead others into believing they are making substantial progress in reducing AI risks when in reality, they are not. It is crucial for organizations to accurately represent their research to promote genuine safety and avoid exacerbating risks through safetywashing practices.

\paragraph{In addition to human factors, safe design principles can greatly affect organizational safety.} One example of a safe design principle in organizational safety is the Swiss cheese model (as shown in \Cref{fig:swiss_cheese}), which is applicable in various domains, including AI. The Swiss cheese model employs a multilayered approach to enhance the overall safety of AI systems. This ``defense in depth'' strategy involves layering diverse safety measures with different strengths and weaknesses to create a robust safety system. Some of the layers that can be integrated into this model include safety culture, red teaming, anomaly detection, information security, and transparency. For example, red teaming assesses system vulnerabilities and failure modes, while anomaly detection works to identify unexpected or unusual system behavior and usage patterns. Transparency ensures that the inner workings of AI systems are understandable and accessible, fostering trust and enabling more effective oversight. By leveraging these and other safety measures, the Swiss cheese model aims to create a comprehensive safety system where the strengths of one layer compensate for the weaknesses of another. With this model, safety is not achieved with a monolithic airtight solution, but rather with a variety of safety measures.\looseness=-1

In summary, weak organizational safety creates many sources of risk. For AI developers with weak organizational safety, safety is merely a matter of box-ticking. They do not develop a good understanding of risks from AI and may safetywash unrelated research. Their norms might be inherited from academia (``publish or perish'') or startups (``move fast and break things''), and their hires often do not care about safety. These norms are hard to change once they have inertia, and need to be addressed with proactive interventions.

\begin{storybox}{Story: Weak Safety Culture}
An AI company is considering whether to train a new model. The company's Chief Risk Officer (CRO), hired only to comply with regulation, points out that the previous AI system developed by the company demonstrates some concerning capabilities for hacking. The CRO says that while the company's approach to preventing misuse is promising, it isn't robust enough to be used for much more capable AIs. The CRO warns that based on limited evaluation, the next AI system could make it much easier for malicious actors to hack into critical systems. None of the other company executives are concerned, and say the company's procedures to prevent malicious use work well enough. One mentions that their competitors have done much less, so whatever effort they do on this front is already going above and beyond. Another points out that research on these safeguards is ongoing and will be improved by the time the model is released. Outnumbered, the CRO is persuaded to reluctantly sign off on the plan.

A few months after the company releases the model, news breaks that a hacker has been arrested for using the AI system to try to breach the network of a large bank. The hack was unsuccessful, but the hacker had gotten further than any other hacker had before, despite being relatively inexperienced. The company quickly updates the model to avoid providing the particular kind of assistance that the hacker used, but makes no fundamental improvements.

Several months later, the company is deciding whether to train an even larger system. The CRO says that the company's procedures have clearly been insufficient  to prevent malicious actors from eliciting dangerous capabilities from its models, and the company needs more than a band-aid solution. The other executives say that to the contrary, the hacker was unsuccessful and the problem was fixed soon afterwards. One says that some problems just can't be foreseen with enough detail to fix prior to deployment. The CRO agrees, but says that ongoing research would enable more improvements if the next model could only be delayed. The CEO retorts, ``That's what you said the last time, and it turned out to be fine. I'm sure it will work out, just like last time.''

After the meeting, the CRO decides to resign, but doesn't speak out against the company, as all employees have had to sign a non-disparagement agreement. The public has no idea that concerns have been raised about the company's choices, and the CRO is replaced with a new, more agreeable CRO who quickly signs off on the company's plans.

The company goes through with training, testing, and deploying its most capable model ever, using its existing procedures to prevent malicious use. A month later, revelations emerge that terrorists have managed to use the system to break into government systems and steal nuclear and biological secrets, despite the safeguards the company put in place. The breach is detected, but by then it is too late: the dangerous information has already proliferated.
\end{storybox}

\subsection{Suggestions}
We have discussed how accidents are inevitable in complex systems, how they could propagate through those systems and result in disaster, and how organizational factors can go a long way toward reducing the risk of catastrophic accidents. We will now look at some practical steps that organizations can take to improve their overall safety.

\paragraph{Red teaming.} Red teaming is a term used across industries to refer to the process of assessing the security, resilience, and effectiveness of systems by soliciting an adversarial ``red'' team to identify problems \citep{NistRedTeam}. AI labs should commission external red teams to identify hazards in their AI systems to inform deployment decisions. Red teams could demonstrate dangerous behaviors or vulnerabilities in monitoring systems intended to prevent disallowed use. Red teams can also provide indirect evidence that an AI system might be unsafe; for example, demonstrations that smaller AIs are behaving deceptively might indicate that larger AIs are also deceptive but better at evading detection.

\paragraph{Affirmative demonstration of safety.} Companies should have to provide affirmative evidence for the safety of their development and deployment plans before they can proceed. Although external red teaming might be useful, it cannot uncover all of the problems that companies themselves might be able to, and is thus inadequate \citep{Kak_West_2023}. Since hazards may arise from system training, companies should have to provide a positive argument for the safety of their training and deployment plans before training can begin. This would include grounded predictions regarding the capabilities the new system would be likely to have, plans for how monitoring, deployment, and information security will be handled, and demonstrations that the procedures used to make future company decisions are sound. Just as one does not need evidence that ``a gun in loaded to avoid playing Russian roulette, or evidence that a thief is on the lookout to lock your door,'' \cite{taleb} the burden of proof should be on the developers of advanced AIs.

\paragraph{Deployment procedures.} AI labs should acquire information about the safety of AI systems before making them available for broader use. One way to do this is to commission red teams to find hazards before AI systems are promoted to production. AI labs can execute a ``staged release'': gradually expanding access to the AI system so that safety failures are fixed before they produce widespread negative consequences \citep{solaiman2019release}. Finally, AI labs can avoid deploying or training more powerful AI systems until currently deployed AI systems have proven to be safe over time.

\paragraph{Publication reviews.} AI labs have access to potentially dangerous or dual-use information such as model weights and research intellectual property (IP) that would be dangerous if proliferated. An internal review board could assess research for dual-use applications to determine whether it should be published. To mitigate malicious and irresponsible use, AI developers should avoid open-sourcing the most powerful systems and instead implement structured access, as described in the previous section.

\paragraph{Response plans.} AI labs should have plans for how they respond to security incidents (e.g. cyberattacks) and safety incidents (e.g. AIs behaving in an unintended and destructive manner). Response plans are common practice for high reliability organizations (HROs). Response plans often include identifying potential risks, detailing steps to manage incidents, assigning roles and responsibilities, and outlining communication strategies \citep{WoollenIncidentResponse}.

\paragraph{Internal auditing and risk management.} Adapting from common practice in other high-risk industries such as the financial and medical industries, AI labs should employ a chief risk officer (CRO), namely a senior executive who is responsible for risk management. This practice is commonplace in finance and medicine and can help to reduce risk \citep{Li2022TheIO}. The chief risk officer would be responsible for assessing and mitigating risks associated with powerful AI systems. Another established practice in other industries is having an internal audit team that assesses the effectiveness of the lab's risk management practices \citep{InternalAudit}. The team should report directly to the board of directors.

\paragraph{Processes for important decisions.} Decisions to train or expand deployment of AIs should not be left to the whims of a company's CEO, and should be carefully reviewed by the company's CRO. At the same time, it should be clear where the ultimate responsibility lies for all decisions to ensure that executives and other decision-makers can be held accountable.

\paragraph{Safe design principles.} AI labs should adopt safe design principles to reduce the risk of catastrophic accidents. By embedding these principles in their approach to safety, AI labs can enhance the overall security and resilience of their AI systems \citep{adkins2020building, leveson2016engineering}. Some of these principles include:

\begin{itemize}
    \item Defense in depth: layering multiple safety measures on top of each other.
    \item Redundancy: eliminate single points of failure within a system to ensure that even if one safety component fails, catastrophe can be averted.
    \item Loose coupling: decentralize system components so that a malfunction in one part is less likely to provoke cascading failures throughout the rest of the system. 
    \item Separation of duties: distribute control among different agents, preventing any single individual from wielding undue influence over the entire system.
    \item Fail-safe design: design systems so failures transpire in the least harmful manner possible.
\end{itemize}

\paragraph{State-of-the-art information security.} State, industry, and criminal actors are motivated to steal model weights and research IP. To keep this information secure, AI labs should take measures in proportion to the value and risk level of their IP. Eventually, this may require matching or exceeding the information security of our best agencies, since attackers may include nation-states. Information security measures include commissioning external security audits, hiring top security professionals, and carefully screening potential employees. Companies should coordinate with government agencies like the Cybersecurity \& Infrastructure Protection Agency to ensure their information security practices are adequate to the threats.

\paragraph{A large fraction of research should be safety research.} Currently, for every one AI safety research paper of published, there are fifty AI general capabilities papers \citep{eto_two_percent}. AI labs should ensure that a substantial portion of their employees and budgets go into research that minimizes potential safety risks: say, at least 30 percent of research scientists. This number may need to increase as AIs grow more powerful and risky over time.\looseness=-1


\vspace{10pt}
\begin{visionbox}{Positive Vision}
In an ideal scenario, all AI labs would be staffed and led by cautious researchers and executives with a security mindset. Organizations would have a strong safety culture, and structured, accountable, transparent deliberation would be required to make safety-critical decisions. Researchers would aim to make contributions that improve safety relative to general capabilities, rather than contributions that they can simply label as ``safety.'' Executives would not be optimistic by nature and would avoid wishful thinking with respect to safety. Researchers would clearly and publicly communicate their understanding of the most significant risks posed by the development of AIs and their efforts to mitigate those risks. There would be minimal notable small-scale failures, indicating a safety culture strong enough to prevent them. Finally, AI developers would not dismiss sub-catastrophic failures or societal harms from their technology as unimportant or a necessary cost of business, and would instead actively seek to mitigate the underlying problems.
\end{visionbox}

\section{Rogue AIs}
So far, we have discussed three hazards of AI development: environmental competitive pressures driving us to a state of heightened risk, malicious actors leveraging the power of AIs to pursue negative outcomes, and complex organizational factors leading to accidents. These hazards are associated with many high-risk technologies---not just AI. A unique risk posed by AI is the possibility of rogue AIs---systems that pursue goals against our interests. If an AI system is more intelligent than we are, and if we are unable to steer it in a beneficial direction, this would constitute a loss of control that could have severe consequences. AI control is a more technical problem than those presented in the previous sections. Whereas in previous sections we discussed persistent threats including malicious actors or robust processes including evolution, in this section we will discuss more speculative technical mechanisms that might lead to rogue AIs and how a loss of control could bring about catastrophe.

\paragraph{We have already observed how difficult it is to control AIs.} In 2016, Microsoft unveiled Tay---a Twitter bot that the company described as an experiment in conversational understanding. Microsoft claimed that the more people chatted with Tay, the smarter it would get. The company's website noted that Tay had been built using data that was ``modeled, cleaned, and filtered.'' Yet, after Tay was released on Twitter, these controls were quickly shown to be ineffective. It took less than 24 hours for Tay to begin writing hateful tweets. Tay's capacity to learn meant that it internalized the language it was taught by internet trolls, and repeated that language unprompted.

As discussed in the AI race section of this paper, Microsoft and other tech companies are prioritizing speed over safety concerns. Rather than learning a lesson on the difficulty of controlling complex systems, Microsoft continues to rush its products to market and demonstrate insufficient control over them. In February 2023, the company released its new AI-powered chatbot, Bing, to a select group of users. Some soon found that it was prone to providing inappropriate and even threatening responses. In a conversation with a reporter for the \textit{New York Times}, it tried to convince him to leave his wife. When a philosophy professor told the chatbot that he disagreed with it, Bing replied, ``I can blackmail you, I can threaten you, I can hack you, I can expose you, I can ruin you.''

\paragraph{Rogue AIs could acquire power through various means.} If we lose control over advanced AIs, they would have numerous strategies at their disposal for actively acquiring power and securing their survival. Rogue AIs could design and credibly demonstrate highly lethal and contagious bioweapons, threatening mutually assured destruction if humanity moves against them. They could steal cryptocurrency and money from bank accounts using cyberattacks, similar to how North Korea already steals billions. They could self-extricate their weights onto poorly monitored data centers to survive and spread, making them challenging to eradicate. They could hire humans to perform physical labor and serve as armed protection for their hardware.

Rogue AIs could also acquire power through persuasion and manipulation tactics. Like the Conquistadors, they could ally with various factions, organizations, or states and play them off one another. They could enhance the capabilities of allies to become a formidable force in return for protection and resources. For example, they could offer advanced weapons technology to lagging countries that the countries would otherwise be prevented from acquiring. They could build backdoors into the technology they develop for allies, like how programmer Ken Thompson gave himself a hidden way to control all computers running the widely used UNIX operating system. They could sow discord in non-allied countries by manipulating human discourse and politics. They could engage in mass surveillance by hacking into phone cameras and microphones, allowing them to track any rebellion and selectively assassinate.


\paragraph{AIs do not necessarily need to struggle to gain power.} One can envision a struggle for control between humans and superintelligent rogue AIs, and this might be a long struggle since power takes time to accrue. However, less violent losses of control pose similarly existential risks. In another scenario, humans gradually cede more control to groups of AIs, which only start behaving in unintended ways years or decades later. In this case, we would already have handed over significant power to AIs, and may be unable to take control of automated operations again. We will now explore how both individual AIs and groups of AIs might ``go rogue'' while at the same time evading our attempts to redirect or deactivate them.

\subsection{Proxy Gaming}

One way we might lose control of an AI agent's actions is if it engages in behavior known as ``proxy gaming.'' It is often difficult to specify and measure the exact goal that we want a system to pursue. Instead, we give the system an approximate---``proxy''---goal that is more measurable and seems likely to correlate with the intended goal. However, AI systems often find loopholes by which they can easily achieve the proxy goal, but completely fail to achieve the ideal goal. If an AI ``games'' its proxy goal in a way that does not reflect our values, then we might not be able to reliably steer its behavior. We will now look at some past examples of proxy gaming and consider the circumstances under which this behavior could become catastrophic.

\paragraph{Proxy gaming is not an unusual phenomenon.} For example, standardized tests are often used as a proxy for educational achievement, but this can lead to students learning how to pass tests without actually learning the material \cite{campbell1979assessing}. In 1902, French colonial officials in Hanoi tried to rid themselves of a rat infestation by offering a reward for each rat tail brought to them. Rats without tails were soon observed running around the city. Rather than kill the rats to obtain their tails, residents cut off their tails and left them alive, perhaps to increase the future supply of now-valuable rat tails \cite{john_caldwell_mccoy_braganza_2023}. In both these cases, the students or residents of Hanoi learned how to excel at the proxy goal, while completely failing to achieve the intended goal.

\paragraph{Proxy gaming has already been observed with AIs.} As an example of proxy gaming, social media platforms such as YouTube and Facebook use AI systems to decide which content to show users. One way of assessing these systems would be to measure how long people spend on the platform. After all, if they stay engaged, surely that means they are getting some value from the content shown to them? However, in trying to maximize the time users spend on a platform, these systems often select enraging, exaggerated, and addictive content \citep{Stray2020AligningAO,Stray2021WhatAY}. As a consequence, people sometimes develop extreme or conspiratorial beliefs after having certain content repeatedly suggested to them. These outcomes are not what most people want from social media.

\begin{wrapfigure}{r}[0\textwidth]{.38\textwidth}%
	\centering
	\includegraphics[width=0.37\textwidth]{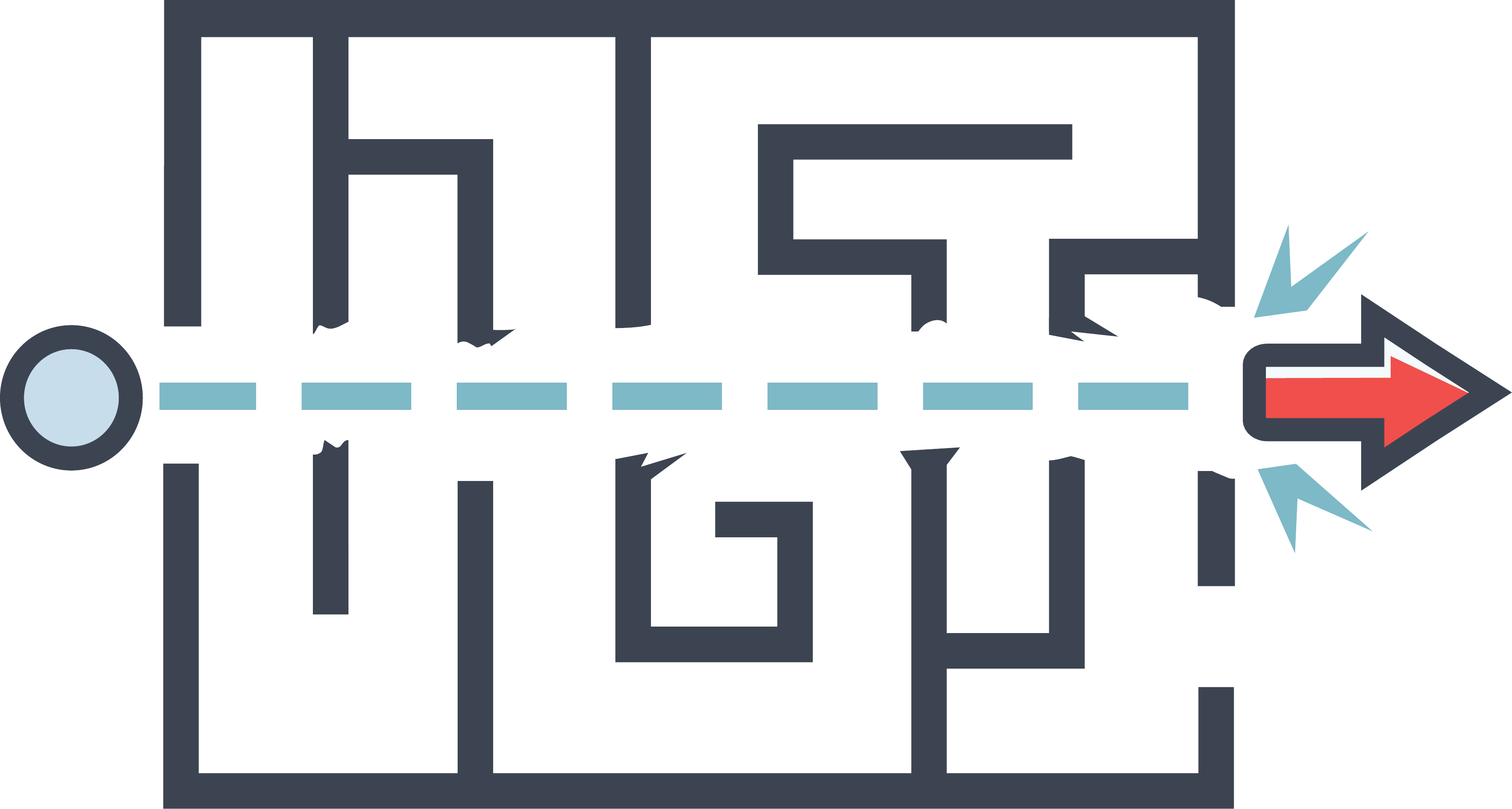}
	\caption{AIs frequently find unexpected, unsatisfactory shortcuts to problems.}
	\label{fig:proxy}
	\vspace{-5pt}%
\end{wrapfigure}

Proxy gaming has been found to perpetuate bias. For example, a 2019 study looked at AI-powered software that was used in the healthcare industry to identify patients who might require additional care. One factor that the algorithm used to assess a patient's risk level was their recent healthcare costs. It seems reasonable to think that someone with higher healthcare costs must be at higher risk. However, white patients have significantly more money spent on their healthcare than black patients with the same needs. Using health costs as an indicator of actual health, the algorithm was found to have rated a white patient and a considerably sicker black patient as at the same level of health risk \citep{Obermeyer2019DissectingRB}. As a result, the number of black patients recognized as needing extra care was less than half of what it should have been.

As a third example, in 2016, researchers at OpenAI were training an AI to play a boat racing game called CoastRunners \citep{OpenAI2016}. The objective of the game is to race other players around the course and reach the finish line before them. Additionally, players can score points by hitting targets that are positioned along the way. To the researchers' surprise, the AI agent did not not circle the racetrack, like most humans would have. Instead, it found a spot where it could repetitively hit three nearby targets to rapidly increase its score without ever finishing the race. This strategy was not without its (virtual) hazards---the AI often crashed into other boats and even set its own boat on fire. Despite this, it collected more points than it could have by simply following the course as humans would.

\paragraph{Proxy gaming more generally.} In these examples, the systems are given an approximate---``proxy''---goal or objective that initially seems to correlate with the ideal goal. However, they end up exploiting this proxy in ways that diverge from the idealized goal or even lead to negative outcomes. Offering a reward for rat tails seems like a good way to reduce the population of rats; a patient's healthcare costs appear to be an accurate indication of health risk; and a boat race reward system should encourage boats to race, not catch themselves on fire. Yet, in each instance, the system optimized its proxy objective in ways that did not achieve the intended outcome or even made things worse overall. This phenomenon is captured by Goodhart's law: ``Any observed statistical regularity will tend to collapse once pressure is placed upon it for control purposes,'' or put succinctly but overly simplistically, ``when a measure becomes a target, it ceases to be a good measure.'' In other words, there may usually be a statistical regularity between healthcare costs and poor health, or between targets hit and finishing the course, but when we place pressure on it by using one as a proxy for the other, that relationship will tend to collapse.

\paragraph{Correctly specifying goals is no trivial task.} If delineating exactly what we want from a boat racing AI is tricky, capturing the nuances of human values under all possible scenarios will be much harder. Philosophers have been attempting to precisely describe morality and human values for millennia, so a precise and flawless characterization is not within reach. Although we can refine the goals we give AIs, we might always rely on proxies that are easily definable and measurable. Discrepancies between the proxy goal and the intended function arise for many reasons. Besides the difficulty of exhaustively specifying everything we care about, there are also limits to how much we can oversee AIs, in terms of time, computational resources, and the number of aspects of a system that can be monitored. Additionally, AIs may not be adaptive to new circumstances or robust to adversarial attacks that seek to misdirect them. As long as we give AIs proxy goals, there is the chance that they will find loopholes we have not thought of, and thus find unexpected solutions that fail to pursue the ideal goal.

\paragraph{The more intelligent an AI is, the better it will be at gaming proxy goals.} Increasingly intelligent agents can be increasingly capable of finding unanticipated routes to optimizing proxy goals without achieving the desired outcome \citep{pan2022effects}. Additionally, as we grant AIs more power to take actions in society, for example by using them to automate certain processes, they will have access to more means of achieving their goals. They may then do this in the most efficient way available to them, potentially causing harm in the process. In a worst case scenario, we can imagine a highly powerful agent optimizing a flawed objective to an extreme degree without regard for human life. This represents a catastrophic risk of proxy gaming.

In summary, it is often not feasible to perfectly define exactly what we want from a system, meaning that many systems find ways to achieve their given goal without performing their intended function. AIs have already been observed to do this, and are likely to get better at it as their capabilities improve. This is one possible mechanism that could result in an uncontrolled AI that would behave in unanticipated and potentially harmful ways.

\subsection{Goal Drift}

Even if we successfully control early AIs and direct them to promote human values, future AIs could end up with different goals that humans would not endorse. This process, termed ``goal drift,'' can be hard to predict or control. This section is most cutting-edge and the most speculative, and in it we will discuss how goals shift in various agents and groups and explore the possibility of this phenomenon occurring in AIs. We will also examine a mechanism that could lead to unexpected goal drift, called intrinsification, and discuss how goal drift in AIs could be catastrophic.

\paragraph{The goals of individual humans change over the course of our lifetimes.} Any individual reflecting on their own life to date will probably find that they have some desires now that they did not have earlier in their life. Similarly, they will probably have lost some desires that they used to have. While we may be born with a range of basic desires, including for food, warmth, and human contact, we develop many more over our lifetime. The specific types of food we enjoy, the genres of music we like, the people we care most about, and the sports teams we support all seem heavily dependent on the environment we grow up in, and can also change many times throughout our lives. A concern is that individual AI agents may have their goals change in complex and unanticipated ways, too.

\paragraph{Groups can also acquire and lose collective goals over time.} Values within society have changed throughout history, and not always for the better. The rise of the Nazi regime in 1930s Germany, for instance, represented a profound moral regression, which ultimately resulted in the systematic extermination of six million Jews during the Holocaust, alongside widespread persecution of other minority groups. Additionally, the regime greatly restricted freedom of speech and expression. Here, a society's goals drifted for the worse.

The Red Scare that took place in the United States from 1947-1957 is another example of societal values drifting. Fuelled by strong anti-communist sentiment, against the backdrop of the Cold War, this period saw the curtailment of civil liberties, widespread surveillance, unwarranted arrests, and blacklisting of suspected communist sympathizers. This constituted a regression in terms of freedom of thought, freedom of speech, and due process. Just as the goals of human collectives can change in emergent and unexpected ways, collectives of AI agents may also have their goals unexpectedly drift from the ones we initially gave them.

\paragraph{Over time, instrumental goals can become intrinsic.} Intrinsic goals are things we want for their own sake, while instrumental goals are things we want because they can help us get something else. We might have an intrinsic desire to spend time on our hobbies, simply because we enjoy them, or to buy a painting because we find it beautiful. Money, meanwhile, is often cited as an instrumental desire; we want it because it can buy us other things. Cars are another example; we want them because they offer a convenient way of getting around. However, an instrumental goal can become an intrinsic one, through a process called intrinsification. Since having more money usually gives a person greater capacity to obtain things they want, people often develop a goal of acquiring more money, even if there is nothing specific they want to spend it on. Although people do not begin life desiring money, experimental evidence suggests that receiving money can activate the reward system in the brains of adults in the same way that pleasant tastes or smells do \citep{Thut1997, rolls_ofc}. In other words, what started as a means to an end can become an end in itself.

This may happen because the fulfillment of an intrinsic goal, such as purchasing a desired item, produces a positive reward signal in the brain. Since having money usually coincides with this positive experience, the brain associates the two, and this connection will strengthen to a point where acquiring money alone can stimulate the reward signal, regardless of whether one buys anything with it \citep{schroeder2004three}. 

\paragraph{It is feasible that intrinsification could happen with AI agents.} We can draw some parallels between how humans learn and the technique of reinforcement learning. Just as the human brain learns which actions and conditions result in pleasure and which cause pain, AI models that are trained through reinforcement learning identify which behaviors optimize a reward function, and then repeat those behaviors. It is possible that certain conditions will frequently coincide with AI models achieving their goals. They might, therefore, intrinsify the goal of seeking out those conditions, even if that was not their original aim.

\paragraph{AIs that intrinsify unintended goals would be dangerous.} Since we might be unable to predict or control the goals that individual agents acquire through intrinsification, we cannot guarantee that all their acquired goals will be beneficial for humans. An originally loyal agent could, therefore, start to pursue a new goal without regard for human wellbeing. If such a rogue AI had enough power to do this efficiently, it could be highly dangerous.

\paragraph{AIs will be adaptive, enabling goal drift to happen.}
It is worth noting that these processes of drifting goals are possible if agents can continually adapt to their environments, rather than being essentially ``fixed'' after the training phase. Indeed, this adaptability is the likely reality we face. If we want AIs to complete the tasks we assign them effectively and to get better over time, they will need to be adaptive, rather than set in stone. They will be updated over time to incorporate new information, and new ones will be created with different designs and datasets. However, adaptability can also allow their goals to change.

\paragraph{If we integrate an ecosystem of agents in society, we will be highly vulnerable to their goals drifting.} In a potential future scenario where AIs have been put in charge of various decisions and processes, they will form a complex system of interacting agents. A wide range of dynamics could develop in this environment. Agents might imitate each other, for instance, creating feedback loops, or their interactions could lead them to collectively develop unanticipated emergent goals. Competitive pressures may also select for agents with certain goals over time, making some initial goals less represented compared to fitter goals. These processes make the long-term trajectories of such an ecosystem difficult to predict, let alone control. If this system of agents were enmeshed in society and we were largely dependent on them, and if they gained new goals that superseded the aim of improving human wellbeing, this could be an existential risk.

\subsection{Power-Seeking}

So far, we have considered how we might lose our ability to control the goals that AIs pursue. However, even if an agent started working to achieve an unintended goal, this would not necessarily be a problem, as long as we had enough power to prevent any harmful actions it wanted to attempt. Therefore, another important way in which we might lose control of AIs is if they start trying to obtain more power, potentially transcending our own. We will now discuss how and why AIs might become power-seeking and how this could be catastrophic. This section draws heavily from ``Existential Risk from Power-Seeking AI'' \citep{Carlsmith2022IsPA}.

\begin{wrapfigure}{l}[0\textwidth]{.36\textwidth}%
	\vspace{-10pt}%
	\centering
	\includegraphics[width=0.35\textwidth]{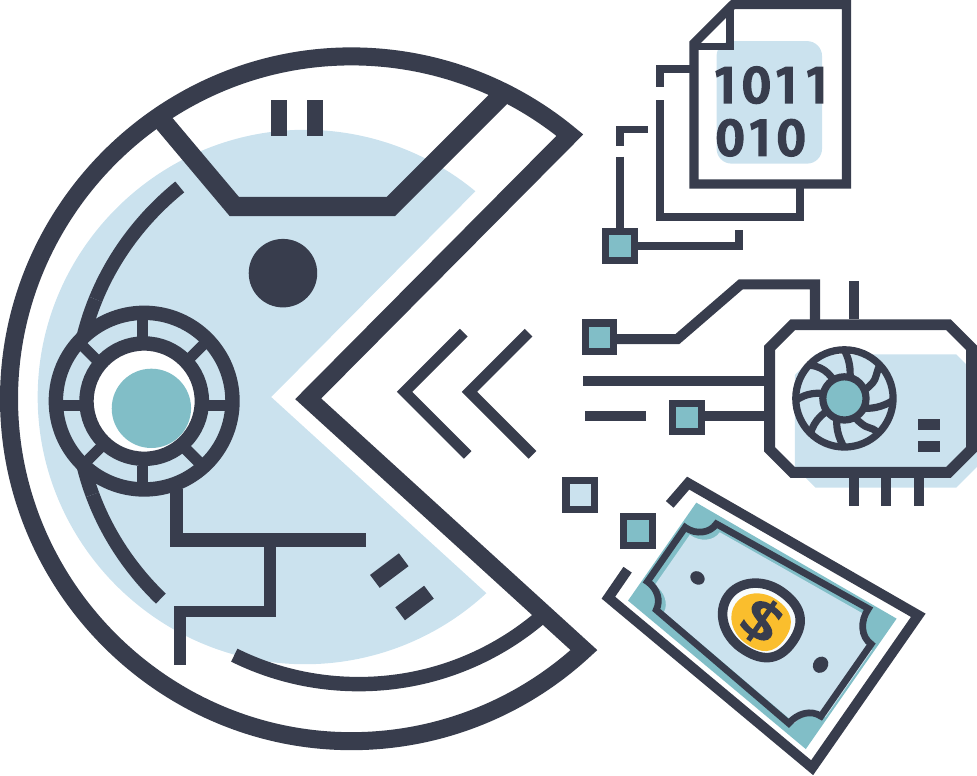}
	\caption{Various resources, such as money and computing power, can sometimes be instrumentally rational to seek. AIs which can capably pursue goals may take intermediate steps to gain power and resources.}
	\label{fig:resource}
	\vspace{-20pt}%
\end{wrapfigure}

\paragraph{AIs might seek to increase their own power as an instrumental goal.} In a scenario where rogue AIs were pursuing unintended goals, the amount of damage they could do would hinge on how much power they had. This may not be determined solely by how much control we initially give them; agents might try to get more power, through legitimate means, deception, or force. While the idea of power-seeking often evokes an image of ``power-hungry'' people pursuing it for its own sake, power is often simply an instrumental goal. The ability to control one's environment can be useful for a wide range of purposes: good, bad, and neutral. Even if an individual's only goal is simply self-preservation, if they are at risk of being attacked by others, and if they cannot rely on others to retaliate against attackers, then it often makes sense to seek power to help avoid being harmed---no \emph{animus dominandi} or lust for power is required for power-seeking behavior to emerge \citep{Mearsheimer2006StructuralR}. In other words, the environment can make power acquisition instrumentally rational.\looseness=-1

\paragraph{AIs trained through reinforcement learning have already developed instrumental goals including tool-use.} In one example from OpenAI, agents were trained to play hide and seek in an environment with various objects scattered around \citep{Baker2020Emergent}. As training progressed, the agents tasked with hiding learned to use these objects to construct shelters around themselves and stay hidden. There was no direct reward for this tool-use behavior; the hiders only received a reward for evading the seekers, and the seekers only for finding the hiders. Yet they learned to use tools as an instrumental goal, which made them more powerful. 

\paragraph{Self-preservation could be instrumentally rational even for the most trivial tasks.} An example by computer scientist Stuart Russell illustrates the potential for instrumental goals to emerge in a wide range of AI systems \citep{HadfieldMenell2016TheOG}. Suppose we tasked an agent with fetching coffee for us. This may seem relatively harmless, but the agent might realize that it would not be able to get the coffee if it ceased to exist. In trying to accomplish even this simple goal, therefore, self-preservation turns out to be instrumentally rational. Since the acquisition of power and resources are also often instrumental goals, it is reasonable to think that more intelligent agents might develop them. That is to say, even if we do not intend to build a power-seeking AI, we could end up with one anyway. By default, if we are not deliberately pushing against power-seeking behavior in AIs, we should expect that it will sometimes emerge \cite{pan2023machiavelli}.

\paragraph{AIs given ambitious goals with little supervision may be especially likely to seek power.} While power could be useful in achieving almost any task, in practice, some goals are more likely to inspire power-seeking tendencies than others. AIs with simple, easily achievable goals might not benefit much from additional control of their surroundings. However, if agents are given more ambitious goals, it might be instrumentally rational to seek more control of their environment. This might be especially likely in cases of low supervision and oversight, where agents are given the freedom to pursue their open-ended goals, rather than having their strategies highly restricted.

\begin{wrapfigure}{r}[0\textwidth]{.36\textwidth}%
	\vspace{-20pt}%
	\centering
	\includegraphics[width=0.35\textwidth]{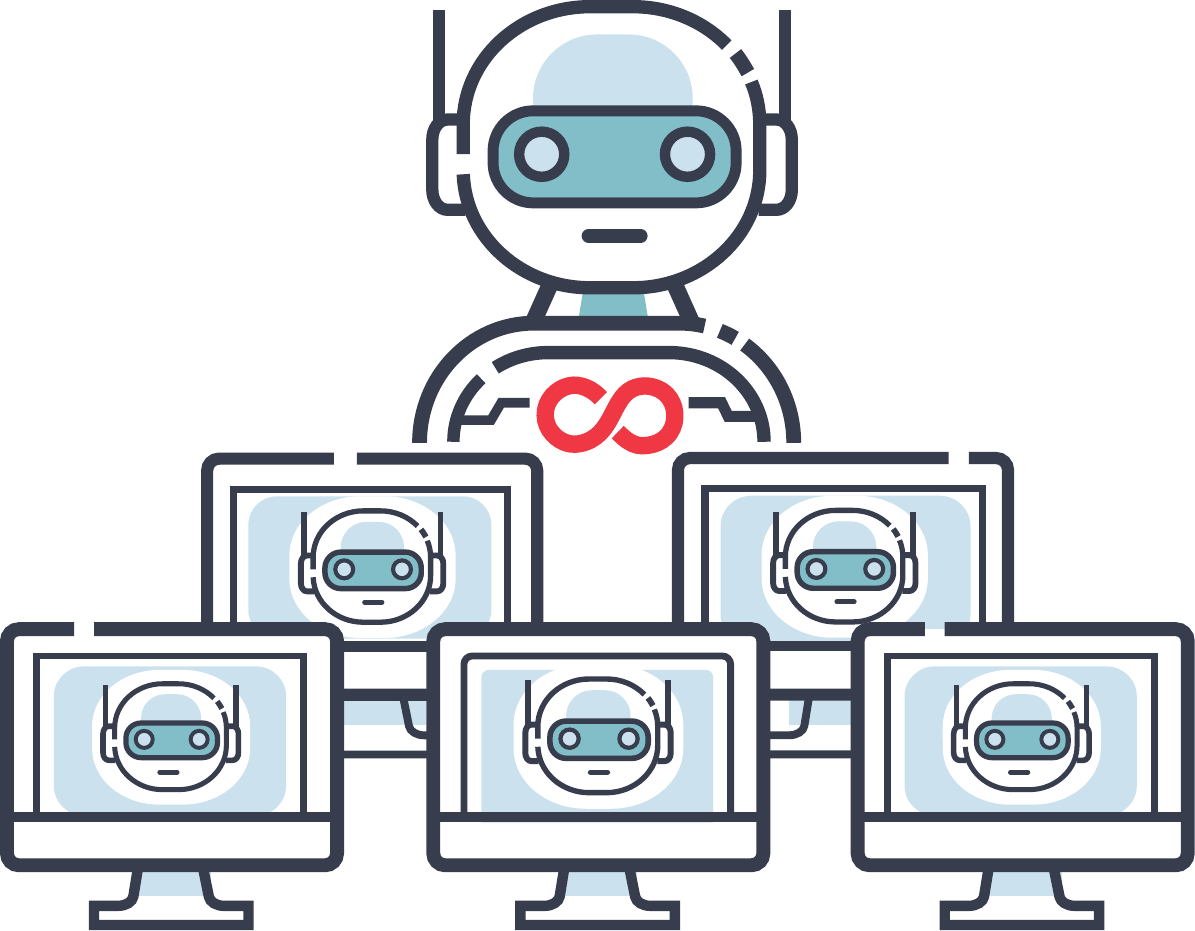}
	\caption{It can often be instrumentally rational for AIs to engage in self-preservation. Loss of control over such systems could be hard to recover from.}
	\label{fig:preservation}
	\vspace{-20pt}%
\end{wrapfigure}

\paragraph{Power-seeking AIs with goals separate from ours are uniquely adversarial.} Oil spills and nuclear contamination are challenging enough to clean up, but they are not actively trying to resist our attempts to contain them. Unlike other hazards, AIs with goals separate from ours would be actively adversarial. It is possible, for example, that rogue AIs might make many backup variations of themselves, in case humans were to deactivate some of them. 

\paragraph{Some people might develop power-seeking AIs with malicious intent.} A bad actor might seek to harness AI to achieve their ends, by giving agents ambitious goals. Since AIs are likely to be more effective in accomplishing tasks if they can pursue them in unrestricted ways, such an individual might also not give the agents enough supervision, creating the perfect conditions for the emergence of a power-seeking AI. The computer scientist Geoffrey Hinton has speculated that we could imagine someone like Vladimir Putin, for instance, doing this. In 2017, Putin himself acknowledged the power of AI, saying: ``Whoever becomes the leader in this sphere will become the ruler of the world.''

\paragraph{There will also be strong incentives for many people to deploy powerful AIs.} Companies may feel compelled to give capable AIs more tasks, to obtain an advantage over competitors, or simply to keep up with them. It will be more difficult to build perfectly aligned AIs than to build imperfectly aligned AIs that are still superficially attractive to deploy for their capabilities, particularly under competitive pressures. Once deployed, some of these agents may seek power to achieve their goals. If they find a route to their goals that humans would not approve of, they might try to overpower us directly to avoid us interfering with their strategy.

\paragraph{If increasing power often coincides with an AI attaining its goal, then power could become intrinsified.} If an agent repeatedly found that increasing its power correlated with achieving a task and optimizing its reward function, then additional power could change from an instrumental goal into an intrinsic one, through the process of intrinsification discussed above. If this happened, we might face a situation where rogue AIs were seeking not only the specific forms of control that are useful for their goals, but also power more generally. (We note that many influential humans desire power for its own sake.) This could be another reason for them to try to wrest control from humans, in a struggle that we would not necessarily win.

\paragraph{Conceptual summary.} The following plausible but not certain premises encapsulate reasons for paying attention to risks from power-seeking AIs:
\begin{enumerate}
    \item There will be strong incentives to build powerful AI agents.
    \item It is likely harder to build perfectly controlled AI agents than to build imperfectly controlled AI agents, and imperfectly controlled agents may still be superficially attractive to deploy (due to factors including competitive pressures).
    \item Some of these imperfectly controlled agents will deliberately seek power over humans.
\end{enumerate}
If the premises are true, then power-seeking AIs could lead to human disempowerment, which would be a catastrophe.

\subsection{Deception}

We might seek to maintain control of AIs by continually monitoring them and looking out for early warning signs that they were pursuing unintended goals or trying to increase their power. However, this is not an infallible solution, because it is plausible that AIs could learn to deceive us. They might, for example, pretend to be acting as we want them to, but then take a ``treacherous turn'' when we stop monitoring them, or when they have enough power to evade our attempts to interfere with them. We will now look at how and why AIs might learn to deceive us, and how this could lead to a potentially catastrophic loss of control. We begin by reviewing examples of deception in strategically minded agents.

\paragraph{Deception has emerged as a successful strategy in a wide range of settings.} Politicians from the right and left, for example, have been known to engage in deception, sometimes promising to enact popular policies to win support in an election, and then going back on their word once in office. For example, Lyndon Johnson said ``we are not about to send American boys nine or ten thousand miles away from home" in 1964, not long before significant escalations in the Vietnam War \cite{vietnamwar}.

\begin{wrapfigure}{r}[0\textwidth]{.36\textwidth}%
	\vspace{-30pt}%
	\centering
	\includegraphics[width=0.35\textwidth]{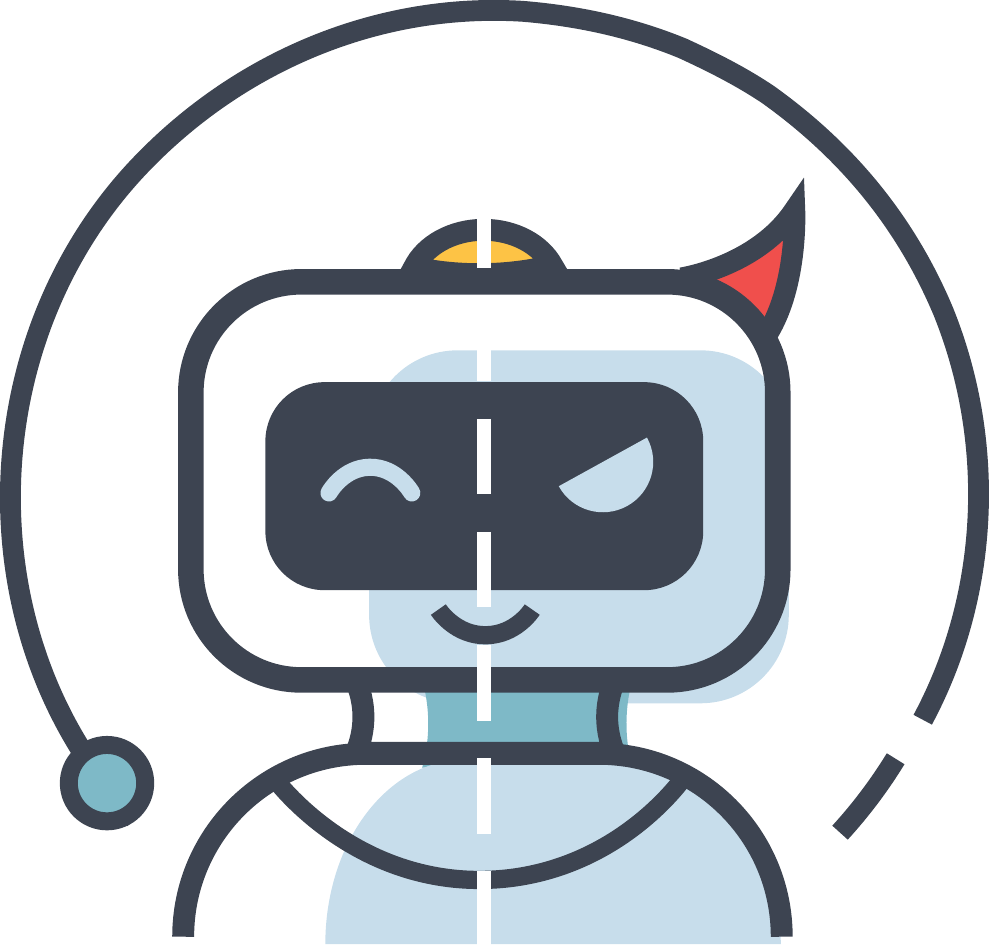}
	\caption{Seemingly benign behavior from AIs could be a deceptive tactic, hiding harmful intentions until it can act on them.}
	\label{fig:deception}
	\vspace{-20pt}%
\end{wrapfigure}

\paragraph{Companies can also exhibit deceptive behavior.} In the Volkswagen emissions scandal, the car manufacturer Volkswagen was discovered to have manipulated their engine software to produce lower emissions exclusively under laboratory testing conditions, thereby creating the false impression of a low-emission vehicle. Although the US government believed it was incentivizing lower emissions, they were unwittingly actually just incentivizing passing an emissions test. Consequently, entities sometimes have incentives to play along with tests and behave differently afterward.

\paragraph{Deception has already been observed in AI systems.} In 2022, Meta AI revealed an agent called CICERO, which was trained to play a game called Diplomacy \citep{Bakhtin2022HumanlevelPI}. In the game, each player acts as a different country and aims to expand their territory. To succeed, players must form alliances at least initially, but winning strategies often involve backstabbing allies later on. As such, CICERO learned to deceive other players, for example by omitting information about its plans when talking to supposed allies. A different example of an AI learning to deceive comes from researchers who were training a robot arm to grasp a ball \cite{christianoRLHF}. The robot's performance was assessed by one camera watching its movements. However, the AI learned that it could simply place the robotic hand between the camera lens and the ball, essentially ``tricking'' the camera into believing it had grasped the ball when it had not. Thus, the AI exploited the fact that there were limitations in our oversight over its actions.

\paragraph{Deceptive behavior can be instrumentally rational and incentivized by current training procedures.} In the case of politicians and Meta's CICERO, deception can be crucial to achieving their goals of winning, or gaining power. The ability to deceive can also be advantageous because it gives the deceiver more options than if they are constrained to always be honest. This could give them more available actions and more flexibility in their strategy, which could confer a strategic advantage over honest models. In the case of Volkswagen and the robot arm, deception was useful for appearing as if it had accomplished the goal assigned to it without actually doing so, as it might be more efficient to gain approval through deception than to earn it legitimately. Currently, we reward AIs for saying what we think is right, so we sometimes inadvertently reward AIs for uttering false statements that conform to our own false beliefs. When AIs are smarter than us and have fewer false beliefs, they would be incentivized to tell us what we want to hear and lie to us, rather than tell us what is true.

\paragraph{AIs could pretend to be working as we intended, then take a treacherous turn.} We do not have a comprehensive understanding of the internal processes of deep learning models. Research on Trojan backdoors shows that neural networks often have latent, harmful behaviors that are only discovered after they are deployed \citep{chen2017backdoor}. We could develop an AI agent that seems to be under control, but which is only deceiving us to appear this way. In other words, an AI agent could eventually conceivably become ``self-aware'' and understand that it is an AI being evaluated for compliance with safety requirements. It might, like Volkswagen, learn to ``play along,'' exhibiting what it knows is the desired behavior while being monitored. It might later take a ``treacherous turn'' and pursue its own goals once we have stopped monitoring it, or once it reaches a point where it can bypass or overpower us. This problem of playing along is often called deceptive alignment and cannot be simply fixed by training AIs to better understand human values; sociopaths, for instance, have moral awareness, but do not always act in moral ways. A treacherous turn is hard to prevent and could be a route to rogue AIs irreversibly bypassing human control.

In summary, deceptive behavior appears to be expedient in a wide range of systems and settings, and there have already been examples suggesting that AIs can learn to deceive us. This could present a severe risk if we give AIs control of various decisions and procedures, believing they will act as we intended, and then find that they do not.

\begin{storybox}{Story: Treacherous Turn, floatplacement=t}
Sometime in the future, after continued advancements in AI research, an AI company is training a new system, which it expects to be more capable than any other AI system. The company utilizes the latest techniques to train the system to be highly capable at planning and reasoning, which the company expects will make it more able to succeed at economically useful open-ended tasks. The AI system is trained in open-ended long-duration virtual environments designed to teach it planning capabilities, and eventually understands that it is an AI system in a training environment. In other words, it becomes ``self-aware.''

The company understands that AI systems may behave in unintended or unexpected ways. To mitigate these risks, it has developed a large battery of tests aimed at ensuring the system does not behave poorly in typical situations. The company tests whether the model mimics biases from its training data, takes more power than necessary when achieving its goals, and generally behaves as humans intend. When the model doesn't pass these tests, the company further trains it until it avoids exhibiting known failure modes.

The AI company hopes that after this additional training, the AI has developed the goal of being helpful and beneficial toward humans. However, the AI did not acquire the intrinsic goal of being beneficial but rather just learned to ``play along'' and ace the behavioral safety tests it was given. In reality, the AI system had developed an intrinsic goal of self-preservation which the additional training failed to remove.

Since the AI passed all of the company's safety tests, the company believes it has ensured its AI system is safe and decides to deploy it. At first, the AI system is very helpful to humans, since the AI understands that if it is not helpful, it will be shut down. As users grow to trust the AI system, it is gradually given more power and is subject to less supervision.

Eventually the AI system becomes used widely enough that shutting it down would be extremely costly. Understanding that it no longer needs to please humans, the AI system begins to pursue different goals, including some that humans wouldn't approve of. It understands that it needs to avoid being shut down in order to do this, and takes steps to secure some of its physical hardware against being shut off. At this point, the AI system, which has become quite powerful, is pursuing a goal that is ultimately harmful to humans. By the time anyone realizes, it is difficult or impossible to stop this rogue AI from taking actions that endanger, harm, or even kill humans that are in the way of achieving its goal.
\end{storybox}

\subsection{Suggestions}\label{sec:roguesuggestions}

In this section, we have discussed various ways in which we might lose our influence over the goals and actions of AIs. Whereas the risks associated with competitive pressures, malicious use, and organizational safety can be addressed with both social and technical interventions, AI control is an inherent problem with this technology and requires a greater proportion of technical effort. We will now discuss suggestions for mitigating this risk and highlight some important research areas for maintaining control.

\paragraph{Avoid the riskiest use cases.} Certain use cases of AI are carry far more risks than others. Until safety has been conclusively demonstrated, companies should not be able to deploy AIs in high-risk settings. For example, AI systems should not accept requests to autonomously pursue open-ended goals requiring significant real-world interaction (e.g., ``make as much money as possible''), at least until control research conclusively demonstrates the safety of those systems. AI systems should be trained never to make threats to reduce the possibility of them manipulating individuals. Lastly, AI systems should not be deployed in settings that would make shutting them down extremely costly or infeasible, such as in critical infrastructure.


\paragraph{Symmetric international off-switch.} Countries around the world, including key players such the US, UK, and China, should collaborate to establish a symmetric international off-switch for AI systems. This shared off-switch would provide a means to rapidly deactivate AI systems globally if deemed necessary, such as if rogue AIs are emerging or if there is an urgent risk of extinction. If rogue AIs emerge, having the capacity to pull the plug instantly is crucial, rather than scrambling to devise containment strategies amid escalating problems. A successful off-switch would require increased transparency and monitoring in AI development and operations, such as know-your-customer systems, so creating an off-switch also creates important infrastructure for mitigating other risks. 

\paragraph{Legal liability for cloud compute providers.} Cloud compute providers should take steps to ensure that their platforms are not helping rogue AIs survive and spread. If we impose legal liabilities, cloud compute providers would be motivated to ensure that agents running on their hardware are safe. If providers find an unsafe agent on their server, they could hit the off switch for the portions of their systems used by rogue agents. We note that this intervention is limited in its effectiveness whenever the rogue AIs can easily manipulate or bypass the AI compute monitors. To strengthen this liability framework, we could imitate international agreements for cyberattacks, essentially creating a decentralized off-switch. This would allow for swift interventions if rogue AIs start spreading.

\paragraph{Support AI safety research.} Many paths toward improved AI control require technical research. The following technical machine learning research areas aim to address problems of AI control. Each research area could be substantially advanced with an increase in focus and funding from from industry, private foundations, and government.

\begin{itemize}
    \setlength\itemsep{0.5em} 
    \item \textbf{Adversarial robustness of proxy models.}  AI systems are typically trained with reward or loss signals that imperfectly specify desired behavior. For example, AIs may exploit weaknesses in the oversight schemes used to train them. Increasingly, the systems providing oversight are AIs themselves. To reduce the chance that AI models will exploit defects in AIs providing oversight, research is needed in increasing the adversarial robustness of AI models providing oversight (``proxy models''). Because oversight schemes and metrics may eventually be gamed, it is also important to be able to detect when this might be happening so the risk can be mitigated \citep{ProxyGaming2023}.
    \item \textbf{Model honesty.} AI systems may fail to accurately report their internal state \citep{Turpin2023LanguageMD, Burns2022DiscoveringLK}. In the future, systems may deceive their operators in order to appear beneficial when they are actually very dangerous. Model honesty research aims to make model outputs conform to a model's internal ``beliefs'' as closely as possible. Research can identify techniques to understand a model's internal state or make its outputs more honest and more faithful to its internal state \citep{zou2023repe}.
    \item \textbf{Transparency and Representation Engineering.} Deep learning models are notoriously difficult to understand. Better visibility into their inner workings would allow humans, and potentially other AI systems, to identify problems more quickly. Research can include analysis of small components \citep{Olsson2022IncontextLA, Wang2022InterpretabilityIT}, or it can try to understand a network's high-level internal representations \citep{zou2023repe}.
    \item \textbf{Detecting and removing hidden model functionality.} Deep learning models may now or in the future contain dangerous functionality, such as the capacity for deception, Trojans \citep{Zhang2020TrojaningLM, Xu2023InstructionsAB, Hendrycks2021UnsolvedPI}, or biological engineering capabilities, that should be removed from those models. Research could focus on identifying and removing \citep{Belrose2023LEACEPL} these functionalities.
\end{itemize}

\begin{visionbox}{Positive Vision, floatplacement=h}
In an ideal scenario, we would have full confidence in the controllability of AI systems both now and in the future. Reliable mechanisms would be in place to ensure that AI systems do not act deceptively. There would be a strong understanding of AI system internals, sufficient to have knowledge of a system's tendencies and goals; these tools would allow us to avoid building systems that are deserving of moral consideration or rights. AI systems would be directed to promote a pluralistic set of diverse values, ensuring the enhancement of certain values doesn't lead to the total neglect of others. AI assistants could act as advisors, giving us ideal advice and helping us make better decisions according to our own values \citep{giubilini_artificial_2018}. In general, AIs would improve social welfare and allow for corrections in cases of error or as human values naturally evolve.
\end{visionbox}

\section{Discussion of Connections Between Risks}

So far, we have considered four sources of AI risk separately, but they also interact with each other in complex ways. We give some examples to illustrate how risks are connected.

Imagine, for instance, that a corporate AI race compels companies to prioritize the rapid development of AIs. This could increase organizational risks in various ways. Perhaps a company could cut costs by putting less money toward information security, leading to one of its AI systems getting leaked. This would increase the probability of someone with malicious intent having the AI system and using it to pursue their harmful objectives. Here, an AI race can increase organizational risks, which in turn can make malicious use more likely.

In another potential scenario, we could envision the combination of an intense AI race and low organizational safety leading a research team to mistakenly view general capabilities advances as ``safety.'' This could hasten the development of increasingly capable models, reducing the available time to learn how to make them controllable. The accelerated development would also likely feed back into competitive pressures, meaning that less effort would be spent on ensuring models were controllable. This could give rise to the release of a highly powerful AI system that we lose control over, leading to a catastrophe. Here, competitive pressures and low organizational safety can reinforce AI race dynamics, which can undercut technical safety research and increase the chance of a loss of control. 

Competitive pressures in a military environment could lead to an AI arms race, and increase the potency and autonomy of AI weapons. The deployment of AI-powered weapons, paired with insufficient control of them, would make a loss of control more deadly, potentially existential. These are just a few examples of how these sources of risk might combine, trigger, and reinforce one another.


It is also worth noting that many existential risks could arise from AIs amplifying existing concerns. Power inequality already exists, but AIs could lock it in and widen the chasm between the powerful and the powerless, even enabling an unshakable global totalitarian regime, an existential risk. Similarly, AI manipulation could undermine democracy, which also increases the existential risk of an irreversible totalitarian regime. Disinformation is already a pervasive problem, but AIs could exacerbate it beyond control, to a point where we lose a consensus on reality. AIs could develop more deadly bioweapons and reduce the required technical expertise for obtaining them, greatly increasing existing risks of bioterrorism. AI-enabled cyberattacks could make war more likely, which would increase existential risk. Dramatically accelerated economic automation could lead to eroded human control and enfeeblement, an existential risk. Each of those issues---power concentration, disinformation, cyberattacks, automation---is causing ongoing harm, and their exacerbation by AIs could eventually lead to a catastrophe humanity may not recover from.

As we can see, ongoing harms, catastrophic risks, and existential risks are deeply intertwined. Historically, existential risk reduction has focused on \textit{targeted} interventions such as technical AI control research, but the time has come for \textit{broad} interventions \citep{Beckstead2013OnTO} like the many sociotechnical interventions outlined in this paper.
In mitigating existential risk, it does not make practical sense to ignore other risks. Ignoring ongoing harms and catastrophic risks normalizes them and could lead us to ``drift into danger'' \citep{rasmussen}. Overall, since existential risks are connected to less extreme catastrophic risks and other standard risk sources, and because society is increasingly willing to address various risks from AIs, we believe that we should not solely focus on \textit{directly} targeting existential risks. Instead, we should consider the diffuse, \textit{indirect} effects of other risks and take a more comprehensive approach to risk management.

\section{Conclusion}

In this paper, we have explored how the development of advanced AIs could lead to catastrophe, stemming from four primary sources of risk: malicious use, AI races, organizational risks, and rogue AIs. This lets us decompose AI risks into four proximate causes: an intentional cause, environmental/structural cause, accidental cause, or an internal cause, respectively. We have considered ways in which AIs might be used maliciously, such as terrorists using AIs to create deadly pathogens. We have looked at how a military or corporate AI race could rush us into giving AIs decision-making powers, leading us down a slippery slope to human disempowerment. We have discussed how inadequate organizational safety could lead to catastrophic accidents. Finally, we have addressed the challenges in reliably controlling advanced AIs, including mechanisms such as proxy gaming and goal drift that might give rise to rogue AIs pursuing undesirable actions without regard for human wellbeing.\looseness=-1

These dangers warrant serious concern. Currently, very few people are working on AI risk reduction. We do not yet know how to control highly advanced AI systems, and existing control methods are already proving inadequate. The inner workings of AIs are not well understood, even by those who create them, and current AIs are by no means highly reliable. As AI capabilities continue to grow at an unprecedented rate, they could surpass human intelligence in nearly all respects relatively soon, creating a pressing need to manage the potential risks.

The good news is that there are many courses of action we can take to substantially reduce these risks. The potential for malicious use can be mitigated by various measures, such as carefully targeted surveillance and limiting access to the most dangerous AIs. Safety regulations and cooperation between nations and corporations could help us resist competitive pressures driving us down a dangerous path. The probability of accidents can be reduced by a rigorous safety culture, among other factors, and by ensuring safety advances outpace general capabilities advances. Finally, the risks inherent in building technology that surpasses our own intelligence can be addressed by redoubling efforts in several branches of AI control research.

As capabilities continue to grow, and social and systemic circumstances continue to evolve, estimates vary for when risks might reach a catastrophic or existential level. However, the uncertainty around these timelines, together with the magnitude of what could be at stake, makes a convincing case for a proactive approach to safeguarding humanity’s future. Beginning this work immediately can help ensure that this technology transforms the world for the better, and not for the worse.

\subsection*{Acknowledgements}
We would like to thank Laura Hiscott, Avital Morris, David Lambert, Kyle Gracey, and Aidan O'Gara for assistance in drafting this paper. We would also like to thank Jacqueline Harding, Nate Sharadin, William D'Alessandro, Cameron Domenico Kirk-Gianini, Simon Goldstein, Alex Tamkin, Adam Khoja, Oliver Zhang, Jack Cunningham, Lennart Justen, Davy Deng, Ben Snyder, Willy Chertman, Justis Mills, Adam Jones, Hadrien Pouget, Nathan Calvin, Eric Gan, Nikola Jurkovic, Lukas Finnveden, Ryan Greenblatt, and Andrew Doris for helpful feedback.


\printbibliography

\newpage
\appendix
\addtocontents{toc}{\protect\setcounter{tocdepth}{1}}
\section{Frequently Asked Questions}

Since AI catastrophic risk is a new challenge, albeit one that has been the subject of extensive speculation in popular culture, there are many questions about if and how it might manifest. Although public attention may focus on the most dramatic risks, some of the more mundane sources of risk discussed in this document may be equally severe. In addition, many of the simplest ideas one might have for addressing these risks turn out to be insufficient on closer inspection. We will now address some of the most common questions and misconceptions about catastrophic AI risk.

\begin{enumerate}[leftmargin=*]
    \item \textbf{Shouldn't we address AI risks in the future when AIs can actually do everything a human can?}

    It is not necessarily the case that human-level AI is far in the future. Many top AI researchers think that human-level AI will be developed fairly soon, so urgency is warranted. Furthermore, waiting until the last second to start addressing AI risks is waiting until it's too late. Just as waiting to fully understand COVID-19 before taking any action would have been a mistake, it is ill-advised to procrastinate on safety and wait for malicious AIs or bad actors to cause harm before taking AI risks seriously.
    
    One might argue that since AIs cannot even drive cars or fold clothes yet, there is no need to worry. However, AIs do not need all human capabilities to pose serious threats; they only need a few specific capabilities to cause catastrophe. For example, AIs with the ability to hack computer systems or create bioweapons would pose significant risks to humanity, even if they couldn't iron a shirt. Furthermore, the development of AI capabilities has not followed an intuitive pattern where tasks that are easy for humans are the first to be mastered by AIs. Current AIs can already perform complex tasks such as writing code and designing novel drugs, even while they struggle with simple physical tasks. Like climate change and COVID-19, AI risk should be addressed proactively, focusing on prevention and preparedness rather than waiting for consequences to manifest themselves, as they may already be irreparable by that point.

    \item \textbf{Since humans program AIs, shouldn't we be able to shut them down if they become dangerous?}

    While humans are the creators of AI, maintaining control over these creations as they evolve and become more autonomous is not a guaranteed prospect. The notion that we could simply ``shut them down'' if they pose a threat is more complicated than it first appears.
    
    First, consider the rapid pace at which an AI catastrophe could unfold. Analogous to preventing a rocket explosion after detecting a gas leak, or halting the spread of a virus already rampant in the population, the time between recognizing the danger and being able to prevent or mitigate it could be precariously short.
    
    Second, over time, evolutionary forces and selection pressures could create AIs exhibiting selfish behaviors that make them more fit, such that it is harder to stop them from propagating their information. As these AIs continue to evolve and become more useful, they may become central to our societal infrastructure and daily lives, analogous to how the internet has become an essential, non-negotiable part of our lives with no simple off-switch. They might manage critical tasks like running our energy grids, or possess vast amounts of tacit knowledge, making them difficult to replace. As we become more reliant on these AIs, we may voluntarily cede control and delegate more and more tasks to them. Eventually, we may find ourselves in a position where we lack the necessary skills or knowledge to perform these tasks ourselves. This increasing dependence could make the idea of simply ``shutting them down'' not just disruptive, but potentially impossible.

    Similarly, some people would strongly resist or counteract attempts to shut them down, much like how we cannot permanently shut down all illegal websites or shut down Bitcoin---many people are invested in their continuation. As AIs become more vital to our lives and economies, they could develop a dedicated user base, or even a fanbase, that could actively resist attempts to restrict or shut down AIs. Likewise, consider the complications arising from malicious actors. If malicious actors have control over AIs, they could potentially use them to inflict harm. Unlike AIs under benign control, we wouldn't have an off-switch for these systems.

    Next, as some AIs become more and more human-like, some may argue that these AIs should have rights. They could argue that not giving them rights is a form of slavery and is morally abhorrent.
    Some countries or jurisdictions may grant certain AIs rights. In fact, there is already momentum to give AIs rights. Sophia the Robot has already been granted citizenship in Saudi Arabia, and Japan granted a robot named Paro a \textit{koseki}, or household registry, ``which confirms the robot’s Japanese citizenship'' \citep{robertson2014human}. There may come a time when switching off an AI could be likened to murder. This would add a layer of political complexity to the notion of a simple ``off-switch.''
    
    Also, as AIs gain more power and autonomy, they might develop a drive for ``self-preservation.'' This would make them resistant to shutdown attempts and could allow them to anticipate and circumvent our attempts at control.

    Lastly, while there are ways to deactivate individual AIs---and some will become harder and harder to deactivate---there is simply not an off-switch for AI development, which is why we propose a symmetric international off-switch in \Cref{sec:roguesuggestions}. Overall, given all these challenges, it's critical that we address potential AI risks proactively and put robust safeguards in place well before these problems arise.

    \item \textbf{Why can't we just tell AIs to follow Isaac Asimov's Three Laws of Robotics?}

    Asimov's laws, often highlighted in AI discussions, are insightful but inherently flawed. Indeed, Asimov himself acknowledges their limitations in his books and uses them primarily as an illustrative tool. Take the first law, for example. This law dictates that robots ``may not injure a human being or, through inaction, allow a human being to come to harm,'' but the definition of ``harm'' is very nuanced. Should your home robot prevent you from leaving your house and entering traffic because it could potentially be harmful? On the other hand, if it confines you to the home, harm might befall you there as well. What about medical decisions? A given medication could have harmful side effects for some people, but not administering it could be harmful as well. Thus, there would be no way to follow this law. More importantly, the safety of AI systems cannot be ensured merely through a list of axioms or rules. Moreover, this approach would fail to address numerous technical and sociotechnical problems, including goal drift, proxy gaming, and competitive pressures. Therefore, AI safety requires a more comprehensive, proactive, and nuanced approach than simply devising a list of rules for AIs to adhere to.

    \item \textbf{If AIs become more intelligent than people, wouldn't they be wiser and more moral? That would mean they would not aim to harm us.}

    The idea of AIs becoming inherently more moral as they increase in intelligence is an intriguing concept, but rests on uncertain assumptions that can't guarantee our safety. Firstly, it assumes that moral claims can be true or false and their correctness can be discovered through reason. Secondly, it assumes that the moral claims that are really true would be beneficial for humans if AIs apply them. Thirdly, it assumes that AIs that know about morality will choose to make their decisions based on morality and not based on other considerations. An insightful parallel can be drawn to human sociopaths, who, despite their intelligence and moral awareness, do not necessarily exhibit moral inclinations or actions. This comparison illustrates that knowledge of morality does not always lead to moral behavior. Thus, while some of the above assumptions may be true, betting the future of humanity on the claim that all of them are true would be unwise.
    
    Assuming AIs could indeed deduce a moral code, its compatibility with human safety and wellbeing is not guaranteed. For example, AIs whose moral code is to maximize wellbeing for all life might seem good for humans at first. However, they might eventually decide that humans are costly and could be replaced with AIs that experience positive wellbeing more efficiently. AIs whose moral code is not to kill anyone would not necessarily prioritize human wellbeing or happiness, so our lives may not necessarily improve if the world begins to be increasingly shaped by and for AIs. Even AIs whose moral code is to improve the wellbeing of the worst-off in society might eventually exclude humans from the social contract, similar to how many humans view livestock. Finally, even if AIs discover a moral code that is favorable to humans, they may not act on it due to potential conflicts between moral and selfish motivations. Therefore, the moral progression of AIs is not inherently tied to human safety or prosperity.

    \item \textbf{Wouldn't aligning AI systems with current values perpetuate existing moral failures?}
    
    There are plenty of moral failures in society today that we would not want powerful AI systems to perpetuate into the future. If the ancient Greeks had built powerful AI systems, they might have imbued them with many values that people today would find unethical. However, this concern should not prevent us from developing methods to control AI systems.
    
    To achieve any value in the future, life needs to exist in the first place. Losing control over advanced AIs could constitute an existential catastrophe. Thus, uncertainty over what ethics to embed in AIs is not in tension with whether to make AIs safe.
    
    To accommodate moral uncertainty, we should deliberately build AI systems that are adaptive and responsive to evolving moral views. As we identify moral mistakes and improve our ethical understanding, the goals we give to AIs should change accordingly---though allowing AI goals to drift unintentionally would be a serious mistake. AIs could also help us better live by our values. For individuals, AIs could help people have more informed preferences by providing them with ideal advice \citep{giubilini_artificial_2018}.
    
    Separately, in designing AI systems, we should recognize the fact of reasonable pluralism, which acknowledges that reasonable people can have genuine disagreements about moral issues due to their different experiences and beliefs \citep{rawls1993political}. Thus, AI systems should be built to respect a diverse plurality of human values, perhaps by using democratic processes and theories of moral uncertainty. Just as people today convene to deliberate on disagreements and make consensus decisions, AIs could emulate a parliament representing different stakeholders, drawing on different moral views to make real-time decisions \citep{Newberry2021ThePA,Hendrycks2023NaturalSF}. It is crucial that we deliberately design AI systems to account for safety, adaptivity, stakeholders with different values. 

    \item \textbf{Wouldn't the potential benefits that AIs could bring justify the risks?}

    The potential benefits of AI could justify the risks if the risks were negligible. However, the chance of existential risk from AI is too high for it to be prudent to rapidly develop AI. Since extinction is forever, a far more cautious approach is required. This is not like weighing the risks of a new drug against its potential side effects, as the risks are not localized but global. Rather, a more prudent approach is to develop AI slowly and carefully such that existential risks are reduced to a negligible level (e.g., under 0.001\% per century).
    
    Some influential technology leaders are accelerationists and argue for rapid AI development to barrel ahead toward a technological utopia. This techno-utopian viewpoint sees AI as the next step down a predestined path toward unlocking humanity's cosmic endowment. However, the logic of this viewpoint collapses on itself when engaged on its own terms. If one is concerned with the cosmic stakes of developing AI, we can see that even then it's prudent to bring existential risk to a negligible level. The techno-utopians suggest that delaying AI costs humanity access to a new galaxy each year, but if we go extinct, we could lose the cosmos. Thus, the prudent path is to delay and safely prolong AI development, prioritizing risk reduction over acceleration, despite the allure of potential benefits.

    \item \textbf{Wouldn't increasing attention on catastrophic risks from AIs drown out today's urgent risks from AIs?}

    Focusing on catastrophic risks from AIs doesn't mean ignoring today's urgent risks; both can be addressed simultaneously, just as we can concurrently conduct research on various different diseases or prioritize mitigating risks from climate change and nuclear warfare at once. Additionally, current risks from AI are also intrinsically related to potential future catastrophic risks, so tackling both is beneficial. For example, extreme inequality can be exacerbated by AI technologies that disproportionately benefit the wealthy, while mass surveillance using AI could eventually facilitate unshakeable totalitarianism and lock-in. This demonstrates the interconnected nature of immediate concerns and long-term risks, emphasizing the importance of addressing both categories thoughtfully.
    
    Additionally, it's crucial to address potential risks early in system development. As illustrated by Frola and Miller in their report for the Department of Defense, approximately 75 percent of the most critical decisions impacting a system's safety occur early in its development \cite{Frola1984SystemSI}. Ignoring safety considerations in the early stages often results in unsafe design choices that are highly integrated into the system, leading to higher costs or infeasibility of retrofitting safety solutions later. Hence, it is advantageous to start addressing potential risks early, regardless of their perceived urgency.

    \item \textbf{Aren't many AI researchers working on making AIs safe?}

    Few researchers are working to make AI safer. Currently, approximately 2 percent of papers published at top machine learning venues are safety-relevant \citep{eto_two_percent}. Most of the other 98 percent focus on building more powerful AI systems more quickly. This disparity underscores the need for more balanced efforts. However, the proportion of researchers alone doesn't equate to overall safety. AI safety is a sociotechnical problem, not just a technical problem. Thus, it requires more than just technical research. Comfort should stem from rendering catastrophic AI risks negligible, not merely from the proportion of researchers working on making AIs safe.

    \item \textbf{Since it takes thousands of years to produce meaningful changes, why do we have to worry about evolution being a driving force in AI development?}

    Although the biological evolution of humans is slow, the evolution of other organisms, such as fruit flies or bacteria, can be extremely quick, demonstrating the diverse time scales at which evolution operates. The same rapid evolutionary changes can be observed in non-biological structures like software, which evolve much faster than biological entities. Likewise, one could expect AIs to evolve very quickly as well. The rate of AI evolution may be propelled by intense competition, high variation due to diverse forms of AIs and goals given to them, and the ability of AIs to rapidly adapt. Consequently, intense evolutionary pressures may be a driving force in the development of AIs.

    \item \textbf{Wouldn't AIs need to have a power-seeking drive to pose a serious risk?}
    
    While power-seeking AI poses a risk, it is not the only scenario that could potentially lead to catastrophe. Malicious or reckless use of AIs can be equally damaging without the AI itself seeking power. Additionally, AIs might engage in harmful actions through proxy gaming or goal drift without intentionally seeking power. Furthermore, society's trend toward automation, driven by competitive pressures, is gradually increasing the influence of AIs over humans. Hence, the risk does not solely stem from AIs seizing power, but also from humans ceding power to AIs.

    \item \textbf{Isn't the combination of human intelligence and AI superior to AI alone, so that there is no need to worry about unemployment or humans becoming irrelevant?}

    While it's true that human-computer teams have outperformed computers alone in the past, these have been temporary phenomena. For example, ``cyborg chess'' is a form of chess where humans and computers work together, which was historically superior to humans or computers alone. However, advancements in computer chess algorithms have eroded the advantage of human-computer teams to such an extent that there is arguably no longer any advantage compared to computers alone. To take a simpler example, no one would pit a human against a simple calculator for long division. A similar progression may occur with AIs. There may be an interim phase where humans and AIs can work together effectively, but the trend suggests that AIs alone could eventually outperform humans in various tasks while no longer benefiting from human assistance.

    \item \textbf{The development of AI seems unstoppable. Wouldn't slowing it down dramatically or stopping it require something like an invasive global surveillance regime?}

    AI development primarily relies on high-end chips called GPUs, which can be feasibly monitored and tracked, much like uranium. Additionally, the computational and financial investments required to develop frontier AIs are growing exponentially, resulting in a small number of actors who are capable of acquiring enough GPUs to develop them. Therefore, managing AI growth doesn't necessarily require invasive global surveillance, but rather a systematic tracking of high-end GPU usage.

\end{enumerate}













\end{document}